\documentstyle[12pt,rotate]{article}

\input epsf.sty

\newcommand{\insertfig}[2]{\mbox{\epsfxsize=#1cm \epsfbox{#2.eps}}}

\newcommand{\ri}{{\rm i}}
\newcommand{\re}{{\rm e}}
\newcommand{\g}{{\sl g}}

\newcommand{\cD}{{\cal D}}
\newcommand{\cK}{{\cal K}}
\newcommand{\cG}{{\cal G}}

\newcommand{\cI}{{\cal I}}
\newcommand{\cO}{{\cal O}}

\newcommand{\cM}{{\cal M}}
\newcommand{\cP}{{\cal P}}
\newcommand{\cN}{{\cal N}}
\newcommand{\cR}{{\cal R}}
\newcommand{\cQ}{{\cal Q}}

\newcommand{\OO}{\mathop{\otimes}}
\font\cmss=cmss12 
\def\1{\hbox{{1}\kern-.25em\hbox{l}}}
\def\bfZ{\relax{\hbox{\cmss Z\kern-.4em Z}}}
\newcommand{\qqv}{{^{QQ}\! f}}
\newcommand{\qqbv}{{^{QQ}\! \overline{f}}}
\newcommand{\HQQ}{{^{QQ}\! h}}
\newcommand{\bHQQ}{{^{QQ}\! \overline{h}}}

\newcommand{\qgv}{{^{QG}\! f^A}}
\newcommand{\qgbv}{{^{QG}\! \overline{f}^A}}
\newcommand{\qgvv}{{^{QG}\! f^V}}
\newcommand{\qgbvv}{{^{QG}\! \overline{f}^V}}
\newcommand{\HQGo}{{^{QG}\! h}^A}
\newcommand{\bHQGo}{{^{QG}\! \overline{h}^A}}
\newcommand{\HQGe}{{^{QG}\! h}^V}
\newcommand{\bHQGe}{{^{QG}\! \overline{h}^V}}

\newcommand{\gqv}{{^{GQ}\! f^A}}
\newcommand{\gqbv}{{^{GQ}\! \overline{f}^A}}
\newcommand{\gqvv}{{^{GQ}\! f^V}}
\newcommand{\gqbvv}{{^{GQ}\! \overline{f}^V}}
\newcommand{\HGQo}{{^{GQ}\! h^A}}
\newcommand{\bHGQo}{{^{GQ}\! \overline{h}^A}}
\newcommand{\HGQe}{{^{GQ}\! h^V}}
\newcommand{\bHGQe}{{^{GQ}\! \overline{h}^V}}

\newcommand{\ggv}{{^{GG}\! f^A}}
\newcommand{\ggbv}{{^{GG}\! \overline{f}^A}}
\newcommand{\ggvv}{{^{GG}\! f^V}}
\newcommand{\ggbvv}{{^{GG}\! \overline{f}^V}}
\newcommand{\ggvc}{{^{GG}\! f^c}}

\newcommand{\HGGo}{{^{GG}\! h^A}}
\newcommand{\bHGGo}{{^{GG}\! \overline{h}}^A}
\newcommand{\HGGe}{{^{GG}\! h^V}}
\newcommand{\bHGGe}{{^{GG}\! \overline{h}}^V}

\newcommand{\beq}{\begin{equation}}
\newcommand{\eeq}{\end{equation}}
\newcommand{\bea}{\begin{eqnarray}}
\newcommand{\eea}{\end{eqnarray}}

\newcommand{\bz}{\bar z}
\newcommand{\be}{\bar \zeta}
\newcommand{\zoe}{\frac{z}{\zeta}}
\newcommand{\bzoe}{\frac{\bar z}{\bar \zeta}}

\newcommand{\emze}{\left ( 1-\frac{z}{\zeta}\right )}
\newcommand{\embze}{\left ( 1-\frac{\bar z}{\bar \zeta}\right )}
\newcommand{\lozembze}{\ln^2 \left ( 1-\frac{\bar z}{\bar \zeta}\right )}
\newcommand{\loembze}{\ln \left ( 1-\frac{\bar z}{\bar \zeta}\right )}
\newcommand{\lozme}{\ln (z-\zeta)}

\newcommand{\loe}{\ln(\bar \zeta)}
\newcommand{\lozz}{\ln^2(z)}
\newcommand{\loz}{\ln(z)}
\newcommand{\lobz}{\ln(\bar z)}
\newcommand{\lozbz}{\ln^2(\bar z)}
\newcommand{\lobzoz}{\ln\left(\frac{\bar z}{z}\right)}
\newcommand{\lozbzoz}{\ln^2\left(\frac{\bar z}{z}\right)}
\newcommand{\lobzobe}{\ln\left(\frac{\bar z}{\bar \zeta}\right)}
\newcommand{\lozbzobe}{\ln^2\left(\frac{\bar z}{\bar \zeta}\right)}

\newcommand{\pqgpo}{{^{QG}p^A}}
\newcommand{\pgqpo}{{^{GQ}p^A}}
\newcommand{\pggpo}{{^{GG}p^A}}

\newcommand{\pqgpe}{{^{QG}p^V}}
\newcommand{\pgqpe}{{^{GQ}p^V}}
\newcommand{\pggpe}{{^{GG}p^V}}
\newcommand{\pgga}{{^{GG}p^a}}
\newcommand{\pqgc}{{^{QG}p^c}}
\newcommand{\pggc}{{^{GG}p^c}}

\begin{document}
\begin{titlepage}

\centerline{\large \bf Evolution kernels of skewed parton distributions:}
\centerline{\large \bf method and two-loop results.}

\vspace{12mm}

\centerline{\bf A.V. Belitsky}

\vspace{3mm}

\centerline{\it C.N.\ Yang Institute for Theoretical Physics}
\centerline{\it State University of New York at Stony Brook}
\centerline{\it NY 11794-3800, Stony Brook, USA}

\vspace{7mm}

\centerline{\bf A. Freund}

\vspace{3mm}

\centerline{\it INFN, Sezione di Firenze, Largo E. Fermi 2}
\centerline{\it 50125 Firenze, Italy}

\vspace{7mm}

\centerline{\bf D. M\"uller}

\vspace{3mm}

\centerline{\it Institute f\"ur Theoretische Physik,
                Universit\"at Regensburg}
\centerline{\it D-93040 Regensburg, Germany}

\vspace{20mm}

\centerline{\bf Abstract}

\hspace{0.5cm}

We present a formalism and explicit results for two-loop flavor singlet
evolution kernels of skewed parton distributions in the minimal subtraction
scheme. This approach avoids explicit multiloop calculations in QCD
and is based on the known pattern of conformal symmetry breaking in this
scheme as well as constraints arising from the graded algebra of the $\cN =
1$ super Yang-Mills theory. The conformal symmetry breaking part of the
kernels is deduced from commutator relations between scale and special
conformal anomalies while the symmetric piece is recovered from the
next-to-leading order splitting functions and $\cN = 1$ supersymmetry
relations.

\vspace{0.8cm}


\vspace{0.5cm}


\end{titlepage}

\section{Introduction}

A main source of information about the structure of matter is gained from
reactions involving hadrons, with the theoretical description being most
reliable for processes with one or more hard scales as for example the
photon momentum transfer square, $Q^2$, in $ep$ reactions. The lowest
approximation for a process, when only the
Fock components with the minimal number of constituents in the hadron wave
function at tree level are accounted for, corresponds to a simple picture
of Feynman's parton model, which neglects QCD dynamics. Once this is taken
into account, it changes (sometimes drastically) the results. The specific
predictions of QCD are given by scaling violations phenomena which arise due
to two sources which play a complementary role depending on the magnitude of
the hard momentum transfer. For moderately large $Q^2$ it is enough to
include higher order perturbative corrections which scale as $\log Q^2$. On
the other hand approaching the low $Q^2 \sim {\rm few}\, {\rm GeV}^2$ domain
one has to take into account contributions relatively suppressed by powers
of the momentum transfer $1/Q^{\tau - 2}$, with $\tau$ being the twist of
the operators contributing to the amplitude of the process. Obviously, in
this region this latter $Q^2$ behavior overwhelms the weaker $\log Q^2$ due
to perturbative evolution. However, even at rather small $Q^2$ the
perturbative dependence on $Q^2$ cannot be discarded: being small at
large $Q^2$ it becomes prominent at small $Q^2$ and modifies in a
significant way the shape of the leading parton configuration of the
hadronic wave function. Therefore, the knowledge of perturbative evolution
is indispensable for the construction of the leading twist component from
the genuinely non-perturbative distribution at low scales. Although this
evolution can be weaker than the real non-perturbative evolution --- which
has to be used at very low $Q^2 \sim m^2_{\rm hadron}$ --- it gives the
right direction of change and can be seen as a builder of the rough features
of hadronic distributions or amplitudes. The same applies for higher twists,
but these contributions are usually discarded.

In the present study we concentrate on the first source of scaling
violation, namely, focus on the description of the perturbative
evolution in the first two orders of the perturbation series in the
QCD coupling at leading twist level, for the so-called skewed parton
distributions (SPD) which arise in a number of processes such as
deeply virtual Compton scattering \cite{MueRobGeyDitHor94,Ji96,Rad96},
electroproduction of mesons \cite{ColFraStr96,Rad96} etc. The
arguments in \cite{Rad97,ColFraStr96,ColFre98,JiOsb98} convince us
that these processes indeed factorize at this level into a perturbatively
calculable hard scattering amplitude and a SPD. At twist two level, the
dominant parton configuration in a hadronic wave function is that of two
quarks or gluons. The main difference in the partonic description of these
processes as compared to conventional deep inelastic scattering
(DIS) is due to the fact that the longitudinal momentum fractions of
particles propagating in the $t$-channel are different from one another
and this difference is called skewedness $\eta$.

Unfortunately, a direct extraction of the SPDs from experimental data will
hardly be possible in general, since the processes are predicted as
a convolution even at leading order (LO). Therefore, to deduce them one has
to confront data with different models which are usually given at a much
lower scale than factorization is expected to work and data are taken in
experiments. Moreover, one of the central issues in the future should be to
test the scaling violation phenomena in the cross section via evolution,
similar to DIS. Thus, the main task is to describe evolution effects of
SPDs as precise as possible. This includes the understanding of perturbative
corrections at LO and beyond.

The SPD is defined as a Fourier transform to the momentum fraction space of
a light-ray operator constructed from $\varphi$-parton fields and sandwiched
between hadronic states non-diagonal in momenta, schematically given by
\begin{equation}
\label{DefSPD}
\phi (x, \eta, Q) = \frac{1}{2\pi} \int d z_- e^{i x z_-}
\langle h (p') | \cO (z_-, Q) | h (p) \rangle
\quad\mbox{with}\quad
\cO (z_-, Q)
=
\left. \varphi^\dagger ( 0 ) \varphi ( z_- ) \right|_{Q} .
\end{equation}
Such a light-cone operator, $\cO (z_-, Q)$, is a formal resummation of
the usual local ones with definite twist. Its scale dependence is
governed by a renormalization group equation (RGE), but with anomalous
dimensions replaced by integral kernels depending on the positions
on the light cone. Therefore, a SPD obeys an evolution equation with
kernels given by integral transformations of the kernel in the
light-cone position representation. However, the generalized skewed
kinematics for the perturbative kernels can unambiguously be restored
\cite{GeyDitHorMueRob88} from the conventional exclusive one, known as
Efremov-Radyushkin-Brodsky-Lepage (ER-BL) region $\eta = 1$. Throughout
the paper we formally deal with ER-BL type equations
\cite{EfrRad78,BroLep79}
\begin{equation}
\label{ER-BLequation}
\frac{d}{d \ln Q^2} \mbox{\boldmath$\phi$} (x, Q) = \int_{0}^{1} dy
\mbox{\boldmath$V$} \left(x, y | \alpha_s(Q) \right)
 \mbox{\boldmath$\phi$} (y, Q) ,
\end{equation}
where $\mbox{\boldmath$\phi$} = { {^Q\phi} \choose {^G\phi} }$ is the
two-dimensional vector of the quark and gluon distribution amplitudes which
mix under renormalization and $\mbox{\boldmath$V$} (x, y | \alpha_s )$
for vector or axial-vector distribution amplitudes is a $2 \times 2$-matrix
of evolution kernels given by a series in the coupling:
\begin{equation}
\mbox{\boldmath$V$} (x, y | \alpha_s )
= \sum_{\ell = 1}^{\infty}
\left( \frac{\alpha_s}{2\pi} \right)^\ell
\mbox{\boldmath$V$}^{(\ell)} (x, y) .
\end{equation}

Until recently, only one-loop exclusive evolution kernels were available
\cite{EfrRad78,BroLep79,Cha80a,Ohr81,BaiGro81,BukFroKurLip85} and apart form
the nonsinglet two-loop result \cite{Sar84,DitRad84,MikRad85} nothing has
been known about higher order kernels. In our previous studies we have been
able to calculate the two-loop anomalous dimensions matrices of the so-called
conformal operators which are in one-to-one correspondence with the given
distributions and finally obtain the singlet kernels in the form of exclusive
convolutions. In the present study we conclude with giving, in great
detail, the formalism we have used as well as presenting explicit results for
the next-to-leading (NLO) kernels. As has been said above, a simple
continuation \cite{GeyDitHorMueRob88} allows one to easily deduce the
generalized non-forward evolution kernels from the ER-BL ones.

Different methods have been offered so far to solve the nonforward evolution
equation. Two of them rely on direct numerical integration
\cite{FraFreGuzStr97,MusRad99} of integro-differential equations. However,
the first of them was designed for the treatment of the
Dokshitzer-Gribov-Lipatov-Altarelli-Parisi (DGLAP) region only, while the
second one does not resum the $\ln Q^2$ terms. Next, one can
reconstruct a SPD from its expansion w.r.t.\ an appropriate basis of
polynomials \cite{Beletal97,ManPilWei97}, with the expansion coefficients
expressed in terms of conformal moments, which do not mix under
renormalization in the one-loop approximation. This latter method can be
employed to describe the evolution for general kinematics, however, it
becomes inefficient in the small $\eta$ region and in the $t \sim \eta$
domain where SPDs rapidly change their shape. Note that the accuracy of this
method is under control, however, the summation of a sufficient number of
polynomials is time consuming. Another possibility is to use the conformal
covariance in order to map the SPD to a forward distribution \cite{Shu99},
which has in general no physical meaning, and then solve the corresponding
DGLAP evolution equations with one of the standard methods. Yet another way
is to solve the evolution equation in configuration space in terms of
non-local conformal operators \cite{BalBra89,KivMan99}, however, in this
case, to evolve a given function explicitly the authors have used the same
method of orthogonal polynomial reconstruction as used by us in previous
studies.

Beyond LO all methods except for the first two can be used in a
straightforward manner only in an unrealistic conformal limit of QCD when
the Gell-Man--Low $\beta$-function is set equal to zero and by making use
of the conformal subtraction (CS) scheme \cite{Mue97a,BelMue97a} which
removes the special conformal symmetry breaking anomaly appearing in
the minimal subtraction scheme, or they can be approximately applied for
small skewedness, when the conformal non-covariant piece will die out.
Thus, in order to be able to describe a realistic situation, we are
left with a direct numerical integration of the evolution equations
or with the orthogonal polynomial reconstruction method. Since all
anomalous dimensions in NLO of conformal operators have been available
for some time \cite{BelMue98a,Mue94,BelMue98c}, we studied the
evolution of several models of SPDs and have gained a first insight
into the magnitude of evolution effects beyond LO for both the flavor
non-singlet \cite{BelMueNieSch98a} and singlet sector \cite{BelMueNieSch98b}.
However, due to complications as mentioned above, one should develop a more
efficient numerical treatment which can be achieved by direct numerical
integration routines. Therefore, the corresponding evolution kernels,
whose Gegenbauer moments define the anomalous dimensions
\cite{BelMue98a,Mue94,BelMue98c} mentioned earlier, are needed in
two-loop approximation.

The analytical structure of these kernels is expected to be more complex
than for the DGLAP ones. Therefore, a direct diagrammatical calculation in
NLO is envisaged to be very cumbersome. Although, there is experience in the
calculation of the ER-BL kernel in the flavor non-singlet sector at
two-loop order \cite{Sar84,DitRad84,MikRad85}, no appropriate technology
has been developed so far for the remaining calculations. A technical
analysis of the mathematical structure, which appears to be completely
analogous in all channels, is given in \cite{MikRad85,MikRad86}. Some
particularly simple parts, i.e. renormalon chain contributions, of the
non-singlet kernels were analyzed anew in Ref.\ \cite{Mik97}.

The goal of this paper is to present the method we employ for the
reconstruction of the evolution kernels in NLO. It is based on the
analysis of general properties of the kernels and known Gegenbauer
moments. Assuming the CS scheme, which implies that the (local) conformal
twist-two operators do not mix under renormalization\footnote{If we
neglect non-covariant terms proportional to the Gell-Mann--Low
$\beta$-function which are induced by the trace anomaly, the conformal
covariance can always be ensured by appropriate normalization conditions
for the renormalized operators.}, the anomalous dimensions are
the same (up to a normalization factor) as in the forward case for DIS.
This relation implies a one-to-one mapping between the evolution kernels
of SPDs and the DGLAP-kernels. In the conformal limit of this special
scheme, the reconstruction of the ER-BL kernels is reduced to an integral
transformation from the known DGLAP-kernels \cite{BelMue98a}. The
realistic situation is not as simple as that and beyond LO conformal
covariance is broken in the conventional scheme, i.e.\ using
dimensional regularization and (modified) minimal subtraction
($\overline{\rm MS}$), but it presents only a slight complication
as compared to the restoration of the conformal covariant part of the
kernels from the splitting functions.

The outline of the paper is the following. We collect the properties of the
ER-BL kernels in section \ref{sec-Anatomy}: their support as well as
relations to the DGLAP kernels, consequences of conformal constraints in QCD
and conformal symmetry in ${\cal N}=1$ supersymmetric Yang-Mills (SYM)
theory. We analyze the implications of these results for the structure of
the flavor non-singlet ER-BL kernel in the two-loop approximation and extend
it to all other channels. In section \ref{sec-ReconS} we reconstruct first
the whole crossed-ladder diagram contributions by means of conformal and
supersymmetric constraints from the known non-singlet results, then we
restore the remaining diagonal pieces making use of the known splitting
functions and present the results for the evolution kernels in the chiral
even sector as convolutions of simple kernels having a one-loop structure.
In section \ref{sec-rep} we perform the convolutions and present our results
in an explicit form for the ER-BL type representation and extract from them
the so-called skewed DGLAP kernels, which are needed for the numerical
treatment of evolution in the DGLAP region. In section \ref{sec-Con} we
give a summary and conclusions. Different appendices are attached to give
technical details.

\section{Structure of evolution kernels.}
\label{sec-Anatomy}

In the following four subsections we give a survey of general properties
of the twist-two singlet evolution kernels. Their support properties are
pointed out in the next subsection \ref{subsec-support}, which contains a
generalization of the results obtained by Geyer, Robaschik, and
collaborators in the past for the flavor non-singlet case
\cite{MueRobGeyDitHor94,GeyDitHorMueRob88,GeyRobBorHor85,BraGeyHorRob87}.
Let us now give a few results which will be used later on:
(i) The non-forward evolution kernels
of SPDs are uniquely defined by an extension procedure from the ER-BL
kernels. (ii) A simple limiting procedure reduces these generalized ER-BL
kernels to the DGLAP kernels. In subsection \ref{subsec-conformal} we
demonstrate that if conformal symmetry holds true in perturbative QCD, the
ER-BL kernels can be deduced from the DGLAP kernels by an integral
transformation with a well-defined resolvent. We then discuss the breaking
of conformal symmetry in the $\overline{\rm MS}$ scheme and explain the
third important issue, namely, (iii) how one part of the total evolution
kernel is induced by special conformal anomalies. It is shown how to
evaluate these anomalies in the ER-BL representation. In subsection
\ref{subsec-SUSY}, we explain the use of the ${\cal N} = 1$ SYM theory as a
meaningful tool for the construction of the ER-BL kernels. Based on the
results in the previous three subsections, we give a method for the
construction of the two-loop ER-BL kernels in subsection
\ref{subsec-reconstr} and demonstrate it in the flavor non-singlet case. The
main result of this section is a representation of the NLO kernels in terms
of convolutions of simple kernels possessing a one-loop structure, which
turns out to be valid in all channels and allows one to restore the missing
information from the known DGLAP kernels.

\subsection{Support properties of evolution kernels.}
\label{subsec-support}

In this subsection we recall the support properties of evolution kernels
for SPDs. Let us start with the definition of these distributions at
leading twist-two level in terms of light-ray operators that are defined
for the vector (V) and axial-vector (A) case by:
\begin{eqnarray}
\label{def-lrOpe-Q}
\left\{{ {^Q\!{\cal O}_i^V} \atop{^Q\!{\cal O}^A_i}}\right\}
(\kappa_1,\kappa_2)
&=& \bar \psi_i (\kappa_2 n)
\left\{ { \gamma_+ \atop \gamma_+ \gamma_5 } \right\}
\psi_i (\kappa_1 n),
\\
\label{def-lrOpe-G}
\left\{{ {^G\!{\cal O}^V} \atop{^G\!{\cal O}^A} }\right\}
(\kappa_1,\kappa_2)
&=&
G^a_{+ \mu} (\kappa_2 n)
\left\{ { g_{\mu\nu} \atop i \epsilon_{\mu\nu-+} } \right\}
G^a_{\nu+} (\kappa_1 n),
\end{eqnarray}
where for brevity we have omitted a path ordered link factor that ensures
gauge invariance. Here $n$ and $n^\ast$ are light-like vectors with $n
n^\ast = 1$ and the plus and minus components of a four-vector are obtained
by contraction with $n$ and $n^\ast$, respectively. Expansion of Eqs.\
(\ref{def-lrOpe-Q},\ref{def-lrOpe-G}) into a series of local operators
immediately guarantees the contribution from the leading twist-two only.
They have even chirality and even (odd) parity for the (axial-)vector  case.

The SPDs are Fourier transforms w.r.t.\ the light-like distance between
the fields of the light-ray operators (\ref{def-lrOpe-Q},\ref{def-lrOpe-G}),
which are sandwiched between off-diagonal
hadronic states:
\begin{eqnarray}
\label{def-SPD}
\left\{{ {^Q\!q_i^\Gamma} \atop {^G\!q^\Gamma}}\right\}(t,\eta;\mu)
= \left\{{1 \atop \frac{4}{P_+} }\right\}
\int \frac{d\kappa}{2\pi} e^{i \kappa t P_+}
\langle P_2 S_2 |
\left\{{ {^Q\!{\cal O}^\Gamma_i}
\atop
{^G\!{\cal O}^\Gamma} }\right\}(\kappa,-\kappa)
| P_1 S_1 \rangle_{\mu}
\end{eqnarray}
given at a renormalization point $\mu$. It can be shown that the support of
the SPDs is $|t| \le 1$. Here the conjugate variable $t$ plays the role of a
momentum fraction and the skewedness parameter $\eta =\Delta_+/P_+ $ is
defined as the $+$-component of the momentum transfer $\Delta = P_1 - P_2$
normalized to the $+$-component of $P_1 + P_2$, i.e.\ $P_+ = n.(P_1 + P_2)$.
More precisely, the momentum fractions of the incoming and outgoing partons
are $\frac{t + \eta}{1 + \eta} P_+$ and $\frac{t - \eta}{1 + \eta} P_+$,
respectively. Note that the SPDs also depend on the momentum transfer square
$\Delta^2$, which, however, is irrelevant for the evolution and thus will be
suppressed in what follows. A form factor decomposition would give us the
functions $H$, $E$ and $\widetilde H$, $\widetilde E$ for the parity even
and odd sectors, respectively, as introduced in ref.\ \cite{Ji96}, which are
governed by the same evolution equations as the distributions
(\ref{def-SPD}). However, we choose the normalization\footnote{For the
gluon distribution we have included an extra pre-factor of 2 in comparison
with the definition given in Ref.\ \cite{Ji96}.} in such a way that in the
forward limit ${^A\!q_i^\Gamma}(t = z)$ they coincide for $z \ge 0$ with
the parton densities ${q_i^\Gamma}(z)$ and ${z g^\Gamma(z)}$ for $A = Q, G$,
respectively, while $\mp{^Q\!q_i^\Gamma}(-z)$ with $z \ge 0$ are interpreted
as antiquark distributions.

The evolution equations for the SPDs (\ref{def-SPD}) arise from the
renormalization group equations (RGEs) of the light-ray operators
(\ref{def-lrOpe-Q}) and (\ref{def-lrOpe-G}). In the following we will
discuss the flavor singlet case, where the quark and gluon operators mix
with each other. For brevity we introduce the two dimensional vector
\begin{equation}
\mbox{\boldmath${\cal O}$}^\Gamma
\equiv \frac{1}{2} \sum_{i = u,d,s,\dots}
\left({
{^Q\!{\cal O}_i^\Gamma}(\kappa_1,\kappa_2) \mp
{^Q\!{\cal O}_i^\Gamma}(\kappa_2,\kappa_1)
\atop
\frac{2}{N_f} {^G\!{\cal O}^\Gamma}(\kappa_1,\kappa_2)
}\right)
\quad\mbox{with}\quad
\quad  \mp\mbox{\ for\ }
\left\{ { \Gamma = V\atop \Gamma = A} \right.
\end{equation}
and $N_f$ is the number of active quark flavors. Note that due to Bose
symmetry, the gluon operator also has definite symmetry with respect to
the interchange of $\kappa_1 \leftrightarrow \kappa_2$, i.e.\ it is
(anti)symmetric in the case of (axial-)vector couplings. The properties of
(the non-local version of) the anomalous dimensions were intensively studied
in the past on general grounds. For instance, the scaling and translation
properties of the operators tell us that the general form of the RGE reads
\cite{BraGeyHorRob87}:
\begin{eqnarray}
\label{def-RGE-LCP}
\mu \frac{d}{d\mu} \mbox{{\boldmath ${\cal O}$}} (\kappa_1, \kappa_2)
=
- \int dy \int dz\,
\mbox{{\boldmath $\gamma$}} (y, z; \kappa_2 - \kappa_1)\,
\mbox{{\boldmath ${\cal O}$}}
\left(
\kappa_1 [1 - y] + \kappa_2 y , \kappa_2 [1 - z] + \kappa_1 z
\right),
\end{eqnarray}
where the $2 \times 2$ matrix valued kernel $\mbox{{\boldmath $\gamma$}}$
explicitly depends on the light-cone position $\kappa_2 - \kappa_1$ in the
mixed channels:
\begin{eqnarray}
\label{def-gamma}
\mbox{{\boldmath $\gamma$}}(y,z;\kappa_2-\kappa_1)
= \left(
\begin{array}{cc}
{^{QQ}\!\gamma}(y,z) & \frac{\kappa_2-\kappa_1}{i}{^{QG}\!\gamma}(y,z)\\
\frac{i}{\kappa_2 - \kappa_1}{^{GQ}\!\gamma}(y,z)&{^{GG}\!\gamma}(y,z)
\end{array}
\right).
\end{eqnarray}

\begin{figure}[htb]
\unitlength1cm
\centering
\begin{picture}(15,7)(0,0)
\put(0,0.6){\insertfig{5}{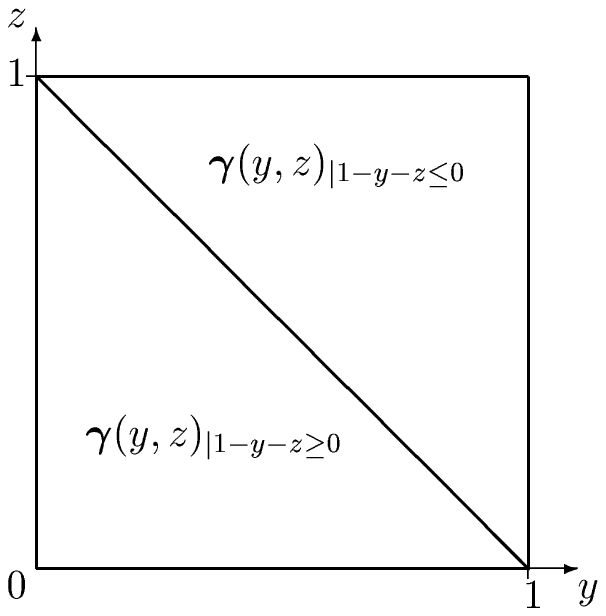}}
\put(8,0.6){\insertfig{6}{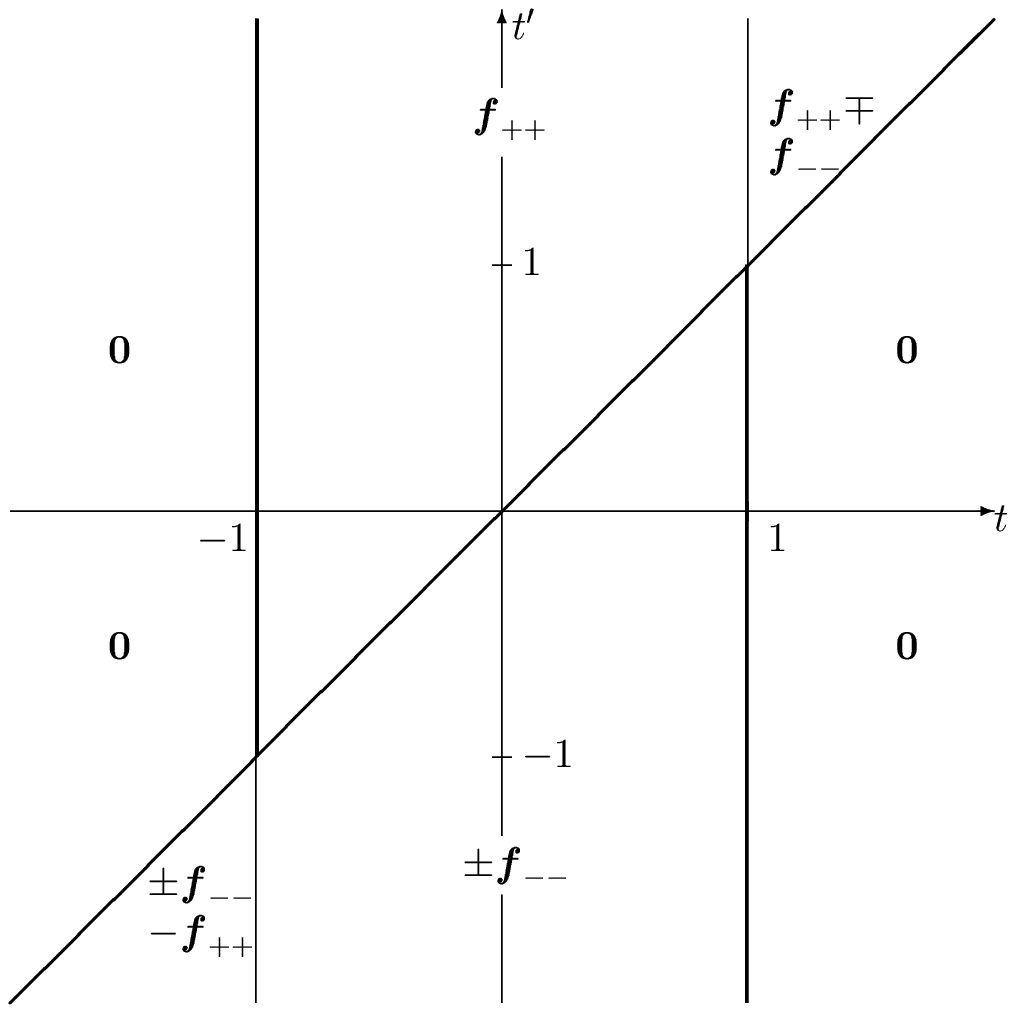}}
\put(2,0.2){(a)}
\put(10.8,0.2){(b)}
\end{picture}
\caption{Support  of the singlet anomalous dimensions
$\mbox{{\boldmath $\gamma$}}$ in light-cone position (a) and fraction (b)
representation. In (b) we only show the support which arises from $1-y-z \ge
0$ in the light-cone position representation. A second contribution that
comes from the region $1-y-z \le 0$ can be formally obtained by $t'\to -t'$,
however, in the flavor singlet case it can be reduced to the first one by
means of symmetry. Here the entries of $ \mbox{{\boldmath $f$}}_{\pm\pm} =
\mbox{{\boldmath $f$}}(\pm t,\pm t')$ are defined by Eqs. (\ref{INTRE-AA})
and (\ref{INTRE-AB}).}
\label{fig-sup-lr}
\end{figure}

The first important issue to understand is the anatomy of the evolution
kernels for SPDs, which arise from the support property of the kernels
{\boldmath $\gamma$}. This problem can be solved by means of the $\alpha$
or Feynman-parameter representation of Green functions with a non-local
operator insertion. It is sufficient to work in light-cone gauge and
to formally generalize the $\alpha$-representation for the gluon propagator
\cite{MueRobGeyDitHor94}. From these studies one can deduce the support of
the kernels shown in Fig.\ \ref{fig-sup-lr} (a):
\begin{eqnarray}
\mbox{{\boldmath ${\gamma}$}}(y, z; \kappa_2 - \kappa_1)
\neq 0, \ \ \mbox{for} \ \ 0 \leq y, z \leq 1; \ \ 0, \ \ \mbox{otherwise}.
\end{eqnarray}
Invariance under charge conjugation implies the following symmetry relation
\begin{eqnarray}
\mbox{{\boldmath ${\gamma}$}}(y,z;\kappa_2 - \kappa_1)
=
\mbox{{\boldmath ${\gamma}$}} (z,y;\kappa_2- \kappa_1).
\end{eqnarray}
It is also worth noting that the symmetry properties of the flavor singlet
operators w.r.t.\ the interchange of their light cone arguments, i.e.\
$\kappa_1 \leftrightarrow \kappa_2$, can be used to map the region $y + z
\ge 1$ into $1 \ge y + z$ by the substitution $y\to 1 - z$ and $z \to
1 - y$. Here the region $1 \ge y + z$ corresponds in the forward case to
quark-quark mixing as it is the case in LO, while $y + z \ge 1$ appears
due to a quark-antiquark interaction.

From the definition of a SPD (\ref{def-SPD}) and the RGE (\ref{def-RGE-LCP})
we easily derive the evolution equation:
\begin{eqnarray}
\label{def-EvoEqu-S}
\mu \frac{d}{d\mu} \mbox{{\boldmath $q$}} (t, \eta; \mu)
= - \int dt' \mbox{{\boldmath $\gamma$}}\!
\left( t,t',\eta;\alpha_s(\mu) \right)
\mbox{{\boldmath $q$}} (t', \eta; \mu).
\end{eqnarray}
Here we use a definition of the singlet distributions, which is motivated
by the one of normal parton densities,
\begin{eqnarray}
\label{def-SPD-sing}
\mbox{\boldmath $q$}^\Gamma (t, \eta; \mu)
\equiv \sum_{i = u, d, s, \dots}\left\{{
{^Q\!q_i^\Gamma (t, \eta; \mu)} \mp {^Q\!q_i^\Gamma(-t, \eta; \mu)}
\atop
\frac{1}{N_f} {^G\!q^\Gamma (t, \eta; \mu)}
}\right\}
\quad\mbox{with}\quad
\quad \mp\mbox{\ for\ }
\left\{ { \Gamma = V \atop \Gamma = A } \right.,
\end{eqnarray}
where the quark sector contains the sum of ``parton'' and ``anti-parton''
distributions. It is a simple exercise to see that the kernels for the
SPDs are obtained by the integration
\begin{eqnarray}
\label{conv-LCP->LCF}
\mbox{{\boldmath ${\gamma}$}} (t, t', \eta)
=
\int_0^1 dy \int_0^1 dz
\left(
\begin{array}{cc}
{^{QQ}\!\gamma}(y,z) &  {^{QG}\!\gamma}(y,z) d_t \\
{^{GQ}\!\gamma}(y,z) d_t^{- 1} & {^{GG}\!\gamma}(y,z)
\end{array}
\right)
\delta \left( t - t' (1 - y - z) - (y - z) \eta \right) ,
\end{eqnarray}
where $d_t \equiv d/dt$ and $d_t^{- 1} \equiv \int^t dt$. Here the
indefinite integration limits in the $GQ$ channel induces an ambiguity
which affects the unphysical moments only and has to be fixed by hand,
e.g.\ by comparison of moments calculated in both representations.
Note that this ambiguity implicitly appears also in the diagrammatic
calculation of Feynman diagrams in the light-cone fraction representation
and is responsible for different results given in the literature. It is
worth mentioning that the representation (\ref{conv-LCP->LCF}) implies
a simple scaling relation:
\begin{eqnarray}
\label{scaling-rel}
\mbox{{\boldmath ${\gamma}$}}(t,t',\eta)
=
\frac{1}{|\eta|} \left(
\begin{array}{cc}
{^{QQ}\!\gamma}\left(\frac{t}{\eta},\frac{t'}{\eta}\right)
&
\eta^{-1} {^{QG}\!\gamma}\left(\frac{t}{\eta},\frac{t'}{\eta}\right)\\
\eta{^{GQ}\!\gamma}\left(\frac{t}{\eta},\frac{t'}{\eta}\right)
&
{^{GG}\!\gamma}\left(\frac{t}{\eta},\frac{t'}{\eta}\right)
\end{array}
\right),
\end{eqnarray}
so that the entries are in fact two-variable functions of
ratios.

The invariance under charge conjugation implies now the symmetry for
diagonal ${^{AA}\!\gamma} (t, t') = {^{AA}\!\gamma} (-t, -t')$ and
off-diagonal $A \neq B$, ${^{AB}\!\gamma} (t, t') = -{^{AB}\!\gamma}
(-t, -t')$ elements, while the connection between the parton-parton
and parton-antiparton regions makes itself apparent in the substitution $t' \to - t'$.
As explained, for the flavor singlet case it is enough to consider the
region $0\le 1 - y - z$ of the support $0 \le y, z \le 1$. Then the
integral representation (\ref{conv-LCP->LCF}) implies the support shown
in Fig.\ \ref{fig-sup-lr} (b), or formally
\begin{equation}
\label{GAMTRE}
\mbox{{\boldmath $ \gamma$}} (t, t')
= \Theta (t, t') \mbox{{\boldmath $f$}} (t, t')
\pm \left\{ {t \to -t \atop t' \to -t'} \right\} ,
\qquad
\Theta (t,t')
= \theta (t' - t) \theta (1 + t) - \theta (t - t') \theta(- 1 - t),
\end{equation}
where $(-)+$ stands for (off-) diagonal entries. The entries of the
matrix $\mbox{{\boldmath $f$}}(t,t')$ are given by
\begin{eqnarray}
\label{INTRE-AA}
{^{AA}\!f}(t,t')
&=& \int_0^{1 + t \over 1 + t'} dw\, {^{AA}\!\gamma}\!
\left(
y = \frac{1 + t - (1 + t') w}{2}, z = \frac{1 - t - (1 - t') w}{2}
\right), \\
\label{INTRE-AB}
{^{AB}\!f}(t,t')
&=& \left\{ d_t \atop d_t^{- 1} \right\}
\int_0^{1 + t \over 1 + t'} dw\, {^{AB}\!\gamma}\!
\left(
y = \frac{1 + t - (1 + t') w}{2}, z = \frac{1 - t - (1 - t') w}{2}
\right) ,
\end{eqnarray}
with $d_t$ ($d_t^{- 1}$) corresponding to $QG$ ($GQ$).

A first glance at Fig.\ \ref{fig-sup-lr}, immediately gives the impression
that the whole kernel can be obtained from the region $|t|,|t'| \le 1$.
Indeed, it was proved in \cite{MueRobGeyDitHor94} that the continuation is
unique (see appendix \ref{app-extension}). For practical purposes it is
sufficient to replace the $\theta$ structure:
\begin{eqnarray}
\theta(t-t')_{|t|,|t'| \le 1} \to \Theta(t,t').
\end{eqnarray}
Thus, the evolution kernel for the corresponding SPD can be considered as
a generalized ER-BL kernel and its restoration from a given ER-BL kernel
is simple.

There are further consequences arising from the evolution equation
(\ref{def-EvoEqu-S}). If we replace in the definition (\ref{def-SPD}) for
SPDs the off-diagonal hadronic states by diagonal ones, we immediately obtain
the definition of the usual forward parton distributions (up to an
additional $1/z$ for the gluon density) with $t = z$ and $\eta = 0$.
Comparing the evolution equations, the DGLAP kernel appears as a limit of
the generalized ER-BL kernel:
\begin{equation}
\label{def-LIM}
\mbox{\boldmath{$P$}} (z)
= - \mbox{\boldmath{$\gamma$}} (t, t', \eta = 0)
= - {\rm LIM} \mbox{\boldmath{$\gamma$}} (t, t')
\equiv - \lim_{\eta\to 0} \frac{1}{|\eta|}
\left(
{\ \ {^{QQ}\!\gamma}\ \frac{1}{\eta}{^{QG}\!\gamma}
\atop
\frac{\eta}{z} {^{GQ}\!\gamma}\  \frac{1}{z}{^{GG}\!\gamma}}
\right)^{\rm ext}
\left( \frac{z}{\eta}, \frac{1}{\eta} \right).
\end{equation}
An alternative derivation of this limit using moments is
presented in appendix \ref{app-LIM}.

As we have just shown, it is sufficient for the following considerations
to work in the exclusive kinematics. Thus, from now on we will
deal only with the momentum fraction $x = (1 + t)/2$ and the common
definition of ER-BL type kernels:
\begin{equation}
\mbox{{\boldmath $ V$}} (x, y)
= - \mbox{{\boldmath $ \gamma$}}
( x - \bar x, y - \bar y )_{0 \le x,y \le 1},
\end{equation}
here and everywhere else we use $\bar x \equiv 1 - x$. The general structure
of the entries is
\begin{equation}
{^{AB} v}(x, y)
= \theta(y - x) {^{AB}\! f} (x, y)
+ \theta(\bar{y} - x) {^{AB}\! g} (x, y)
\pm \left\{ {x \to \bar x \atop y \to \bar y } \right\}
\quad
\mbox{for}
\quad
\left\{ {A = B \atop A \not = B } \right. .
\end{equation}
As discussed above, the second $\theta$-structure, i.e.\
$\theta(\bar{y} - x)$, in the singlet case, can be removed by the (anti-)
symmetry of the parton distributions w.r.t.\ $y\to\bar{y}$. Again, the
extension of the kernel in the whole region is done by a simple
replacement of $\theta$ functions, e.g.\
\begin{eqnarray}
\label{Extension}
\theta (y - x)
\to
\theta \left( 1 - \frac{x}{y} \right)
\theta \left( \frac{x}{y} \right)
\mbox{sign} (y).
\end{eqnarray}

\subsection{Conformal properties of evolution kernels.}
\label{subsec-conformal}

To understand the classification of different contributions to the
evolution kernel {\boldmath $V$} with respect to the conformal
transformation most clearly, we deal in this subsection with the
so-called conformal operators ${^I\!{\cal O}^\Gamma_{jl}}$ and their
anomalous dimension matrix {\boldmath $\gamma$}. These operators build an
infinite dimensional, irreducible representation of the collinear conformal
algebra ${\rm so} (2,1)$ in the space spanned by the bilinear field
operators. This algebra arises from the full conformal algebra ${\rm so}
(4, 2)$ by projection onto the light cone and consists of the generators
of dilatation $\cD$, special conformal transformation\footnote{The special
conformal transformation is given by the product $\cR \cP_c \cR$, where the
inversion $\cR$ acts as $\cR x_\mu = x_\mu/ x^2$ and $\cP_c$ is a
translation with a vector $c$.} $\cK_-$, boost along the light-cone
$\cM_{-+}$, and translation $\cP_+$. Conformal operators are generated from
the light-ray operators (\ref{def-lrOpe-Q}) and (\ref{def-lrOpe-G}) by
differentiation w.r.t.\ $\kappa_1, \kappa_2$:
\begin{equation}
\label{def-conOpe}
{^A\!{\cal O}^\Gamma_{jj}}
= \left(
i \partial_{\kappa_1} + i \partial_{\kappa_2}
\right)^{j + \nu(A) - 3/2}
C^{\nu(A)}_{j + \nu(A) - 3/2}
\!\left(\frac{
\partial_{\kappa_1} - \partial_{\kappa_2}}{
\partial_{\kappa_1} + \partial_{\kappa_2} }
\right)
{^A\!{\cal O}^\Gamma}
(\kappa_1,\kappa_2)_{| \kappa_1 = \kappa_2 = 0},
\end{equation}
where $\nu (Q) = 3/2$ and $\nu (G) = 5/2$. The index of the Gegenbauer
polynomial $C^{\nu}_j$ is determined by group theory: $2 \nu(A) = 2 d(A)
+ 2 s(A) - 1$, where $d(A)$ and $s(A)$ are the canonical dimension and the
spin of the field of species $A$. The operators in Eq.\ (\ref{def-conOpe})
are the highest weight vectors $\cK_- {^A\!{\cal O}^\Gamma_{jj}} = 0$ and
they carry the conformal spin $j + 1$ and the angular momentum $l + 1$.
Acting with the generator of translation, i.e.\ applying $\cP_+^{(j - l)}$
on ${^A\!{\cal O}^\Gamma_{jj}}$, we generate the whole conformal tower of
operators. In LO all members of a tower do not mix under renormalization,
i.e.\ they have the same anomalous dimension $\gamma_j$. This is a
consequence of classical conformal symmetry and arises from the commutator
constraints between the generators of dilatation $\cD$ and special conformal
transformation $\cK_-$: $\left[ \cD , \cK_- \right]_- = i \cK_-$, where
as we have established above $\cK_-$ acts in the conformal tower as a step
down operator.

Scale symmetry is known to be broken by the trace anomaly in the
energy momentum tensor \cite{AdlColDun77,ColDunJog77,Nie77} and
is proportional to the Gell-Mann--Low $\beta$-function. For the
Green functions involving only local field operators at different
space-time points there is a one-to-one correspondence between
breaking of special conformal and scale symmetries \cite{Symanzik71}.
However, as we will discuss in detail below, the renormalization
of the Green functions with composite operator insertions there is an
extra source of breaking of the special conformal covariance besides
terms proportional to $\beta$. Poincar\'e invariance implies that
the mixing matrix is triangular, thus, the RGE has the general form
\begin{eqnarray}
\mu \frac{d}{d\mu} \mbox{\boldmath ${\cal O}$}_{jl}
= - \sum_{k = 0}^{j}
\mbox{\boldmath $\gamma$}_{jk} \mbox{\boldmath ${\cal O}$}_{kl},
\qquad\mbox{with}\qquad
\mbox{\boldmath ${\cal O}$}_{jl}
= \sum_{i=u,d,s,\dots}
\left\{{^Q\!{\cal O}_{jl}^i}
\atop \frac{1}{N_f} {^G\!{\cal O}_{jl}}
\right\} .
\end{eqnarray}
To make contact with the (generalized) ER-BL kernel, we use the fact
that the conformal moments of the SPDs (\ref{def-SPD}) are given by the
expectation values of conformal operators (\ref{def-conOpe}). For the
flavor singlet case [see Eq.\ (\ref{def-SPD-sing})] we have
\begin{eqnarray}
\eta^{j + \nu(A) - 3/2} \int_{-1}^1 dt\,
C^{\nu(A)}_{j + \nu(A) - 3/2} \!\left( \frac{t}{\eta} \right)
{^A\!q^\Gamma}(t, \eta; \mu)
= \left( \frac{2}{P_+} \right)^{j + 1}
\langle P_2, S_2 | {^A\!{\cal O}}^\Gamma_{jj} | P_1, S_1 \rangle_{\mu}.
\end{eqnarray}
Moreover, since the Gegenbauer polynomials form a complete basis in the
region $[-1,1]$, we can represent the kernel in the ER-BL region by the
following sum
\begin{eqnarray}
\label{rep-v-GC-ND}
{^{AB}\!v^i} (x, y)
= \sum_{j = \nu(A)-3/2}^\infty \ \sum_{k = \nu(B)-3/2}^j
\frac{w \left( x | \nu(A) \right)}{N_j \left( \nu (A) \right)}
C^{\nu(A)}_{j + 3/2 - \nu(A)} (x - \bar x) {^{AB}\!v^i_{jk}} \,
C^{\nu(B)}_{k + 3/2 - \nu(B)} (y - \bar y),
\end{eqnarray}
where $w(x|\nu)=(x\bar{x})^{\nu - 1/2}$ is the weight function and
$N_j(\nu) = 2^{ - 4 \nu + 1 } \frac{ \Gamma^2 (\frac{1}{2}) \Gamma
( 2 \nu + j )}{\Gamma^2 (\nu) ( \nu + j ) j! }$ is the normalization
coefficient. Up to an overall normalization, the conformal moments are
identical with the anomalous dimensions, i.e. ${^{AB}\!v^i_{jk}} =
-{^{AB}\!\gamma^i_{jk}}/2$. Note that invariance under charge conjugation
implies  ${^{AB}\!v^i_{jk}} =0$ for odd $j-k$.

Fortunately, the additional special conformal symmetry breaking alluded to
above is scheme dependent and as mentioned in the introduction it can be
avoided in a special CS scheme. In such a CS scheme and in the formal
conformal limit of QCD, when we set the Gell-Mann--Low function to zero,
conformal operators do not mix with each other. However, once the
$\beta$-function is kept non-zero the RGE to all orders of
perturbation theory is now
\begin{eqnarray}
\mu \frac{d}{d\mu} \mbox{\boldmath $\widetilde\cO$}_{jl}
= - \mbox{\boldmath $\gamma$}_{jj} \mbox{\boldmath $\widetilde\cO$}_{jl}
- \frac{\beta}{\g} \sum_{k = 0}^{j - 2}
\mbox{\boldmath $\Delta$}_{jk} \mbox{\boldmath $\widetilde\cO$}_{kl}.
\end{eqnarray}
Thus, in the CS scheme and up to a term proportional to the $\beta$-function
the ER-BL kernel is diagonal w.r.t.\ Gegenbauer polynomials:
\begin{eqnarray}
\label{rep-v-GC-D}
{^{AB}\!v^i} (x, y)
= \sum_{j = 0,1}^\infty
\frac{w \left( x | \nu(A) \right)}{N_j \left( \nu (A) \right)}
C^{\nu(A)}_{j + 3/2 - \nu(A)} (x - \bar x) {^{AB}\!v^i_{j}} \,
C^{\nu(B)}_{j + 3/2 - \nu(B)} (y - \bar y),
\end{eqnarray}
where the sum starts at $j=0$ for $AB=QQ$ in the parity even case and
$j=1$ otherwise.
The eigenvalues ${^{AB}\!v^i_{j}}$ are given by the diagonal entries,
i.e.\ ${^{AB}\!v^i_{j}} \equiv {^{AB}\!v^i_{jj}}$. From the representation
(\ref{rep-v-GC-D}) it necessarily follows that the diagonal entries have
the symmetry properties
\begin{eqnarray}
\label{sym-AA}
(y \bar y)^{\nu(A)-1/2}\; {^{AA}\! v}^{\rm D} (x, y)
= (x \bar x)^{\nu(A)-1/2}\; {^{AA}\! v}^{\rm D} (y, x).
\end{eqnarray}
Taking into account the equalities among Gegenbauer polynomials with
shifted index $\nu$,
\begin{eqnarray}
\label{rel-GC}
\frac{d}{dx} C_j^{3/2}(x - \bar x)
= 6 C_{j-1}^{5/2}(x - \bar x),
\quad
\frac{d}{dx} \frac{w(x|5/2)}{N_j(5/2)} C_{j - 1}^{5/2}(x - \bar x)
= - 6 \frac{w(x | 3/2)}{N_j (3/2)} C_{j}^{3/2}(x - \bar x),
\end{eqnarray}
we can easily derive necessary conditions for the kernels in the mixed
channels, too:
\begin{eqnarray}
\label{sym-AB}
y \bar y \frac{\partial}{\partial x} {^{GQ}\! v}^{\rm D} (x, y)
=
x \bar x \frac{\partial}{\partial y} {^{GQ}\! v}^{\rm D} (y, x),
\quad
\frac{\partial}{\partial x} \frac{{^{QG}\! v}^{\rm D} (x, y)}{x \bar x}
=
\frac{\partial}{\partial y} \frac{{^{QG}\! v}^{\rm D} (y, x)}{y \bar y}.
\end{eqnarray}

Since the matrix of conformal moments $\mbox{\boldmath ${v}$}_{j}$ are
given by the forward anomalous dimensions known from DIS up to NLO,
\begin{eqnarray}
\label{con-v&gamma}
\mbox{\boldmath ${v}$}_{j}
= -\frac{1}{2}
\left(
{ {^{QQ}\!\gamma_j} \quad \frac{6}{j} {^{QG}\!\gamma_j}
\atop
\frac{j}{6} {^{GQ}\!\gamma_j} \quad  {^{GG}\!\gamma_j} }
\right) ,
\end{eqnarray}
we may reconstruct the kernels from these information with the help of
Eq.\ (\ref{rep-v-GC-D}). This representation as an infinite sum can be
converted into an integral by means of the generating function
\begin{eqnarray}
\label{DGLAPtoERBL-1}
G(x, y; z| \nu) &=&
\sum_{j = 0}^{\infty}
\frac{w \left( x | \nu \right)}{N_j \left(\nu \right)}
C^{\nu}_j(x-\bar x) z^j\, C^{\nu}_{j} (y-\bar y)
\nonumber\\
&=&
\frac{\Gamma(\nu)\Gamma(\nu + 1)}{\Gamma^2(\frac{1}{2})\Gamma(2\nu)}
\frac{2^{4\nu - 1}(x \bar x)^{\nu -1/2}(1 - z^2)}{
\left[ 1 - 2 \left( (x - \bar x)(y - \bar y)
- 4 \sqrt{x \bar x y \bar y} \right) z + z^2
\right]^{\nu + 1}}
\\
&&\times
{_2F_1}
\left( { \nu + 1, \nu \atop 2 \nu }
\left| \frac{16 \sqrt{x \bar x y \bar y} z}{1
- 2 \left( (x - \bar x)(y - \bar y)
- 4 \sqrt{x \bar x y \bar y} \right) z + z^2 } \right. \right),
\nonumber
\end{eqnarray}
which was obtained by Gegenbauer's summation theorem
\cite{BE53_1}. For completeness we give here a simple reduction
formula that is obtained from $G(x, y; z|3/2)$ together with Eq.\
(\ref{rel-GC}):
\begin{eqnarray}
\label{DGLAPtoERBL-2}
\mbox{\boldmath$V$}^{\rm D} (x,y) = \int_0^1 dz
\left(
{{^{QQ}\!V} \quad
\frac{1}{6}\frac{\partial}{\partial y}{^{QG}\!V}
\atop
-\frac{1}{6}\int^x_{-1} dx\, {^{GQ}\! V}
\quad -\frac{1}{36}\int^x_{-1} dx\, \frac{\partial}{\partial y} {^{GG}\!V}}
\right)\!(z)\, G(x, y; z|3/2),
\end{eqnarray}
where ${^{AB}\! V}(z)$ are related to the DGLAP kernels, however,
defined as a Mellin transformation of conformal moments:
\begin{eqnarray}
\label{DGLAPtoERBL-3}
\mbox{\boldmath $V$}(z)
= \frac{1}{2\pi i} \int_{c - i \infty}^{c + i \infty} dj\,
\mbox{\boldmath $v$}_{j} z^{- j + 1}.
\end{eqnarray}
Note that in the $GQ$ and $GG$ channel the first term of the expansion
w.r.t.\ Gegenbauer polynomials does not exist and must be removed by
hand. A different version of the reduction formula avoiding this problem
has been applied at LO for the restoration of the ER-BL kernels from the
DGLAP ones \cite{BelMue98a}. The evaluation of the integrals has been
reduced by the Cauchy theorem to the evaluation of residues of simple poles,
therefore, it was essential that the one-loop DGLAP
kernels do not generate cuts in the complex plane. Beyond LO this property
is lost and the derived reduction formula (\ref{DGLAPtoERBL-2})
is extremely hard to handle.

To calculate the complete ER-BL kernel in any scheme, e.g. in the
$\overline{\rm MS}$ scheme with dimensional regularization, we have to take
into account the breaking of conformal covariance induced by the conformal
anomaly in the action\footnote{Beside the trace anomaly there also appear
equation of motion terms and exact BRST operator insertions. In $4-2\epsilon$
dimensions the trace anomaly is proportional to $\beta_\epsilon(\epsilon,g)
= - g\epsilon + \beta(g)$ and we have to renormalize the product of trace
anomaly and composite operators. Thus, this procedure together with the
$\epsilon$ dependent term causes anomalous contributions, e.g.\ anomalous
dimensions in the case of dilatation.}. There are different ways to account
for this breaking:
\begin{itemize}
\item
An explicit calculation of the anomalous dimensions or kernels. However,
depending on the representation, the calculation and/or extraction of
the conformal non-covariant piece is extremely difficult or at least not
straightforward.

\item
Taking normalization conditions that ensure the conformal covariance of
the renormalized operator w.r.t.\ the special conformal transformation for
vanishing Gell-Man--Low function. For instance, doing so in LO, the anomalous
dimensions are diagonal in NLO and the result in the $\overline{\rm MS}$
scheme can be obtained by a finite renormalization extracted in LO.

\item
One can analyze the conformal symmetry breaking with the help of
conformal Ward identities (CWI) and derive constraints for the appearing
anomalies which in turn will fix the piece we are interested in. This
approach provides us also information on terms proportional to the
Gell-Man--Low function.
\end{itemize}

All of the approaches sketched above are equivalent in the formal conformal
limit. In order to reduce the effort in obtaining the desired result we
prefer to work with the third one. Moreover, it enables us to get a deeper
insight into the structure of conformal symmetry breaking counterterms. It
offers as well the technical tools to get the off-diagonal anomalous
dimensions and the corresponding kernels in $n^{\rm th}$ order by
calculating Feynman graphs at $(n - 1)^{\rm th}$ order. For the reader's
convenience we quickly outline this approach. The exact technical steps are
published in great detail elsewhere
\cite{BelMue97a,BelMue98a,Mue94,BelMue98c,Mue91a}.

The major steps consist of:\\
\noindent (i) Derivation of conformal Ward identities for the Green
functions with conformal operator insertion
\begin{equation}
\label{GreenFunction}
\mbox{\boldmath $\cG$}_{jl} (x_1, x_2, \dots, x_N)
= \langle
\mbox{\boldmath $\cO$}_{jl} (0) \phi (x_1) \phi (x_2) \dots \phi (x_N)
\rangle
\end{equation}
by means of the path integral using dimensional regularization. The
renormalization of the Ward identities provides us with a prescription
for the calculation of the dilatation and the special conformal
anomalies. Here we present these Ward identities in a simplified form
required just to demonstrate the way conformal anomalies appear
\begin{eqnarray}
\label{CWI-D}
&&\sum_{j = 1}^{N}
\left( - x_{\mu_j} \frac{\partial}{\partial x_{\mu_j}} - d_{\phi_j} \right)
\mbox{\boldmath $\cG$}_{jl} (x_1, x_2, \dots, x_N) \nonumber\\
&&\qquad\qquad\qquad\qquad\qquad =
\sum_{k = 0}^{j}
\left\{
(l + 3) \mbox{\boldmath $1$}
+ \mbox{\boldmath ${\cal \gamma}$}
\right\}_{jk}
\mbox{\boldmath $\cG$}_{kl} (x_1, x_2, \dots, x_N) + \dots , \\
\label{CWI-K}
&&\sum_{j = 1}^{N} n^\star_\mu
\left(
- 2 x_{\mu_j} x_{\nu_j} \frac{\partial}{\partial x_{\nu_j}}
+ x_{\nu_j}^2 \frac{\partial}{\partial x_{\mu_j}}
- 2 ( d_{\phi_j} + s_{\phi_j} ) x_{\mu_j}
\right)
\mbox{\boldmath $\cG$}_{jl} (x_1, x_2, \dots, x_N) \nonumber\\
&&\qquad\qquad\qquad\qquad\qquad =
\sum_{k = 0}^{j}
\left\{
a(j,l) \mbox{\boldmath $1$}
+ \mbox{\boldmath ${\cal \gamma}$}^c (l)
\right\}_{jk}
\mbox{\boldmath $\cG$}_{kl} (x_1, x_2, \dots, x_N) + \dots ,
\end{eqnarray}
where the anomalous dimensions matrix $\mbox{\boldmath ${\cal
\gamma}$}_{jk}$ and the so-called special conformal anomaly matrix
$\mbox{\boldmath ${\cal \gamma}$}^c_{jk}$ are induced by quantum
fluctuations. We introduced the conventions $a(j,l) = 2(j - l)(j + l + 3)$
for the coefficient which appears at tree level in the transformation of a
conformal operator under special conformal transformations. The dots in
Eqs.\ (\ref{CWI-D},\ref{CWI-K}) stand for terms involving Green functions
with the renormalized product of a conformal operator and a conformal
variation of the action $\langle [ \mbox{\boldmath $\cO$}_{jl} \delta S]
\phi (x_1) \phi (x_2) \dots \phi (x_N) \rangle$.

\noindent (ii) As a next step we derive matrix constraints for the conformal
anomalies and find their solution. It turns out that there are two
commutator relations stemming from the algebra of the collinear conformal
group which provide non-trivial relations between the conformal anomalies.
For vanishing Gell-Mann--Low function (complete constraints) we have
\begin{eqnarray}
\label{conf-constr-KD-1}
& &\left[ \cD , \cK_- \right]_- = i \cK_-
\hspace{2cm}\qquad\Rightarrow\qquad
\left[ \mbox{\boldmath{$a$}} (l)
+ \mbox{\boldmath{$\gamma$}}^c (l),
\mbox{\boldmath{$\gamma$}} \right]_- = 0,
\\
\label{conf-constr-KD-2}
& &\left[ \cP_+ , \cK_- \right]_- = 2i( \cD + \cM_{-+} )
\qquad\Rightarrow\qquad
\mbox{\boldmath{$\gamma$}}^c (l + 1)
- \mbox{\boldmath{$\gamma$}}^c (l)
=  - 2   \mbox{\boldmath{$\hat \gamma$}} .
\end{eqnarray}
Since the matrix $\mbox{\boldmath{$a$}}$ is diagonal,
$\mbox{\boldmath{$a$}}_{jk} = a(j, l) \delta_{jk}$, the constraint
(\ref{conf-constr-KD-1}) tells us that the off-diagonal matrix elements of
the anomalous dimension {\boldmath{$\gamma$}} are indeed induced by the
special conformal anomaly matrix $\mbox{\boldmath{$\hat \gamma$}}^c(l)$,
while the spin dependence of $\mbox{\boldmath{$\hat \gamma$}}^c(l)$,
induced by the breaking of Poincar\'e invariance due to the special
conformal transformation, is governed by Eq.\ (\ref{conf-constr-KD-2}).
Note that the diagonal form of the special conformal anomaly
$\mbox{\boldmath{$\hat\gamma$}}^c(l)$
is necessary and sufficient for the vanishing  of the
off-diagonal part of the anomalous dimension matrix
$\mbox{\boldmath{$\hat \gamma$}}$. This ensures the existence of the CS
scheme, in which the conformal covariance of the operators holds true in
the conformal limit, $\beta = 0$, at any order of perturbation theory
\cite{Mue97a}.

The extension of these constraints (\ref{conf-constr-KD-1}) to nonvanishing
$\beta$-function is straightforward, however, they require additional
algebra \cite{BelMue98a,Mue94,BelMue98c}, which results in the following
change of Eq.\ (\ref{conf-constr-KD-1})
\begin{equation}
\label{conf-constr-KD-full}
\left[  \mbox{\boldmath{$a$}} (l)
+ \mbox{\boldmath{$\gamma$}}^c (l)
+ 2 {\beta \over \g} \mbox{\boldmath{$b$}} (l) ,
\mbox{\boldmath{$\gamma$}} \right]_- = 0 ,
\end{equation}
where\footnote{ We use the following definition of the step-function:
$\theta_{jk} = \left\{ 1, \ \mbox{for} \ j - k \geq 0 ; 0, \ \mbox{for}
\ j - k < 0 \right\}$.} $\mbox{\boldmath $b$}_{jk} (l) = \theta_{jk}
\left\{ 2 (l + k + 3) \delta_{jk} - [1 + (-1)^{ j - k}] (2 k + 3)
\right\} \mbox{\boldmath $1$}$. Decomposing the anomalous dimension matrix
in its diagonal (D) and off-diagonal (ND) part, the solution of this
constraint can be constructed by successive approximations:
\begin{eqnarray}
\label{opco-di3}
\mbox{\boldmath{$\gamma$}}^{\rm ND}(g)
= - \frac{\mbox{\boldmath{$\cG$}}}{\mbox{\boldmath{$1$}}
+ \mbox{\boldmath{$\cG $}}} \mbox{\boldmath{ $\gamma$}}^{\rm D} (g)
= - \mbox{\boldmath{$\cG$}} \mbox{\boldmath{$\gamma$}}^{\rm D}(g)
+ \cdots,
\quad\mbox{with}\quad
{\mbox{\boldmath{$\cG$}}\mbox{\boldmath{$A$}}}_{jk}
= \frac{\left[\mbox{\boldmath{$\gamma$}}^c (l)
+ 2{\beta\over \g} \mbox{\boldmath{$b$}} (l),
\mbox{\boldmath{$A$}}\right]_{jk}}{a(j,k)} .
\end{eqnarray}

\noindent (iii) The last step consists of the explicit evaluation of the
anomalies. The LO anomalous dimensions of conformal operators have been
known for a long time \cite{Ohr81,Cha80,ShiVys81}
\begin{eqnarray}
\label{def-AD-LO-QQ-e}
{^{QQ}\!\gamma}_{j}^{(0)}
&=&
- C_F \left(
3 + \frac{2}{( j + 1 )( j + 2 )} - 4 \psi( j + 2 ) + 4 \psi(1)
\right)
\\
\label{def-AD-LO-QG-e}
{^{QG}\!\gamma}_{j}^{(0)}
&=&
\frac{-24 N_f T_F}{j( j + 1 )( j + 2 )( j + 3 )}
\times \left\{
j^2 + 3 j + 4,
\atop
j( j + 3 ),
\right.
\\
\label{def-AD-LO-GQ-e}
{^{GQ}\!\gamma}_{j}^{(0)}
&=&
\frac{-C_F}{3( j + 1 )( j + 2 )}
\times \left\{
j^2 + 3 j + 4 ,
\atop
j ( j + 3 ),
\right.
\\
\label{def-AD-LO-GG-e}
{^{GG}\!\gamma}_{j}^{(0)}
&=&
- C_A \left(
- 4 \psi( j + 2 ) + 4 \psi(1) - \frac{\beta_0}{C_A}
\right) \\
&&\hspace{3cm}-
\frac{8 C_A}{j( j + 1 )( j + 2 )( j + 3 )}
\times \left\{
j^2 + 3 j + 3,
\atop
j( j + 3 ),
\right. . \nonumber
\end{eqnarray}
where the upper (lower) row corresponds to even (odd) parity and $\beta_0 =
\frac{4}{3} T_f N_F - \frac{11}{3} C_A$. For the special conformal
anomalies we found at one-loop level \cite{BelMue98a,Mue94,BelMue98c}:
\begin{equation}
\label{def-SCA}
\mbox{\boldmath $\gamma$}^{c(0)} = -
 \mbox{\boldmath $b$}\mbox{\boldmath $\gamma$}^{(0)}
+ \mbox{\boldmath $w$}.
\end{equation}
The first term on the r.h.s.\ appears as a result of the one-loop
renormalization of a conformal operator in a subgraph and we observe that
it is induced by the breaking of scale invariance. For the Green functions
constructed from local field operators this exhausts the sources of
symmetry breaking, however, for the case at hand it is no longer true and
we have an addendum {\boldmath $w$}. The {\boldmath $w$} matrix contains
new information from the renormalization of the operator product of
conformal operators and the conformal variation of the action in $4 -
2\epsilon$ dimensions. These renormalization constants are calculated by
means of modified Feynman rules. It turns out that the {\boldmath $w$}
matrix is universal for vector and axial-vector operators and its matrix
elements read
\begin{eqnarray}
\label{w-QQ}
{^{QQ}\!w}_{jk}&=&
- 2 C_F \left[ 1 + (-1)^{j-k} \right] \theta_{j-2,k}
( 3 + 2 k ) \nonumber\\
&\times&\left\{
2 A_{jk} + ( A_{jk} - \psi( j + 2 ) + \psi(1) )
\frac{(j - k)(j + k + 3)}{( k + 1 )( k + 2 )}
\right\} , \\
\label{w-QG}
{^{QG}\!w}_{jk}
&=& 0
\\
\label{w-GQ}
{^{GQ}\!w}_{jk}&=&
- 2 C_F \left[ 1 + (-1)^{j-k} \right] \theta_{j-2,k} ( 3 + 2k )
\frac{1}{6}
\frac{(j - k)(j + k + 3)}{( k + 1 ) ( k + 2 )} ,
\\
\label{w-GG}
{^{GG}\!w_{jk}} &=&
- 2 C_A [ 1 + ( - 1)^{j - k} ] \theta_{j - 2,k}
( 3 + 2k ) \\
&\times&\!\!\!\left\{
2 A_{jk} + ( A_{jk} - \psi (j+2) + \psi(1) )
\left[
\frac{\Gamma (j + 4)\Gamma (k)}{\Gamma (j)\Gamma (k + 4)} - 1
\right]
+ 2 (j - k)( j + k + 3 )
\frac{\Gamma (k)}{\Gamma (k + 4)}
\right\} .
\nonumber
\end{eqnarray}
The elements of the matrix $\hat A$ are rather complicated
\begin{equation}
A_{jk} = \psi\left( \frac{j + k + 4}{2} \right)
- \psi\left( \frac{j - k}{2} \right)
+ 2 \psi ( j - k ) - \psi ( j + 2 ) - \psi(1) .
\nonumber
\end{equation}
It is instructive to note that the special conformal anomalies
(\ref{def-SCA}) obey certain constraints which originate from the anomalous
superconformal WI and a commutator of the generators of the special
conformal transformation and restricted supersymmetry, and imply that one
can restore all entries in Eqs.\ (\ref{w-QQ})-(\ref{w-GG}) from the
knowledge of e.g.\ ${^{QQ}\gamma}^{c(0)}$ and superconformal anomalies
\cite{BelMul99s}.

The form of the special conformal anomaly (\ref{def-SCA}) implies the
following structure for the off-diagonal part of the anomalous dimension
matrix
\begin{equation}
\mbox{\boldmath{$\gamma$}}^{{\rm ND}(0)}
=
\left[ \mbox{\boldmath{$\gamma$}}^{(0)},
\left(
\beta_0 \mbox{\boldmath{$1$}} - \mbox{\boldmath{$\gamma$}}^{(0)}
\right)
\mbox{\boldmath{$d$}}
+ \mbox{\boldmath{$g$}}
\right]_- ,
\end{equation}
where we defined the matrices $\mbox{\boldmath{$d$}}_{jk} =
\mbox{\boldmath{$b$}}_{jk}/a(j,k)$ and $\mbox{\boldmath{$g$}}_{jk} =
\mbox{\boldmath{$w$}}_{jk}/a(j,k)$. This is the final result of the
conformal approach which give together with the known anomalous dimensions
from DIS, see for instance \cite{FloKouLac81,MerNee96}, the complete
anomalous dimension matrix to NLO in the $\overline{\rm MS}$ scheme relevant to
non-forward processes.

As was explained in the introduction it is paramount to know the ER-BL
kernels. To obtain them, one can transform, in a straightforward way,
the conformal anomalies into the momentum fraction representation.
For instance, for the $QQ$ channel this kernel reads
\begin{eqnarray}
\label{def-QQw}
{[{^{QQ}\! w}(x,y)]}_+
&=& {^{QQ}\! w} (x,y) - \delta(x-y) \int_0^1 dz {^{QQ}\! w} (z,y)
+ {d\over dx} \delta(x - y) \int_0^1 dz (z - y) {^{QQ}\! w} (z,y),
\nonumber\\
{^{QQ}\! w}(x,y) &=& - C_F \theta(y - x) {x \over y}
{ 2 \over (x - y)^2} + \left\{ {x \to 1-x \atop y \to 1-y} \right\}.
\end{eqnarray}
In the next step we construct the ${^{QQ}\! g}(x,y)$ kernel with
its conformal moments satisfying the equality $ 2[(j+1)(j+2)-(k+1)(k+2)]
{^{QQ}\! g}_{jk} =  {^{QQ}\! w}_{jk}$. Using the eigenvalue equation
for Gegenbauer polynomials,
\begin{eqnarray}
\frac{d^2}{dx^2}
\left[ x \bar x C_j^{3/ 2} (x - \bar x) \right]
= - (j + 1)(j + 2) C_j^{3/ 2} (x - \bar x),
\end{eqnarray}
and the representation (\ref{rep-v-GC-ND}), we can easily write down a
second order differential equation for the ${^{QQ}\! g}(x,y)$ kernel
\begin{eqnarray}
\label{equ-for-QQg}
x \bar x \frac{\partial^2}{\partial x^2} {^{QQ}\! g} (x, y)
- \frac{\partial^2}{\partial y^2} \left[ y \bar y {^{QQ}\! g}(x,y) \right]
= - \frac{1}{2} {^{QQ}\! w} (x, y), \quad \mbox{for} \quad x \neq y .
\end{eqnarray}
The solution of the homogeneous equation is purely diagonal w.r.t.\
Gegenbauer polynomials, thus, we find from Eq.\ (\ref{equ-for-QQg}) the
${^{QQ}\! g}$ kernel, which contains beside the desired off-diagonal also
(arbitrary) diagonal conformal moments. Similarly one constructs the other
channels. Our results for the whole singlet sector reads
\begin{eqnarray}
\label{set-g-kernels}
\mbox{\boldmath$g$} (x, y)
= \theta(y - x)
\left(
\begin{array}{cc}
- C_F \left[ \frac{ \ln \left( 1 - \frac{x}{y} \right) }{y - x} \right]_+
& 0 \\
C_F \frac{x}{y}
& - C_A\left[ \frac{ \ln \left( 1 - \frac{x}{y} \right) }{y - x} \right]_+
\end{array}
\right)
\pm
\left\{ x \to \bar x \atop y \to \bar y \right\},
\end{eqnarray}
with the ($-$)$+$ sign corresponding to (non-) diagonal entries
and the ``+''-prescription defined as $[V(x,y)]_+ = V(x,y)
- \delta(x-y) \int_0^1 dz V(z,y) + {\rm const} \cdot \delta(x - y)$,
where the constant term is fixed in appendix \ref{app-LIM}.

Now we come to the dotted kernels $\mbox{\boldmath{$\dot v$}}(x,y)$ which
posses the conformal moments $\left[ \mbox{\boldmath{$\gamma$}}^{(0)},
\mbox{\boldmath{$d$}} \right]_-$. The matrix $d_{jk}$ can be generated
by a derivative w.r.t.\ the index of Gegenbauer polynomials
\begin{eqnarray}
\label{GC-der-index}
\frac{d}{d\nu} C_j^\nu(t)_{|\nu=3/2} =
-2\sum_{k=0}^j d_{jk}  C_k^{3/2}(t),\quad
\frac{d}{d\nu} C_{j-1}^\nu(t)_{|\nu=5/2} =
-2\sum_{k=1}^{j} d_{jk}  C_{k-1}^{5/2}(t),
\end{eqnarray}
and analogous relations, but with a different sign  and range of summation
on the r.h.s., follow immediately from
the orthogonality relation, for the function $ w \left( x | \nu\right)
C^{\nu}_{j} (x -\bar x)/ N_j \left(\nu \right)$. Thus, it is obvious that
the dotted kernels can be generated from
\begin{eqnarray}
\label{gener-dv}
{^{AB}\!v^i} (x, y|\epsilon)
=\sum_{j = 0,1}^\infty
\frac{w \left( x|\nu(A)+\epsilon\right)}{N_j \left(\nu(A)+\epsilon \right)}
C^{\nu(A)+\epsilon}_{j + 3/2 - \nu(A)} (x -\bar x)
{^{AB}\!v^{i(0)}_{jj}} \,
C^{\nu(B)+\epsilon}_{j+ 3/2 - \nu(B)} (y - \bar y)
\end{eqnarray}
by differentiation w.r.t.\ the parameter $\epsilon$ at $\epsilon = 0$. Of
course, for diagonal entries the generating kernel ${^{AA}\!v^i} (x, y| 0)$
is symmetric w.r.t.\ the weight function ${^{AA}\!w} (x|\nu)$ [see
relation (\ref{sym-AA})]. Thus Eq.\ (\ref{gener-dv}) provides essentially
a logarithmic modification of the LO kernel
\begin{equation}
\label{def-v}
{^{AB} v^i}(x, y)
= \theta(y - x) {^{AB}\! f^i}(x, y)
\pm \left\{ {x \to \bar x \atop y \to \bar y } \right\}
\quad
\mbox{for}
\quad
\left\{ {A = B \atop A \neq B } \right. .
\end{equation}
However, using the symmetry
relation (\ref{sym-AB}) to get the mixed channels one has to fix
the integration constant appropriately. Thus, the generic form of the
dotted kernel is
\begin{equation}
\label{def-DotKernel}
{^{AB} \dot v^i} (x, y) =
\theta(y - x) {^{AB}\! f^i} (x, y) \ln \frac{x}{y}
+ \Delta{^{AB}\! \dot{f}^i} (x, y)
\pm \left\{ {x \to \bar x \atop y \to \bar y } \right\}
\quad
\mbox{for}
\quad
\left\{ { A = B \atop A \neq B } \right. .
\end{equation}
To obtain the addendum $\Delta{^{AB}\! \dot{f}^i}$, it is necessary to
have a closer look at the structure of LO kernels, which will be given
in the next subsection.

The factorized structure of NLO off-diagonal anomalous dimensions
is transferred to the momentum fraction kernels and can be constructed
out of the conformal anomalies we have just found. In convolution form
they read
\begin{equation}
\label{def-ND-kernel}
\mbox{\boldmath$V$}^{{\rm ND}(1)} (x, y)
= - ( \cI - \cD )\,
\left\{
\mbox{\boldmath$\dot V$} \OO^\re
\left(
\mbox{\boldmath$V$}^{(0)} + \frac{\beta_0}{2}\, \1
\right)
+
\left[
\mbox{\boldmath$g$} \OO^\re_{,} \mbox{\boldmath$V$}^{(0)}
\right]_-
\right\} (x, y) ,
\end{equation}
where $\tau_1 \displaystyle{ \mathop{\otimes}^{\rm e}} \tau_2 (x,y)
\equiv \int_0^1 dz\, \tau_1(x,z) \tau_2 (z,y)$ defines the exclusive
convolution and $( {\cal I} - {\cal D} )$ projects out the diagonal
part.

\subsection{${\cal N}=1$ supersymmetric constraints.}
\label{subsec-SUSY}

Since the two-loop off-diagonal kernels (\ref{def-ND-kernel}) are
available we are left with the construction of their diagonal
counterparts. Unfortunately, the direct use of the integral transformation
(\ref{DGLAPtoERBL-2}) with the generating function given in Eq.\
(\ref{DGLAPtoERBL-1}) is difficult due to the complicated analytical
structure of the integrand in the complex plane: cuts appear on top of
poles. Therefore, we have to look for complimentary sources of
restrictions on the form of the NLO kernels. Sufficient information comes
from the constraints deduced from the graded commutator algebra of
$\cN = 1$ SYM theory \cite{BukFroKurLip85,BelMueSch98}.
There is a one-to-one correspondence between QCD and SYM Lagrangians
provided one identifies quarks with gluinos and put the former in the
adjoint representation of the colour group. We can map thus the QCD
result to SYM theory by equating the colour factors: $C_F = 2 T_F = C_A
\equiv N_c$. The main use of the previously mentioned constraints will be
in the reconstruction of the crossed ladder diagram contributions which
have the most complicated analytical structure. For this type of Feynman
graphs the restoration of colour factors is unique.

Since the conformal symmetry breaking part has been previously fixed,
we can legitimately assume in what follows that conformal covariance
holds for the anomalous dimensions. In addition to this, we have to use
the renormalization procedure preserving supersymmetry. In
reality none of these statements hold true in the $\overline{\rm MS}$
scheme beyond LO. From the commutator of the dilatation and restricted
supersymmetry generators $[\cQ, \cD]_- = \frac{i}{2} \cQ$ applied to
the Green function (\ref{GreenFunction}) one finds six constraints for
eight anomalous dimensions in the chiral even case
\begin{eqnarray}
\label{con-SUSY-1}
&&{^{QQ}\!\gamma}^i_j + \frac{6}{j} {^{GQ}\!\gamma}^i_j
= \frac{j}{6} {^{QG}\!\gamma}^i_j + {^{GG}\!\gamma}^i_j ,
\quad i = V,A \\
\label{con-SUSY-2}
&&{^{QQ}\!\gamma}^V_{j + 1}
+ \frac{6}{j + 1} {^{GQ}\!\gamma}^V_{j + 1}
=
{^{QQ}\!\gamma}^A_j
- \frac{j}{6} {^{QG}\!\gamma}^A_j,
\quad\mbox{and}\quad V \Leftrightarrow  A, \\
\label{con-SUSY-3}
&& \frac{6}{j} {^{GQ}\!\gamma}^V_j
- \frac{j + 3}{6} {^{QG}\!\gamma}^V_j
= 0,
\quad
\frac{6}{j} {^{GQ}\!\gamma}^A_j
- \frac{j + 3}{6} {^{QG}\!\gamma}^A_j
=0,
\end{eqnarray}
which we call type-I (or Dokshitzer relation \cite{Dok77}), II and III
constraints, respectively. It is worth to mention that the first two
constraints are valid to all orders of perturbation theory in the
supersymmetry preserving regularization and renormalization scheme while
the last one gets modified beyond one-loop order by conformal
non-covariant, off-diagonal elements of the anomalous dimension matrix
\cite{BelMueSch98}.

As an example of the power of these constraints we demonstrate their
use at LO. We show that all entries in the singlet sector
can be reconstructed from the $QQ$ channels by solving the constraints
(\ref{con-SUSY-1})-(\ref{con-SUSY-3}). At first we consider the anomalous
dimensions, and eliminate with the help of the type-III relation the $QG$
entries in the type-II relation. By elimination of the $GQ$ entry either
in the vector or axial-vector channel we obtain a recurrence relation
for the remaining mixed channel. For instance, the solution in the parity
odd sector reads
\begin{eqnarray}
\label{con-SUSY-sol-AD}
{^{GQ}\!\gamma}^A_j &=& \frac{2{^{GQ}\!\gamma}^A_0}{(j + 1)(j + 2)} \\
&&+ \sum_{i = 2, 4, \dots}^{j} \frac{i(i + 1)}{6(j + 1)(j + 2)}
\left[ (i + 2)
\left(
{^{QQ}\!\gamma}_{i - 1}^V - {^{QQ}\!\gamma}_i^A
\right)
+ (i - 1)
\left(
{^{QQ}\!\gamma}_{i - 1}^V - {^{QQ}\!\gamma}_{i - 2}^A
\right)
\right], \nonumber
\end{eqnarray}
where in addition the (analytic continuation to the non-physical region
of the) lowest moment has to be known, i.e.\ ${^{GQ}\!\gamma}^A_0$. The
remaining five entries are obviously linear combinations of the known
ones: The type-II relation gives the $QG$ entry for even parity, the typ-III
constraints provide the remaining two mixed channel entries, and the
$GG$ channel follows from the Dokshitzer relation. In this way we obtain
the whole anomalous dimension matrix for SYM theory
in LO. Let us add for completeness that the restoration of colour factors
is unique up to the $GG$ channels, where the self-energy contribution
provides a constant term proportional to $N_f$. This constant can be
fixed either by explicit calculation of self-energy insertions in QCD or
by the fact that the energy-momentum tensor is conserved and its anomalous
dimension vanishes.

The remaining goal is to derive and solve constraints for the ER-BL
kernels by means of the representation (\ref{rep-v-GC-D}). Unfortunately,
we can only write down the type-II constraints in terms of integral
transformations such as given in Eq.\ (\ref{DGLAPtoERBL-2}). Therefore,
we expect a cumbersome solution of these constraints, which is maybe useless
for practical purposes. Fortunately, the knowledge of the eigenvalues
give us a hint to overcome this problem as described below. The
representation (\ref{rep-v-GC-D}) together with the type-III relation
(\ref{con-SUSY-3}) provides us with the relation between the kernels in
the mixed channels
\begin{eqnarray}
\label{con-SUSY-sol-K3}
{^{GQ}\!v^i} (x, y)
= \frac{(\bar x x)^2}{\bar y y} {^{QG}\!v^i} (y, x).
\end{eqnarray}
Next we can eliminate the $GQ$ entry in the Dokshitzer relation and
take into account the relations (\ref{rel-GC}) between the Gegenbauer
polynomials, to find a differential equation for the $GG$ kernels:
\begin{eqnarray}
\label{con-SUSY-sol-K1+3}
\frac{\partial}{\partial y}
{^{QQ}\!v^i} (x, y)
+ \frac{\partial}{\partial x} {^{GG}\!v^i} (x, y)
= - 3 {^{QG}\!v^i} (x, y)\quad \mbox{for} \quad i=V,A.
\end{eqnarray}
Thus, it remains to find equations which determine the mixed channel
contribution through the knowledge of the quark one. A closer look at the
anomalous dimensions (\ref{def-AD-LO-QQ-e})-(\ref{def-AD-LO-GG-e}) shows
that the $QG$ entry for parity even and the difference between parity even
and odd in the $GQ$ channel, i.e.\ ${^{GQ}\!\gamma}^\delta =
{^{GQ}\!\gamma}^V - {^{GQ}\!\gamma}^V $, are proportional
$1/\bigg((j + 1)(j + 2)\bigg)$, which coincide with the eigenvalues of the
scalar $\phi^3_{(D = 6)}$ theory kernel ${^{QQ}\!v^a} (x,y) =
\theta(y - x) x/y + \theta(x - y) \bar x / \bar y$, which appears as a
part of the whole $QQ$ kernel. Thus, we can write down for the kernels
${^{QG}\!v^A}$ as well as for ${^{QG}\!v^\delta} = {^{QG}\!v^V} -
{^{QG}\!v^A}$ the following differential equations
\begin{eqnarray}
\label{con-SUSY-obs1}
&& {^{QG}\!v^A} (x, y) = \frac{\partial}{\partial y} {^{QQ}\!v^a} (x,y), \\
\label{con-SUSY-obs2}
&& \frac{\partial}{\partial x} {^{GQ}\!v^\delta}(x,y)
= - 4  {^{QQ}\! v^a} (x,y) + {\rm const} \cdot x \bar x
\quad\mbox{and}\quad
\frac{\partial}{\partial y}{^{GQ}\!v^\delta} (x,y)
= 2\, {^{GG}\! v^a} (x,y),
\end{eqnarray}
where the kernel ${^{GG}\! v^a}(x,y)$ has the eigenvalues
$2/\bigg((j + 1)(j + 2)\bigg)$ and can be constructed from the $QQ$ ones
by means of the differential equation (\ref{con-SUSY-sol-K1+3}) with the
r.h.s. set equal to zero. In the first equation of the set
(\ref{con-SUSY-obs2}) we included a term proportional to $x \bar x$, which
reflect the fact that the conformal expansion of ${^{GQ}\!v^\delta} (x,y)$
starts with $j=1$ [compare with Eq.\ (\ref{rep-v-GC-D})]. Consequently, the
lowest conformal moment of the r.h.s.\ has to vanish. Thus, in our case this
constant is ${\rm const} = 4 {^{QQ}\!v}_0^a /N_0(3/2) = 12$.

Let us now discuss the construction of the LO kernels from the knowledge of
the $QQ$ entry
\begin{eqnarray}
\label{v-kernels-QQ}
C_F \left[ {^{QQ}\!v} (x, y) \right]_+
\quad\mbox{with}\quad
{^{QQ}\!v} \equiv {^{QQ}\!v^a} + {^{QQ}\!v^b}, \quad
{^{QQ}\!v^b} (x,y) = \theta(y-x) \frac{x}{y} \frac{1}{y-x} +
\left\{{x\to \bar x \atop y \to \bar y}\right\}.
\end{eqnarray}
In the parity odd sector Eq.\ (\ref{con-SUSY-obs1}) give us the $QG$ entry
and Eq.\ (\ref{con-SUSY-sol-K3}) the $GQ$ one. The integration of the
constraint (\ref{con-SUSY-sol-K1+3}) provides us then with the missing $GG$
kernel. The integration constant $c(y)$ as a function of $y$ is almost
determined by the necessary condition (\ref{sym-AA}) for the kernels to have
a diagonal form. The remaining degree of freedom, i.e.\ $(y\bar y)^2 c'(y)
= (x\bar x)^2 c'(x) = {\rm const}$, can be easily fixed by the requirement
that ${^{GG}\!v}^A_{j1}=0$ for $j>0$. Now we consider the difference between
parity even and odd cases. The set of differential equations
(\ref{con-SUSY-obs2}) gives us the corresponding $GQ$ entry, while the
remaining integration constant is fixed from the requirement that the
conformal moments ${^{GQ}\!v}^\delta_{j0}$ vanish for $j>0$. Then the
difference in the $QG$ channel follows from the symmetry relation
(\ref{con-SUSY-sol-K3}). Finally, the integration of Eq.\
(\ref{con-SUSY-sol-K1+3}) gives us the contribution in the $GG$ channel.

The results obtained, with a minimal calculation of Feynman diagrams, has
the advantage that the kernels are diagonal for all conformal moments. A
direct calculation suffers in the parity even case from subtleties, which
generate in the unphysical sector off-diagonal conformal moments. Now we
present the improved kernels in such a way, that the underlying symmetries
are explicitly manifest. For $I=\{A,V\}$ the kernels  read
\begin{eqnarray}
\label{decomp-V-I}
\mbox{\boldmath$V$}^{(0)I} (x, y)
=
\left(
\begin{array}{ll}
C_F \left[ {^{QQ}\!v} (x, y) \right]_+
& 2 T_F N_f \, {^{QG}\!v^I} (x, y) \\
C_F\, {^{GQ}\!v^I} (x, y)
& C_A {^{GG}\!v^I}(x, y) - \frac{\beta_0}{2} \delta(x - y)
\end{array}
\right),
\end{eqnarray}
where the entries for parity odd are given by only two
types of kernels and the difference to the even case arise from
a third one:
\begin{eqnarray}
\label{v-kernels-A}
&&
{^{QQ}\!v} \equiv {^{QQ}\!v^a} + {^{QQ}\!v^b}, \quad
{^{QG}\!v^A} \equiv -{^{QG}\!v^a} , \quad
{^{GQ}\!v^A} \equiv {^{GQ}\!v^a} ,
\quad
{^{GG}\!v^A} \equiv \left[2\, {^{GG}\!v^a} + {^{GG}\!v^b}\right]_+ ,
\\
\label{v-kernels-V}
&&
{^{QG}\!v^V} \equiv {^{QG}\!v^A} - 2 {^{QG}\!v^c},
\quad
{^{GB}\!v^V} \equiv {^{GB}\!v^A} + 2 {^{GB}\!v^c}
\quad \mbox{for} \quad B = \left\{Q,G\right\}.
\end{eqnarray}
The functions ${^{AB} v^i}$ are defined by Eq.\ (\ref{def-v}) with
\begin{eqnarray}
\label{def-f-functions-ab}
\left\{{ {^{AB}\! f^a} \atop {^{AB}\! f^b} }\right\}
&=& \frac{ x^{\nu(A) - 1/2}}{y^{\nu(B) - 1/2}}
\left\{ { 1 \atop \frac{1}{y - x} } \right\}
\qquad\qquad\qquad\qquad\qquad\mbox{for}\quad
\left\{ {A,B=\{Q,G\} \atop A=B } \right. , \\
\label{def-f-functions-cAA}
{^{AA}\! f^c}
&=& \frac{ x^{\nu(A) - 1/2}}{y^{\nu(A) - 1/2}}
\left\{
{ 2 \bar x y \left[ \frac{4}{3} - \ln( \bar x y ) \right] + y - x
\atop
2 \bar x y + y - x }
\right\}
\quad\mbox{for}\quad
A = \left\{ {Q \atop G } \right. ,
\\
\label{def-f-functions-cAnotB}
{^{AB}\! f^c}
&=& \frac{ x^{\nu(A) - 1/2}}{y^{\nu(B) - 1/2}}
\left\{
{ 2 \bar x  y -  x
\atop 2 \bar x y - \bar y }
\right\}
\qquad\qquad\qquad\qquad\ \mbox{for}\quad
A = \left\{ {Q \atop G} \right\} \not = B .
\end{eqnarray}
The index $\nu(A)$ coincides again with the index of the Gegenbauer
polynomials. It is important to note that we have introduced different
``+''-definitions in the $QQ$ and $GG$ channels in order to have a
one-to-one correspondence between exclusive and inclusive kernels
which require a regularization. For the exclusive $QQ$ kernels we use
the conventional prescription
\begin{eqnarray}
\label{def-con-+pre}
\left[ V (x, y) \right]_+ = V (x, y)
- \delta(x - y) \int_0^1 dz\, V (z, y),
\end{eqnarray}
while for the $GG$ kernel an extra term ${\rm const}(V) \delta(x - y)$
is subtracted, with the constant expressed in terms of an integral of
the kernel itself [see Eq.\ (\ref{def-GG+pre})].

We also introduced in Eq.\ (\ref{def-f-functions-cAA}) the $^{QQ}\!v^c$
kernel, which does not show up in the LO kernel, but is of importance beyond
one-loop approximation. It satisfies the differential equation
\begin{eqnarray}
\frac{\partial}{\partial y}
{^{QQ}\!v}^c(x,y) = - {^{QG}\!v}^c (x,y) ,
\end{eqnarray}
and is diagonal w.r.t.\ Gegenbauer polynomials. The given explicit
representation of this kernel will be essential for the analysis of the
ER-BL kernel in NLO.

It is worth to mention that the eigenvalues of the same $v^i$-kernel in
different channels are related to each other (here $v_{jj} \equiv v_j$)
by the relations
\begin{eqnarray}
\label{eigenvalues-LO-a}
&&{^{QQ}\!v^a_j}
= - \frac{1}{6} {^{QG}\!v^a_j}
= \frac{6}{j ( j + 3 )} {^{GQ}\!v^a_j}
= \frac{1}{2} {^{GG}\!v^a_j}
= \frac{1}{(j + 1)(j + 2)}, \\
\label{eigenvalues-LO-b}
&&{^{QQ}\!v^b_j} = {^{GG}\!v^b_j}
= - 2 \psi( j + 2 ) + 2 \psi( 1 ) + 2, \\
\label{eigenvalues-LO-c}
&&{^{QQ}\!v^c_j}
= - \frac{1}{6}{^{QG}\!v^c_j}
= \frac{6}{j ( j + 3 )}{^{GQ}\!v^c_j}
= \frac{1}{3}{^{GG}\!v^c_j}
= \frac{2}{j ( j + 1 )( j + 2 )( j + 3 )},
\end{eqnarray}
where for the ${^{AA}\!v^b_j}$ kernel the proper ``+''-prescription has
been taken into account.

Even though we do not consider the chiral odd sector in our presentation we
just want to mention that the corresponding constraint can be easily
implemented
for the kernels
\begin{eqnarray}
\label{con-SUSY-sol-T}
\frac{\partial}{\partial y}
{^{QQ}\!v^T} (x, y)
+ \frac{\partial}{\partial x} {^{GG}\!v^T} (x, y) = 0 .
\end{eqnarray}
It allows us to find the $GG$ kernel in a unique way from the $QQ$ entry
as discussed above:
\begin{equation}
\label{kernel-tr-0}
{^{AA}\!V}^{(0)T} (x, y)
= \left\{{C_F \atop C_A}\right\} \left[ {^{AA}\! v}^b (x, y) \right]_+
- \frac{1}{2} \delta(x - y)
\left\{{C_F \atop 4C_A+\beta_0}\right\}
\quad\mbox{for}\quad
A = \left\{ { Q \atop G .} \right.
\end{equation}

\subsection{Method of reconstruction of two-loop kernels.}
\label{subsec-reconstr}

In the previous sections we have established the complete structure
of the ER-BL kernels in LO and also partially in NLO. For the
$\overline{\rm MS}$ scheme the piece containing the off-diagonal part
to NLO is known as a convolution (\ref{def-ND-kernel}), however, we have
no explicit representation of the projection operator $(\cI - \cD)$ at
hand which makes the following analysis more complicated. Let us drop it
and represent the whole contribution to NLO as
\begin{equation}
\label{NDkernel}
\mbox{\boldmath$V$}^{(1)} (x, y)
= -
\mbox{\boldmath$\dot V$} \OO^\re
\left(
\mbox{\boldmath$V$}^{(0)} + \frac{\beta_0}{2}\, \1
\right)(x,y)-
\left[
\mbox{\boldmath$g$} \OO^\re_{,} \mbox{\boldmath$V$}^{(0)}
\right]_- (x, y) +  \mbox{\boldmath$\cal D$}(x,y),
\end{equation}
where $\mbox{\boldmath$\cal D$}(x,y)$ is a pure diagonal part. We adopt
the following strategy for our considerations. To restore this part one
goes first to the forward limit (\ref{def-LIM}) in which all non-diagonal
terms die out and compares the result with the known two-loop DGLAP kernels
$\mbox{\boldmath $P$}^{(1)}(z)$ in order to get the DGLAP representation
of $\mbox{\boldmath$\cal D$}(x,y)$:
\begin{eqnarray}
\mbox{\boldmath $\cal D$}(z)
&=& {\rm LIM} \mbox{\boldmath $\cal D$}(x,y) \\
&=& \mbox{\boldmath $P$}^{(1)}(z)
- {\rm LIM} \left\{
-\mbox{\boldmath $\dot V$} \OO^\re
\left(
\mbox{\boldmath $V$}^{(0)} + \frac{\beta_0}{2}\, \1
\right)(x,y)-
\left[
\mbox{\boldmath $g$} \OO^\re_{,} \mbox{\boldmath $V$}^{(0)}
\right]_- (x, y) \right\} .
\nonumber
\end{eqnarray}
Afterwards one may try to use the reduction formula (\ref{DGLAPtoERBL-2})
to restore the ER-BL representation. Unfortunately, as we mentioned
several times, this last step is too cumbersome to be performed in an
analytical manner. Thus, we are forced to apply supersymmetry to get a
better understanding of the structure and in order to reconstruct
the missing diagonal pieces. If one looks to the DGLAP kernels, one
immediately sees that the most complicated terms are Spence functions
multiplied by the kernels appearing in LO
\begin{eqnarray}
\label{str-G}
\pm 2 {^{AB}\!p}(-z) S_2(- z)
+ {^{AB}\!p}(z) \left[\ln^2 z - 2 \zeta (2)\right],
\quad \mbox{where}\quad
S_2 (z) = \int_{z/(1 + z)}^{1/(1 + z)} \frac{dx}{x}\ln\frac{1 - x}{x}.
\end{eqnarray}
Such contributions arise from the crossed ladder Feynman diagrams. Since
they have no ultraviolet divergent subgraphs and, thus, require no
subtraction, it appears just as in LO only as a $1/\epsilon$ divergence.
Therefore, conformal covariance and supersymmetric constraints hold true
for these contributions. Of course, it is questionable if these six
constraints relate all contributions of this set of Feynman graphs in a
given gauge, e.g. light-cone gauge. Fortunately, the answer is irrelevant
for the reconstruction of the most complicated part containing the Spence
functions. Since the ER-BL kernel is known in the non-singlet case
\cite{Sar84,DitRad84,MikRad85}, we have the one-to-one correspondence of
this piece in the DGLAP and ER-BL representation for the $QQ$ channel at
hand \cite{MueRobGeyDitHor94} which helps in the reconstruction of the other
channels.

It is convenient to introduce the following notation for the NLO result,
\begin{eqnarray}
\label{kernel-NS}
V^{(0)} (x, y) &=& C_F \left[ v(x,y)\right]_+,
\quad
v(x,y) = \theta(y - x) f(x, y)
+
\left\{
{x\to \bar x \atop y \to \bar y}
\right\},
\quad f \equiv {^{QQ}\!f}^a + {^{QQ}\!f}^b,
\nonumber\\
V^{(1)} (x, y) &= &
C_F
\left[
C_F V_F (x, y)
- \frac{\beta_0}{2} V_\beta (x, y)
- \left( C_F - \frac{C_A}{2} \right) V_G (x, y)
\right]_+ ,
\end{eqnarray}
where the functions ${^{QQ}\!f}^i$ with $i=\left\{a,b\right\}$ are given
in Eq.\ (\ref{def-f-functions-ab}). The piece which mainly originates from
the crossed ladder diagram is proportional to $(C_F - C_A/2)$,
\begin{eqnarray}
\label{kernel-NS-CAa}
V_G (x, y)
= 2 v^a (x, y) + \frac{4}{3} v (x, y)
+ \left( G (x, y)
+ \left\{ x \to \bar x \atop y \to \bar y \right\} \right),
\end{eqnarray}
and is diagonal w.r.t.\ the Gegenbauer polynomials. This is obvious for
the first two terms appearing in Eq.\ (\ref{kernel-NS-CAa}). The
function\footnote{We have slightly changed the original definition given
in \cite{Sar84,DitRad84} by $G (x, y) + 2 \theta (y - x) \overline f
\ln y \ln \bar x \to G (x, y)$.} $G (x,y)$ contains both $\theta$-structures
\begin{eqnarray}
\label{kernel-NS-G}
G (x, y) =\theta (y - x) H (x, y)
+ \theta (y - \bar x) \overline H (x, y),
\end{eqnarray}
where the functions $H(x, y)$ and $\overline H (x, y)$ are build from
the LO function $f (x, y)$ and $\overline f (x, y)= f(\bar x,\bar y)$ in
combination with the Spence function ${\rm Li}_2$ and double logs:
\begin{eqnarray}
\label{kernel-NS-H}
H (x, y)
&=& 2 \left[ \overline f
\left( {\rm Li}_2 (\bar x) + \ln y \ln \bar x \right)
- f\, {\rm Li}_2 (\bar y) \right] , \\
\label{kernel-NS-bH}
\overline H (x, y)
&=& 2 \left[ ( f - \overline f )
\left( {\rm Li}_2 \left( 1 - \frac{x}{y} \right)
+ \frac{1}{2} \ln^2 y \right)
+ f \left( {\rm Li}_2 (\bar y) - {\rm Li}_2 (x)- \ln y \ln x \right)
\right],
\end{eqnarray}
where ${\rm Li}_2(y) = -\int_0^y dx\, [\ln(1-x)]/x$. It can be easily checked
that the $G$-contribution is symmetric w.r.t.\ the weight $x \bar x$. Note
that the two different $\theta$-structures do not obey separately this
symmetry and contain therefore off-diagonal conformal moments. At this stage
it is not so important to redefine $H$ and $\overline H$ so that they are
separately symmetric and diagonal. This will be done for the presentation of
the explicit NLO result in section \ref{sec-rep}. Performing the limit
(\ref{def-LIM}) [of course, only the $QQ$ channel is relevant] we obtain the
following correspondence with the non-singlet DGLAP kernel
\cite{MueRobGeyDitHor94}:
\begin{equation}
\label{LIM-NS-G}
G(z) \equiv {\rm LIM} G (x, y)
= \theta(z) \theta(1 - z) H (z)
+ \theta(- z) \theta(1 + z) \overline H (z) ,
\end{equation}
where
\begin{eqnarray}
\label{LIM-NS-H}
H (z) &\equiv& {\rm LIM}\, H (x, y)
= p(z) \left( \ln^2 z - 2 \zeta (2) \right) + T (z) , \\
\label{LIM-NS-bH}
\overline H (z) &\equiv& {\rm LIM}\, \overline H (x, y)
= 2 p(z) S_2(- z) + T (-z),\ T (z) = 2 (1 + z) \ln z + 4(1 - z).
\end{eqnarray}
Up to the simpler term $T (z)$, we recover by mapping of $-1\le z \le0$
into the $0\le z\le 1$ region the desired expression (\ref{str-G}).

As demonstrated in subsection \ref{subsec-SUSY} for the LO kernels, we can
now apply six supersymmetric constraints to find the $G$ kernels in all
other channels from the kernel (\ref{kernel-NS-G}) in the $QQ$ channel.
This task will be achieved in the next sections.

Now we study the structure of the remaining diagonal piece $D(x,y)$ in the
non-singlet sector. Knowing the correspondence of the $G$ kernels in
different representations, the remaining diagonal piece of $V_G(x,y)$ can
be easily restored from the DGLAP kernel
\cite{CurFurPet80}
\begin{eqnarray}
\label{kernelP-NS-CA}
P_G (z) = 2 p^a (z) + \frac{4}{3} p(z)+ G(z),
\end{eqnarray}
where $p^a (z) \equiv {\rm LIM} v^a (x,y)= 1-z $ and
$p(z)\equiv {\rm LIM} v (x,y)= (1+z^2)/(1-z)$  by substitution
\begin{eqnarray}
\label{sub-NS-1}
p^a (z) \to   v^a (x,y),\quad p(z) \to  v (x,y).
\end{eqnarray}

Next, the Feynman diagrams containing vertex and self-energy corrections
provide $V_\beta$ proportional to $\beta_0$. Its off-diagonal part is
induced by the renormalization of the coupling and is contained in the
dotted kernel:
\begin{equation}
V_\beta (x, y) = \dot v (x, y) + D_\beta (x, y).
\end{equation}
This dotted kernel is given in Eq.\ (\ref{def-DotKernel}), where $\Delta
f$ is zero in the $QQ$ sector:
\begin{equation}
\label{def-DotKernel-QQ}
\dot v (x, y) =
\theta(y - x) {f} (x, y) \ln \frac{x}{y}
+ \left\{ {x \to \bar x \atop y \to \bar y } \right\}.
\end{equation}
This can be proved easily by forming conformal moments and using the
triangular form of this matrix. The remaining diagonal piece, $D_\beta$,
is deduced from the known NLO DGLAP kernel
\cite{CurFurPet80}
\begin{equation}
\label{kernelP-NS-beta}
P_\beta (z) = \frac{5}{3} p (z) + p^a (z) + \dot p (z), \quad
\dot p(z)\equiv {\rm LIM} \dot v(x,y) = p(z) \ln z + 1-z,
\end{equation}
by going to the forward kinematics and restoring the missing
contributions from it by the substitutions (\ref{sub-NS-1}):
\begin{equation}
\label{Dbeta-QQ}
D_\beta (x, y) = \frac{5}{3} v (x, y) + v^a (x, y) .
\end{equation}
Indeed, the final result coincides with \cite{Sar84,DitRad84,MikRad85}.

Making use of the known non-diagonal part in the flavor non-singlet
sector [which is the same as appearing in the $QQ$ channel of Eq.\
(\ref{NDkernel})], $V_F$ can be represented up to a pure diagonal term,
denoted as $D_F(x,y)$, by the convolution\footnote{
We remind the reader that all flavor non-singlet kernels are supplemented by a
``+''-prescription. Therefore, we use for simplicity in this section the
convention that $C = A \OO B$ is indeed defined by the convolution of
$ [C]_+ =[A]_+ \OO [B]_+$. }
\begin{equation}
V_F (x, y) =
- \left( \dot{v} \OO^\re v
+ g \OO^\re v - v \OO^\re g \right) (x, y)
+ D_F (x, y),
\end{equation}
where the $g$ kernel is given by the $QQ$ entry of the matrix
(\ref{set-g-kernels}). To find an appropriate representation of this
missing diagonal element we first take the forward limit. Since the
forward limit of the convolution is
\begin{equation}
{\rm LIM}\, \left\{ [\dot v]_+ \OO^\re [v]_+ \right\} =
\left\{ {\rm LIM}\, [\dot v]_+ \right\}
\OO^\ri \left\{ {\rm LIM}\, [v]_+ \right\} ,
\end{equation}
where we have introduced the inclusive convolution
\begin{eqnarray*}
P_1 (z) \OO^\ri P_2 (z)
\equiv \int_0^1 dx \int_0^1 dy \delta( z - xy ) P_1 (x) P_2 (y) ,
\end{eqnarray*}
with the commutator $g \OO V^{(0)} - V^{(0)} \OO g$ droping out in the
forward limit, we obtain
\begin{eqnarray}
\label{LIM-VF}
{\rm LIM}\, V_F (x, y)
= - \dot p \OO^\ri p
+ {\rm LIM}\, D_F (x,y).
\end{eqnarray}
The comparison
of ${\rm LIM}\, V_F (x, y)$ with the corresponding part of the DGLAP
kernel \cite{CurFurPet80}
\begin{eqnarray}
P_F (z)
&=& \left\{ \frac{4}{3} - 2 \zeta (2)
- \frac{3}{2} \ln z + \ln^2 z - 2\ln z \ln(1 - z)
\right\} p(z) \nonumber\\
&+& 1 - z + \frac{1 - 3 z}{2} \ln z  - \frac{1 + z}{2} \ln^2 z ,
\end{eqnarray}
yields the result in which all double log terms are contained
in the convolution
\begin{eqnarray}
\label{con-p-dp-NS}
\dot p \OO^\ri p
&=& \left[- \frac{27}{12} +2 \zeta (2)
+ \frac{3}{2} \ln z -\ln^2 z+ 2 \ln z \ln (1 - z) \right] p(z) \\
&&+ \frac{z}{1 - z}
+ 2(1 - z) \ln \frac{1-z}{z}  + \frac{1 + z}{2} \ln^2 z
\nonumber
\end{eqnarray}
and, therefore, only
single logs survive in $D_F (z) = {\rm LIM} D_F (x, y)$:
\begin{eqnarray}
\label{LIM-DF}
D_F(z) &=& P_F + \dot p \OO^\ri p \nonumber\\
&=& - \frac{1}{2} p^a (-z) \ln z
- p^a (z) \left\{ \ln z - 2 \ln (1 - z) - \frac{1}{2} \right\}
- \frac{5}{12} p (z) .
\end{eqnarray}
Here we have introduced for convenience the kernel $p^a (z) = 1 - z$.
The next important point is that the remaining log terms can be
represented as convolutions of $p^a$ and $p$, i.e.
\begin{eqnarray}
\left[ p^a \OO^\ri p^a\right]_+ &=&  -\left[3 p^a(z) + p^a(-z) \ln z
\right]_+,
\\
\left[ p^a \OO^\ri p\right]_+ &=&
 -\left[ \frac{1}{2} p(z)+ p^a(z) \left\{\frac{1}{2} + \ln z - 2 \ln(1 -
z)\right\}
\right]_+ .
\end{eqnarray}
Thus, we finally have
\begin{eqnarray}
D_F (z)
= \frac{1}{2} p^a \OO^\ri \left\{ 2\, p + p^a \right\} (z)
+ \frac{1}{12} p(z) + \frac{5}{2} p^a(z) .
\end{eqnarray}
Since $D_F (x, y)$ is by definition diagonal, the extension of $D_F (z)$
towards the ER-BL kinematics is trivial:
\begin{eqnarray}
\label{DF-QQ}
D_F (z) \to D_F (x, y)
= \frac{1}{2} v^a \OO^\re
\left( 2\, v + v^a \right) (x, y)
+ \frac{1}{12} v (x, y) + \frac{5}{2} v^a (x, y).
\end{eqnarray}
Evaluating the convolutions one can establish the equivalence of this
equation with the explicit calculated expression in Ref.\
\cite{Sar84,DitRad84,MikRad85}:
\begin{eqnarray}
\label{kernel-NS-CFa}
V_F (x, y)
&=& \theta (y - x)
\Bigg\{ \left( \frac{4}{3} - 2 \zeta (2) \right) f
+ 3 \frac{x}{y}
- \left( \frac{3}{2} f - \frac{x}{2 \bar y} \right) \ln \frac{x}{y}
- ( f - \overline f ) \ln \frac{x}{y}
\ln \left( 1 - \frac{x}{y} \right)
\nonumber \\
&+&
\left( f + \frac{x}{2 \bar y} \right) \ln^2 \frac{x}{y} \Bigg\}
- \frac{x}{2\bar y} \ln x \left( 1 + \ln x - 2 \ln \bar x \right)
+ \left\{ {x \to \bar x \atop y \to \bar y } \right\}.
\end{eqnarray}

Recapitulating our results presented in this section, the ER-BL kernel to
NLO has a rather simple structure governed by conformal anomalies, the
crossed ladder contribution $\mbox{\boldmath$G$}(x,y)$, as well as a
remaining diagonal piece $\mbox{\boldmath$D$}(x,y)$:
\begin{eqnarray}
\label{def-str-NLO}
\mbox{\boldmath$V$}^{(1)I}(x,y)
= -
\mbox{\boldmath$\dot V$}^I \OO^\re
\left(
\mbox{\boldmath$V$}^{(0)I} + \frac{\beta_0}{2}\, \1
\right)(x,y)-
\left[
\mbox{\boldmath$g$} \OO^\re_{,} \mbox{\boldmath$V$}^{(0)I}
\right]_- (x, y)
+ \mbox{\boldmath$G$}^{I}(x,y)
+ \mbox{\boldmath$D$}^I(x,y).
\end{eqnarray}
The first two terms on the r.h.s. are known, the kernel
$\mbox{\boldmath$G$}(x,y)$ can be restored from the $QQ$ entry by means
of supersymmetric constraints, and the missing diagonal piece can be
extracted in the forward case from the DGLAP kernels
$\mbox{\boldmath$P$}^{(1)I}(z)$:
\begin{eqnarray}
\label{get-D}
\mbox{\boldmath$D$}^I (z)
= \mbox{\boldmath$P$}^{(1)I}(z)
- {\rm LIM}
\left\{
- \mbox{\boldmath$\dot V$}^I \OO^\re
\left( \mbox{\boldmath$V$}^{(0)I} + \frac{\beta_0}{2} \1 \right)
- \left[
\mbox{\boldmath$g$} \OO^\re_{,} \mbox{\boldmath$V$}^{(0)}
\right]_-
+ \mbox{\boldmath$G$}^{I}
\right\}.
\end{eqnarray}
In the non-singlet case we saw that this piece does not contain double
logs or Spence functions. Moreover, it can be represented in a rather
straightforward manner by simple kernels, known from the LO result, and
their convolutions. This immediately allows one to obtain the ER-BL
representation. Since the established structure is deeply related to the
topology of Feynman graphs and the renormalization of the subgraphs, we
expect that the missing entry $\mbox{\boldmath$D$}^I(x,y)$ in all other
cases can be build by means of known kernels, too.

\section{Reconstruction of evolution kernels in NLO.}
\label{sec-ReconS}

In this section we present, in detail, the reconstruction of all eight
evolution kernels appearing in the vector and axial-vector case. In the
following two subsections we complete the construction of
$\mbox{\boldmath$\dot V$}^I$ and of $\mbox{\boldmath$G$}^I$ kernels,
respectively. Afterwards we extract the remaining diagonal part and
represent the ER--BL type kernels as convolution of kernels, appearing
in LO or dotted kernels, and the $\mbox{\boldmath$G$}$ kernels.

\subsection{Construction of $\mbox{\boldmath$\dot V$}$.}

In this subsection we present a systematic construction of the dotted
kernels ${^{QQ}\!\dot v^i} (x, y)$, which posses the off-diagonal conformal
moments $({^{AB}\!v}_j^{i} - {^{AB}\!v}_k^{i}) d_{jk}$. Their generic form
is given in Eq.\ (\ref{def-DotKernel}) and the remaining goal is to
determine the addenda $\Delta{^{AB}\! \dot{f}^i} (x, y)$. As we already saw
in the previous sections, in the $QQ$ channel there is no such extra term
for the $a$ and $b$ kernels. From the generating kernel (\ref{gener-dv})
for the $QQ$ channel we can derive all the other ones. Since the eigenvalues
of the $a$-kernel in the $QG$ channel are $\frac{1}{6} {^{QG}\!v^a_j} =
- {^{QQ}\!v^a_j}$, we can obtain the dotted kernel by differentiation
w.r.t.\ $y$:
\begin{eqnarray}
{^{QG}\! \dot v^a} (x, y)=-
\frac{\partial}{\partial y} {^{QQ}\!\dot v^a} (x, y) + \cdots,
\end{eqnarray}
where the ellipsis denotes possible diagonal terms. One can easily
check that this equation is satisfied by Eq.\ (\ref{def-DotKernel})
with $\Delta{^{QG}\! \dot{f}^a} (x, y)=0.$ To find the dotted $a$-kernel
in the $GQ$ entry, we use a symmetry relation for the generating
kernels, which arise from $ -{^{QG}\!v^a_j}/6 =
6 {^{GQ}\!v^a_j}/j (j+3)$:
\begin{eqnarray}
\label{dotSUSY}
{^{GQ}\! \dot v^a} (x, y)
&=&
\frac{d}{d\epsilon}{^{GQ}\! v^a} (x, y|\epsilon)_{|\epsilon=0}
=
-\frac{d}{d\epsilon}\left[
\frac{(x \bar x )^{2+\epsilon}}{(y\bar y)^{1 + \epsilon}}
{^{QG}\!v^a} (y, x|\epsilon)
\right]_{|\epsilon=0} + \cdots,
\nonumber\\
&=& -\frac{(x \bar x )^2}{y\bar y} {^{QG}\!\dot v^a} (y, x)
+ {^{GQ}\!v^a} (x, y) \ln\frac{x\bar x}{y\bar y}+ \cdots .
\end{eqnarray}
From this equation one can easily deduce that $\Delta{^{GQ}\! \dot{f}^a}
(x, y)=0$. The $a$ and $b$-kernels in the $GG$ channel may be obtained
form the $QQ$ ones by means of the differential equation
\begin{eqnarray}
\frac{\partial }{\partial x} {^{GG}\! \dot v^i}(x, y)
= - 2 \frac{\partial }{\partial y} {^{QQ}\! \dot v^i} (x, y),
\quad \mbox{for} \quad i = a, b,
\end{eqnarray}
which immediately provides the generic form (\ref{def-DotKernel}) as a solution
with $\Delta {^{GG}\! \dot f^i}(x,y) = \Delta {^{GG}\! \dot f^i} (y)$. Since
the lowest moment, i.e. $\int_0^1 dx\, {^{GG}\! \dot v^i}(x, y)$, is a
constant, $\Delta {^{GG}\! \dot f^i}(y)$ with $i=a,b$ itself has to be a
constant, which turns out to be zero.

Now we come to the dotted $c$-kernels. Since the eigenvalues of the LO
kernel in the $GQ$ channel are given by the $a$-kernels in the $AA$
channels, i.e.\ ${^{GQ}\!v_j^c} ={^{QQ}\!v_j^c}/3 ={^{GG}\!v_j^c}/6$
[see Eq.\ (\ref{eigenvalues-LO-c})], we can derive the following two
differential equations
\begin{eqnarray}
\label{DiffEq1}
\frac{\partial}{\partial y} {^{GQ}\! \dot v^c} (x, y)
= {^{GG}\! \dot v^a} (x, y) + {^{GG}\! v^a} (x,y),
\quad
\frac{\partial}{\partial x} {^{GQ}\! \dot v^c} (x, y)
= - 2 \,{^{QQ}\! \dot v^a} (x, y) -{^{QQ}\! v^a} (x,y),
\end{eqnarray}
where the added diagonal terms on the r.h.s\ are fixed in the way that the
equations are solvable. Taking the generic form (\ref{gener-dv}), we obtain
two differential equations for the addendum:
\begin{eqnarray}
\frac{\partial}{\partial y} \Delta{^{GQ}\! \dot f^c} (x, y)
=
-\frac{x^2}{y}(2x-3), \quad
\frac{\partial}{\partial x} \Delta {^{GQ}\! \dot f^c} (x, y)
=
x (2x-3)-6x\bar{x} \ln\frac{x}{y}.
\end{eqnarray}
The solution $\Delta{^{GQ}\! \dot{f}^c}
= x^2 (2 x - 3) \ln\frac{x}{y}$
is fixed up to a constant, which by comparison with the lowest conformal
moments turns out to be zero.

Since the $c$-kernels in the mixed channels are related by supersymmetry,
we may use Eq.\ (\ref{dotSUSY}) do find the dotted $c$-kernel in the $QG$
channel from the $GQ$ one.  The $GG$ entry follows then by integration from
\begin{eqnarray}
{^{QG}\! \dot v^c} (x, y)
= \frac{1}{3} \frac{\partial}{\partial x} {^{GG}\! \dot v^c} (x, y) .
\label{DiffEq3}
\end{eqnarray}
Since $\Delta{^{GG}\! \dot{f}^c} (x, y) = 2 \frac{x^2}{y^2} (y - x)$ has
been obtained by a direct calculation of a graph with a quark bubble
insertion [see conformal predictions (\ref{def-ND-kernel})] in Ref.\
\cite{BelMue98c}, we can use Eq.\ (\ref{DiffEq3}) to get
$\Delta{^{QG}\! \dot{f}^c}
= - \frac{x}{3 y^2} \left( 4 x - 5  y + 2 x y \right)$.

Now we are ready to present all dotted kernels. For odd parity we
introduce the dotted matrix
\begin{equation}
\label{dotv-kernels-I}
\mbox{\boldmath$\dot V$}^{I} (x, y)
=
\left(
\begin{array}{ll}
C_F \left[ {^{QQ}\!\dot v} (x, y) \right]_+
&  2 T_F N_f {^{QG}\!\dot v}^I (x, y) \\
C_F {^{GQ}\!\dot v}^I (x, y)
&
C_A  {^{GG}\!\dot v}^I (x, y)
\end{array}
\right)
\quad \mbox{for} \quad I=\{A,V\}.
\end{equation}
Here we use the decomposition as for the
LO kernels (\ref{v-kernels-A}) and (\ref{v-kernels-V}),
\begin{eqnarray}
\label{dotv-kernels-A}
&& {^{QQ}\!\dot v} \equiv {^{QQ}\!\dot v^a} +
 {^{QQ}\!\dot v^b},
\quad
{^{QG}\!\dot v^A} \equiv -{^{QG}\!\dot v^a} , \quad
 {^{GQ}\!\dot v^A} \equiv {^{GQ}\!\dot v^a} ,
\quad
{^{GG}\!\dot v^A} \equiv
	\left[2\, {^{GG}\!\dot v^a} + {^{GG}\!\dot v^b}\right]_+,
\\
\label{dotv-kernels-V}
&&
{^{QG}\!\dot v^V} \equiv {^{QG}\!\dot v^A} - 2 {^{QG}\!\dot v^c},
\quad
{^{GB}\!\dot v^V} \equiv {^{GB}\!\dot v^A} + 2 {^{GB}\!\dot v^c}
\quad \mbox{for} \quad B=\left\{Q,G\right\},
\end{eqnarray}
and included the same ``+''-prescription, although
the dotted kernels are regular at the point $x = y$. The general
structure of ${^{AB} \dot v^i}$ are given in Eq.\ (\ref{gener-dv}),
where the addenda are
\begin{eqnarray}
\label{def-dot-addenda}
&& \Delta{^{AB}\! \dot{f}^a} = \Delta{^{AA}\! \dot{f}^b}\equiv 0,
\quad
\Delta{^{GG}\! \dot{f}^c} (x, y) = 2 \frac{x^2}{y^2} (y - x)
\\
&&\Delta{^{QG}\! \dot{f}^c}
= - \frac{x}{3 y^2} \left( 4 x - 5  y + 2 x y \right),
\quad
\Delta{^{GQ}\! \dot{f}^c}
= x^2 (2 x - 3) \ln\frac{x}{y}.
\nonumber
\end{eqnarray}

\subsection{Construction of $\mbox{\boldmath$G$}$ kernels.}
\label{subsec-Con-G}

The construction of the diagonal $\mbox{\boldmath$G$} (x, y)$ kernel,
related to the crossed ladder diagrams, goes in the same manner as
demonstrated in subsection \ref{subsec-SUSY} for the reconstruction of the
LO kernels from the $QQ$ one by means of conformal and supersymmetric
constraints. Since not all steps are quite obvious, we give here the
construction in detail.

The colour structure of the entries in the $\mbox{\boldmath$G$}$ kernel is
\begin{equation}
\label{G-kernel-S}
\mbox{\boldmath$G$}^I (x, y)
= - \frac{1}{2}
\left(
\begin{array}{cc}
2 C_F \left( C_F - \frac{C_A}{2} \right)
\left[ {^{QQ}\!G}^I (x, y) \right]_+
&
2 C_A T_F N_f \, {^{QG}\!G}^I (x, y)
\\
C_F C_A \, {^{GQ}\!G}^I (x, y)
&
C_A^2 \left[ {^{GG}\!G}^I (x, y) \right]_+
\end{array}
\right),
\end{equation}
for $I=\{A,V\}.$
As explained in subsection \ref{subsec-reconstr}, from the result
(\ref{kernel-NS-G})-(\ref{kernel-NS-bH}) in the flavor non-singlet sector
\cite{Sar84,DitRad84,MikRad85} and its correspondence to the forward case
given in Eqs.\ (\ref{LIM-NS-G})-(\ref{LIM-NS-bH}),
we conclude that all entries in this matrix have the generic form
\begin{equation}
\label{kernel-S-G}
{^{AB} G}^I (x, y)
= \theta (y - x)
\left( {^{AB}\! H}^I + \Delta{^{AB}\! H}^I \right) (x, y)
+ \theta (y - \bar x)
\left( {^{AB} \overline H}^I
+ \Delta{^{AB} \overline H}^I \right) (x, y),
\end{equation}
with the following expressions for $H$ and $\overline{H}$
\begin{eqnarray}
\label{kernel-S-H}
{^{AB} H}^I (x, y)
\!\!&=&\!\! 2 \left[ \pm {^{AB} \overline f}^I
\left( {\rm Li}_2( \bar x ) + \ln y \ln \bar x \right)
- {^{AB}\! f}^I\, {\rm Li}_2( \bar y ) \right],
\\
\label{kernel-S-bH}
{^{AB} \overline{H}}^I (x, y)
\!\!&=&\!\! 2 \left[
\left( {^{AB}\! f}^I \mp {^{AB} \overline f}^I \right)
\left( {\rm Li}_2 \left( 1 - \frac{x}{y} \right)
+ \frac{1}{2} \ln^2 y \right)
+ {^{AB}\! f}^I \left( {\rm Li}_2 ( \bar y )
- {\rm Li}_2 (x) - \ln y \ln x \right) \right].
\nonumber\\
\end{eqnarray}
Here the upper (lower) sign corresponds to the $A = B$ ($A \not= B$)
channels.   In the $QQ$ sector we have $\Delta{^{QQ}\!H} =
\Delta{^{QQ}\!\overline H}=0$. However, in general it turns out that the
non-vanishing addenda are needed to ensure the diagonal form of the kernels.
The forward limit of these ${^{AB}\! G}^I$ kernels
has as a generalization of Eqs.\ (\ref{LIM-NS-G})-(\ref{LIM-NS-bH}) the
following form:
\begin{equation}
\label{LIM-S-G}
{^{AB}\! G}^I(z)
\equiv
\left[{\rm LIM} {^{AB}\!G}^I (x, y)\right]_{z\le 0 \Rightarrow z \ge 0}
=
\theta(z) \theta(1 - z)\left[{^{AB}\!H}^I (z)
\pm {^{AB}\!\overline H}^I (-z)\right]
\quad \mbox{for}\quad \left\{{I=A \atop I=V} \right.,
\end{equation}
where we employed the (anti) symmetry of the singlet parton densities to map
the mixing between partons and anti-partons, given for $-1\le z\le 0$ into
the region $0\le z \le 1$. The ${^{AB}\!H}^I$ and ${^{AB}\!\overline H}^I$
functions are defined as
\begin{eqnarray}
\label{LIM-S-H}
{^{AB}\!H^I} (z)
&\equiv&
{\rm LIM}\left({^{AB}\!H}^I+\Delta {^{AB}\!H}^I \right) (x, y)
=
{^{AB}\!p}^I(z) \left( \ln^2 z - 2 \zeta (2) \right) + {^{AB}\!T}^I (z),
\\
\label{LIM-S-bH}
{^{AB}\!\overline H}^I (z)
&\equiv&
{\rm LIM}\left({^{AB}\!\overline H}^I+
\Delta {^{AB}\!\overline H}^I\right)(x, y)
= 2 {^{AB}\!p}^I(z) S_2(- z) + {^{AB}\! \overline T}^I (-z),
\end{eqnarray}
where we have to require that ${^{AB}\!T}^I$ and ${^{AB}\! \overline T}^I$
are rational functions and/or terms containing single logs of momentum
fractions. Note that the mapping, which took place in Eq.\ (\ref{LIM-S-G}),
can be also performed in the ER-BL representation by the substitution
$y \to \bar y$ and is discussed in section \ref{sec-rep}.

First we reconstruct the missing three entries of $\Delta {^{AB}\!H^A}$ and
$\Delta {^{AB}\! \overline H^A}$ in the axial-vector case. Since the
${^{AB}\!G}^A$ are essentially determined by the LO functions, it is
convenient to decompose them in the same way as the LO kernels in
Eqs.\ (\ref{v-kernels-QQ}) and (\ref{v-kernels-A}):
\begin{eqnarray}
\label{dec-G-A}
{^{QQ}\!G^A} \equiv {^{QQ}\!G^a} + {^{QQ}\!G^b},
\quad
{^{QG}\!G^A} \equiv -{^{QG}\!G^a} ,
\quad
{^{GQ}\!G^A} \equiv {^{GQ}\!G^a} ,
\quad
{^{GG}\!G^A} \equiv 2\, {^{GG}\!G^a} + {^{GG}\!G^b}.
\end{eqnarray}
As for the kernels ${^{AB}\!v^A}(x, y)$ to LO, we can obtain the $QG$ entry
from the $a$-kernel in the $QQ$ channel by a derivative w.r.t.\
$y$ as in Eq.\ (\ref{con-SUSY-obs1}). Since the ${^{QQ}\!G}^a$ kernel
is defined in Eqs.\ (\ref{kernel-S-G})- (\ref{kernel-S-bH}) with
$\Delta{^{QQ}\!H}^a = \Delta{^{QQ}\!\overline H}^a = 0$, we find the
addenda in the $QG$ channel from
\begin{eqnarray*}
\Delta  {^{QG}\!H^A} (x, y)
= - {^{QG}\!H^A} (x, y) + \frac{\partial }{\partial y} {^{QQ}\!H^a} (x, y),
\
\Delta {^{QG}\!\overline H^A} (x, y)
= - {^{QG}\!\overline H^A} (x, y)
+ \frac{\partial }{\partial y} {^{QQ}\!\overline H^a} (x, y).
\end{eqnarray*}
Since conformal covariance and supersymmetry connect these functions in a
simple way with the $GQ$ ones [see Eq.\ (\ref{con-SUSY-sol-K3})],
\begin{eqnarray*}
{^{GQ}\!G^A} (x, y)
= \frac{(\bar x x)^2}{\bar y y} {^{QG}\!G^A} (y, x)
\quad \Rightarrow\quad\left\{
{
\Delta{^{GQ}\!H^A} (x, y)
= - \frac{(\bar x x)^2}{\bar y y} {^{QG}\!H^A} (\bar y, \bar x)
\atop
\Delta{^{GQ}\!\overline H^A} (x, y)
= \frac{(\bar x x)^2}{\bar y y} {^{QG}\! \overline H^A} (y,x)
\hfill }\right\},
\end{eqnarray*}
we can write our findings in the following symmetric manner
\begin{eqnarray}
\label{def-DH-QG-A}
\Delta {^{QG}\!H}^A (x, y)
&=& \Delta {^{QG}\!\overline H}^A (\bar x, y),
\quad
\Delta{^{QG}\!\overline H}^A (x, y)=
\frac{x \bar x}{(y \bar y)^2} \Delta{^{GQ}\!\overline H}^A (y,x),
\\
\label{def-DH-GQ-A}
\Delta{^{GQ}\!H}^A (x, y)
&=&  - \Delta{^{GQ}\!\overline H}^A (\bar x, y),
\quad
\Delta{^{GQ}\!\overline H}^A (x, y)
= - 2 \frac{x \bar x}{y} \ln x + 2 \frac{x \bar x}{\bar y} \ln y .
\end{eqnarray}
To find the remaining $GG$ entries we employ the differential equation
(\ref{con-SUSY-sol-K1+3}), i.e.
\begin{eqnarray*}
 \Delta {^{GG}\! H^A} (x, y)
= - {^{GG}\! H^A} (x, y)
- \int^x dx' \left[\frac{\partial}{\partial y}
{^{QQ}\!H^A}(x', y) + 3 {^{QG}\!H^A}(x', y)
+ 3 \Delta {^{QG}\! H^A}(x', y)\right]
\end{eqnarray*}
and an analogous equation for $ \Delta {^{GG}\!\overline H^A} (x, y)$. Again
the remaining freedom can be fixed by the necessary condition (\ref{sym-AA})
for diagonality w.r.t.\ Gegenbauer polynomials and from the requirement that
the moments ${^{GG}\!G}^A_{j1}$ vanishes for $j>1$. To simplify the result,
we remove a symmetric function\footnote{Here and below for the vector case
this functions slightly differ from that one in Ref.\ \cite{BelFreMue99}.}
(w.r.t.\ the simultaneous interchange $x \to \bar x$ and $y \to \bar y$)
which enters in both $\Delta{^{GG}\!H}^A$ and $\Delta{^{GG}\!\overline H}^A$
kernels, however, with different overall signs and, therefore, disappears
from ${^{GG}\!G}^A$:
\begin{eqnarray}
\label{def-DH-GG-A}
\Delta{^{GG}\!H}^A (x, y)
&=&	-\Delta{^{GG}\!\overline H}^A(\bar x, y),
\\
\Delta{^{GG}\!\overline H}^A (x, y)
&=&
- \frac{{\bar x}^2}{y^2}
-2 \frac{x {\bar x}}{y {\bar y}}
- 2 \frac{x (\bar x + y - 3 \bar x y)}{y^2 \bar y } \ln x
- 2 \frac{\bar x (x + \bar y - 3 x \bar y)}{y \bar y^2} \ln y .
\nonumber
\end{eqnarray}

Now we come to the parity even sector. Instead of dealing with the whole
sector, we can consider only the difference between vector and axial-vector
kernels
\begin{equation}
{^{AB}\!G}^V = {^{AB}\!G}^A + {^{AB}\!G}^\delta
\quad \mbox{with} \quad
{^{QG}\!G}^\delta = -2 {^{QG}\!G}^c,\
{^{GB}\!G}^\delta = 2 {^{GB}\!G}^c \quad\mbox{for}\quad B=\{Q,G\},
\end{equation}
and the same notation for ${^{AB}\!H},{^{AB}\!{\overline H}}$ and their
addenda.

In LO we were able to write down the set (\ref{con-SUSY-obs2}) of
differential equations that determine the entry in the $GQ$ channel
depending on the $a$-kernels. Unfortunately, the analogous equations for
${^{AB}\!G}^\delta$ provide us a solution, which does not preserve the
generic form of ${^{GQ}\!G^\delta}$ in the forward limit given in Eq.\
(\ref{LIM-S-G}) - (\ref{LIM-S-bH}). However, it turns out that we can
restore the generic form by adding convolutions of $c$-kernels of the
diagonal channels given in Eqs.\ (\ref{v-kernels-A}) and
(\ref{def-f-functions-cAA}):
\begin{eqnarray}
\frac{\partial }{\partial x} {^{GQ}\!G^\delta}(x,y)
&=& - 4 \left[ {^{QQ}\! G^a}(x,y) + 9\, {^{QQ}\! v^c}
\OO^\re
{^{QQ}\! v^c}(x,y) - 9\, {^{QQ}\! v^c}
\OO^\re
{^{QQ}\! v^c}(\bar x,y) \right],
\\
\frac{\partial}{\partial y} {^{GQ}\! G^\delta}(x,y)
&=& 2 \left( {^{GG}\! G^a} + 2\, {^{GG}\! v^c}
\OO^\re
{^{GG}\! v^c}(x,y) + 2\, {^{GG}\! v^c}
\OO^\re
{^{GG}\! v^c}(\bar x,y) \right),
\end{eqnarray}
where the kernel ${^{GG}\! G^a}$ are part of the whole parity odd functions
derived in the fashion already explained above. Note that this set of
differential equations represents after separation of the $\theta$-structures
in fact two sets, one for $\Delta{^{QG}\! H}^\delta (x, y)$ and the other
one for $\Delta{^{QG}\!\overline H}^\delta (x, y)$. The two integration
constants can be easily determined from the vanishing of the conformal
moments ${^{GQ}\! G^c}_{j0}=0$ for $j>0$. Finally, we simplify the solution
by adding pure diagonal pieces containing $a$ and $c$ kernels and their
convolution as well as by removing symmetric terms that die out in
${^{GQ}\!G}$. Using again the supersymmetric relation
(\ref{con-SUSY-sol-K3}), we write our findings in the mixed channels in the
following way:
\begin{eqnarray}
\label{def-DH-QG-V}
\Delta{^{QG}\! H}^\delta (x, y)
&=& - \frac{x\bar x}{(y \bar y)^2}
\Delta{^{GQ}\!H}^\delta (\bar y, \bar x),
\quad
\Delta{^{QG}\!\overline H}^\delta (x, y)
= \frac{x \bar x}{(y \bar y)^2}
\Delta{^{GQ}\!\overline H}^\delta (y, x),
\\
\label{def-DH-GQ-V}
\Delta {^{GQ}\!H}^\delta (x, y)
&=& \Delta {^{GQ}\!\overline H}^\delta (\bar x, y)+
  20 \frac{x (x - \bar x)}{3y}
- 4 \frac{\bar x (3 + 2 \bar x)}{3y} \ln\bar x
+ 4 \frac{x (3 + 2 x)}{3\bar y} \ln y
\nonumber\\
\Delta{^{GQ}\!\overline H}^\delta (x, y)
&=&
-\frac{61}{9} {\bar x} + 2 x {\bar x} \left(
      1+ \frac{25}{18}(x-{\bar x})- (3-10{\bar x} )\ln{y}+(3-10x)\ln{x}
	\right)
\\
& &+ \frac{\bar x \left( 6 - 19 \bar x + 6 \bar x^2 \right)}{3y}
- 2\frac{\bar x \left( y + x ( \bar x - x) \right)}{\bar y} \ln y
+ 2 \frac{x \left( \bar y + \bar x ( x - \bar x) \right)}{y} \ln x ,
\nonumber
\end{eqnarray}
The last task is to construct the $GG$ kernel by means of
the constraint (\ref{con-SUSY-sol-K1+3}) where ${^{QQ}\! G^c} \equiv 0$.
As already described, the solution is obtained in a straightforward manner
and reads:
\begin{eqnarray}
\label{def-DH-GG-V}
\Delta{^{GG}\!H}^\delta (x, y)
&=& \Delta{^{GG}\!\overline H}^\delta (\bar x, y)
 +\frac{x (3 - 13 {\bar x})}{y^2} -\frac{2 x(2 + 3 x)}{y{\bar y}}
- 2 \frac{\bar x}{y}
\left( 2 \frac{ \bar x - x }{\bar y}
+ \frac{2 + 3 \bar x}{y} \right) \ln\bar x
\nonumber\\
& &
- 2 \frac{x}{\bar y}
\left(
2 \frac{x - \bar x}{y}
+ \frac{2 + 3 x}{\bar y} \right)\ln y,
\\
\Delta{^{GG}\!\overline H}^\delta (x, y)
&=&
	\frac{(1-x^2)(1-21{\bar x})}{3 y^2} -
 	 \frac{{\bar x}(19(1+x) -36 x^2)}{3y}
	+ \frac{2{\bar x}^3}{3\bar y}
\nonumber\\
& &
+2 \left( \frac{x^3}{3 y^2}
- \frac{x^2 (21 - 20 x) }{3 y}
-2 \frac{x \bar x^2}{\bar y} \right)\ln{x}
+ 2 \left(
\frac{\bar x^3}{3\bar y^2}
- \frac{\bar x^2 (21 - 20 \bar x) }{3\bar y}
- 2 \frac{{\bar x} x^2}{y}
\right)\ln{y}  .
\nonumber
\end{eqnarray}
This completes our construction of the complete $\mbox{\boldmath $G$}$
matrices relevant for parity odd and even sector.

\subsection{Restoration of remaining diagonal terms.}

Now we have to find the remaining diagonal pieces of the ER-BL kernels,
which are expected to have a simple representation in terms of the kernels
known from Eqs.\ (\ref{v-kernels-A})-(\ref{def-f-functions-cAnotB}). The
DGLAP representations of the missing terms are obtained with the help of
Eq.\ (\ref{get-D}), where we can interchange the exclusive convolution with
the forward limit procedure:
\begin{eqnarray}
\label{get-D-as-LIM-1}
\mbox{\boldmath$D$}^I (z)
= \mbox{\boldmath$P$}^{(1)I}(z)
-
\left\{
- \mbox{\boldmath$\dot P$}^I \OO^\ri
\left( \mbox{\boldmath$P$}^{(0)I} + \frac{\beta_0}{2} \1 \right)
- \left[
\mbox{\boldmath$g$} \OO^\ri_{,} \mbox{\boldmath$P$}^{(0)I}
\right]_-
+ \mbox{\boldmath$G$}^{I}
\right\}(z).
\end{eqnarray}
Here we have introduced the analogous notation for the DGLAP
representations as used in the non-forward kinematics. The DGLAP
kernels to LO read
\begin{eqnarray}
\mbox{\boldmath$P$}^{(0)I}
\equiv
{\rm LIM} \mbox{\boldmath$V$}^{(0)I}
=
\left(
\begin{array}{ll}
C_F \left[ {^{QQ}\!p} (z) \right]_+
& 2 T_F N_f \, {^{QG}\!p^I} (z) \\
C_F\, {^{GQ}\!p^I} (z)
& C_A \left[{^{GG}\!p^I} (z) \right]_+
- \frac{\beta_0}{2} \delta(1-z)
\end{array}
\right)\quad\mbox{for}\quad I=\{A,V\},
\end{eqnarray}
where the entries are decomposed in the same manner as in
Eqs.\ (\ref{v-kernels-A}) and (\ref{v-kernels-V}), i.e.
\begin{eqnarray}
\label{p-kernels-A}
&& {^{QQ}\!p} \equiv {^{QQ}\!p^a} + {^{QQ}\!p^b}
\quad
{^{QG}\!p^A} \equiv -{^{QG}\!p^a} , \quad
{^{GQ}\!p^A} \equiv {^{GQ}\!p^a} ,
\quad
{^{GG}\!p^A} \equiv \left[2\, {^{GG}\!p^a} + {^{GG}\!p^b} \right]_+,
\\
\label{p-kernels-V}
&&{^{QG}\!p^V} \equiv {^{QG}\!p^A} - 2 {^{QG}\!p^c},
\quad
{^{GB}\!p^V} \equiv {^{GB}\!p^A} + 2 {^{GB}\!p^c}
\quad \mbox{for} \quad B=\left\{Q,G\right\},
\end{eqnarray}
and the functions ${^{AB}\!p^i}$ are defined in the following way
\begin{eqnarray}
\label{def-p-abc}
&&{^{QQ}\!p^a} = \frac{1}{2} {^{GG}\!p^a} = 1-z, \quad
{^{QQ}\!p^b} = {^{GG}\!p^b}=\frac{2z}{1-z},
\quad
{^{QQ}\!p^c} = \frac{1}{3}{^{GG}\!p^c}= \frac{(1-z)^3}{3z},
\\
&&
{^{QG}\!p^a} = 1-2 z, \quad {^{QG}\!p^c}= -(1-z)^2,
\qquad
{^{GQ}\!p^a} = 2- z, \quad {^{GQ}\!p^c} = \frac{(1-z)^2}{z}.
\nonumber
\end{eqnarray}
The ``+''-prescription is now uniquely defined in the $QQ$ and
$GG$ channel as  the conventional one:
\begin{eqnarray*}
\left[ {^{AA}\!p}(z) \right]_+ = {^{AA}\!p}(z)
- \delta(1 -z) \int_0^1 dy\, {^{AA}\!p}(y).
\end{eqnarray*}
The limit of the dotted kernels defined in Eqs.\ (\ref{dotv-kernels-I})-
(\ref{def-dot-addenda}) give
\begin{eqnarray}
\mbox{\boldmath$\dot P$}^{I}
\equiv
{\rm LIM} \mbox{\boldmath$\dot V$}^{I}
=
\left(
\begin{array}{cc}
C_F  {^{QQ}\!\dot p} (z)
& 2 T_F N_f \, {^{QG}\!\dot p^I} (z) \\
C_F\, {^{GQ}\!\dot p^I} (z)
& C_A {^{GG}\!\dot p^I} (z)
\end{array}
\right),
\end{eqnarray}
with
\begin{eqnarray}
&&{^{AA}\!\dot p}^A = \left[{^{AA}\! p^A} \ln z +{^{AA}\! p^a}\right]_+,
\quad
{^{AB}\!\dot p^A}= {^{AB}\! p^A}  \ln z \mp {^{QQ}\! p^a}
\mbox{\ \ for\ \ } \left\{ { AB=QG \atop AB =GQ} \right.,
\\
&&{^{QG}\!\dot p^V}  =
{^{QG}\! p^V} \left( \ln z+\frac{13}{6}\right)  - \frac{2}{3}{^{GQ}\! p^c}
+ \frac{13}{6} {^{QG}\! p^a}  - {^{QQ}\! p^a},
\quad
{^{GQ}\!\dot p^V}= - {^{GQ}\!\dot p^A},
\\
&&{^{GG}\!\dot p^V} =
{^{GG}\!\dot p^A} + 2{^{GG}\! p^c} \ln z + \frac{11}{3} {^{GG}\! p^c} .
\nonumber
\end{eqnarray}
The limit of the $\mbox{\boldmath$g$}$ matrix defined in Eq.\
(\ref{set-g-kernels}) is
\begin{eqnarray}
\label{set-g-kernels-DGLAP}
\mbox{\boldmath$g$} (z) \equiv {\rm LIM}\, \mbox{\boldmath$g$} (x,y) =
2 \left(
\begin{array}{cc}
- C_F \left[ \frac{ \ln \left( 1 - z\right) }{1 - z} \right]_+
& 0 \\
C_F
& - C_A\left[ \frac{ \ln \left( 1 -z\right) }{z(1-z)} \right]_+
\end{array}
\right).
\end{eqnarray}
Finally, we need the complete forward limit of the $\mbox{\boldmath$G$}$
kernel given in Eqs.\ (\ref{LIM-S-G}) - (\ref{LIM-S-bH}), i.e.\ we have to
determine ${^{AB}\!T^I}$ and ${^{AB}\!\overline T^I}$. From the result
of the $\mbox{\boldmath$G$}$ matrix given in Eqs.\
(\ref{G-kernel-S})-(\ref{kernel-S-bH}) with the addenda
(\ref{def-DH-QG-A})-(\ref{def-DH-GG-A}) and
(\ref{def-DH-QG-V})-(\ref{def-DH-GG-A}), we obtain the following
expressions for the sum or differences in the parity odd and even
sector, respectively:
\begin{eqnarray}
\label{def-T-AB-A}
&&{^{AA}\!T^A}(z)+{^{AA}\!\overline T^A}(-z) =
4 {^{AA}\!p^a}(-z) \ln z + 8 {^{AA}\!p^a}(z),
\\
&&{^{AB}\!T^A}(z)+{^{AB}\!\overline T^A}(-z)
=
\mp 4 {^{BA}\!p^a}(-z) \ln z \mp 12 {^{QQ}\!p^a}(z)
\mbox{\ \ for\ \ } \left\{{AB=QG \atop AB=GQ}\right. ,
\nonumber \\
\label{def-T-AB-V}
&&{^{QQ}\!T^V}(z)-{^{QQ}\!\overline T^V}(-z) = 0,
\quad
{^{GG}\!T^V}(z)-{^{GG}\!\overline T^V}(-z)
= - \frac{8}{3} {^{GG}\!p^c}(-z)\ln z,
\\
&&{^{AB}\!T^V}(z)-{^{BA}\!\overline T^V}(-z)
= \frac{8}{3} {^{QQ}\!p^c}(-z) \left(2\pm 3 \ln z \right)
+ \frac{16}{3}{^{QQ}\!p^a}(z) \ln z
\mbox{\ \ for\ \ } \left\{{AB=QG \atop AB=GQ}\right. .
\nonumber
\end{eqnarray}

With these results we are now able to find $\mbox{\boldmath$D$}^I (z)$ from
the known DGLAP kernels in NLO by means of Eq.\ (\ref{get-D-as-LIM-1}). It
remains a simple exercise to express these findings in terms of convolutions
of $a,b$ and $c$-type kernels defined in Eqs.\ (\ref{def-p-abc}). In the
following two subsubsections we treat the parity odd and even sector
separately. We start with the parity odd sector responsible for the
evolution of axial-vector distribution amplitudes, since the structure is
simpler than the one in the parity even sector.

\subsubsection{Parity odd sector.}

We already have explicit expressions for all kernels entering on the r.h.s.\
of Eq.\ (\ref{get-D-as-LIM-1}). It remains to convolute the
$\mbox{\boldmath$\dot P$}$ and $\mbox{\boldmath$g$}$ kernels with the LO
ones. Since cancellations between these separate terms drastically simplify
the final result, we present it as entries of the matrix
\begin{eqnarray}
\label{def-A-matrix}
\mbox{\boldmath$\cal A$}^I
= - \mbox{\boldmath$\dot P$}^I \OO^\ri
\left( \mbox{\boldmath$P$}^{(0)I} + \frac{\beta_0}{2} \1 \right)
- \left[
\mbox{\boldmath$g$} \OO^\ri_{,} \mbox{\boldmath$P$}^{(0)I}
\right]_- .
\end{eqnarray}
In other words this matrix contains the diagonal part that we could
not separate from the conformal anomalies. The calculation is
straightforward and results in
\begin{eqnarray}
\label{def-A-QQ-A}
{^{QQ}\!{\cal A}^A}
= 2 C_F T_F N_f \left\{ 4(1-z)-(1-3z)\ln z -(1+z)\ln^2 z \right\}
+ {^{\rm NS}\!{\cal A}} ,
\end{eqnarray}
where the convolution in the non-singlet part $\ {^{\rm NS}\!{\cal A}}
= - C_F^2 \, {^{QQ}\!{\dot p}} {\displaystyle \OO^\ri} {^{QQ}\!{p}} -
C_F \beta_0  {^{QQ}\!{\dot p}}/2$ has been worked out in
Eq.\ (\ref{con-p-dp-NS}). The remaining channels read:
\begin{eqnarray}
\label{def-A-QG-A}
{^{QG}\!{\cal A}^A}
&=&
2 C_F T_F N_f \Bigg\{
(1 - 2z)\left[-\frac{7}{4}+ 2\zeta (2)  -\frac{1}{2} \ln^2 z
+ 2 \ln z \ln (1 - z) - \ln^2 (1-z)\right]
\nonumber\\
&& + (1 - z) \left[-4 - 3 \ln z + 4 \ln (1-z) \right] \Bigg\}
\nonumber\\
&& + 2 C_A T_F N_f \Bigg\{ (1 - 2 z)\left[2 \zeta(2) + \ln^2(1 - z)\right]
-2(1 -z)\left[1+2 \ln(1 - z)\right] \\
&&- 6 \ln z -2(1 + z)\ln^2 z \Bigg\},
\nonumber\\
\label{def-A-GQ-A}
{^{GQ}\!{\cal A}^A} &=&
C_F^2 \left\{\frac{1 - 7z}{2} +
\frac{4 + 3z}{2} \ln z  - 4\ln(1 - z)  +
(2 - z)\left[\frac{1}{2} \ln^2 z- \ln^2 (1-z) \right] \right\}
\nonumber\\
&& -C_F \frac{\beta_0}{2} \left\{3-z +(2 - z)\ln z\right\}
+
C_F C_A \Bigg\{
- 5(1 - z) + (4 - 9z) \ln z \\
&&+ \frac{3-2z+3z^2}{z} \ln(1-z)
+ (4 + z) \ln^2 z
+ (2 - z) \ln(1 - z)\left[ \ln(1 - z) - 2\ln z \right] \Bigg\},
\nonumber\\
\label{def-A-GG-A}
{^{GG}\!{\cal A}^A} &=&
C_A^2 \Bigg[
2\frac{2 - 3z + 2z^2}{1 - z}
\left\{
- 2 \zeta(2) + \ln^2 z - 2 \ln z \ln (1 - z)\right\} \\
&&+ 4 (1 - z)\left[4 - \ln(1 - z)\right] + 12 (2 - z)\ln z
+ 4 (1 + z)\ln^2 z \Bigg]_+ \nonumber\\
&& + 2 C_F T_F N_f
\left\{ - 10(1 - z) - (7 + z)\ln z - (1 + z) \ln^2 z \right\}.
\nonumber
\end{eqnarray}

Now we are ready to extract the remaining diagonal terms  ${^{AB}\! D}^A$
from
\begin{eqnarray}
\label{get-D-as-LIM-2}
\mbox{\boldmath$D$}^A (z)
= \mbox{\boldmath$P$}^{(1)A}(z)
- \left\{ \mbox{\boldmath$\cal A$}^A + \mbox{\boldmath$G$}^{A}
\right\}(z).
\end{eqnarray}
As observed in the flavor non-singlet case, all double logs and
Spence functions appearing in the DGLAP kernel in NLO \cite{MerNee96,Vog96}
are contained in the sum of $\mbox{\boldmath$\cal A$}^A$ given in Eqs.\
(\ref{def-A-QQ-A})-(\ref{def-A-GG-A}) and $\mbox{\boldmath$G$}^{A}$
defined in Eqs.\ (\ref{LIM-S-G})-(\ref{LIM-S-bH}) together with Eq.\
(\ref{def-T-AB-A}). The remaining single log terms give us a hint to
write the entries of $\mbox{\boldmath$D$}^A (z)$ as convolution of the
simple kernels (\ref{def-p-abc}), which allows us to restore the ER-BL
representation. In the following we present this issue in detail.

In the $QQ$ channel  the only new information arises from the pure singlet
term. Employing formula (\ref{get-D-as-LIM-2}), we immediately find that
this term is
\begin{eqnarray*}
- 6 \, C_F T_F N_f {^{QQ}\!  p}^a
\Rightarrow
- 6\, C_F T_F N_f {^{QQ}\!  v}^a.
\end{eqnarray*}
Adding the result of the flavor non-singlet sector, the $QQ$ entry reads
\begin{eqnarray}
\label{D-QQ-o}
{^{QQ}\! D}^A
&=& {^{\rm NS}\! D}- 6\, C_F T_F N_f {^{QQ}\!  v^a},
\\
\label{D-NS}
{^{\rm NS}\! D}&=&C_F^2 \left[ D_F \right]_+
- C_F \frac{\beta_0}{2} \left[ D_\beta \right]_+
- C_F \left( C_F - \frac{C_A}{2} \right)
\left[ \frac{4}{3} {^{QQ}\!  v} + 2\, {^{QQ}\! v^a} \right]_+ ,
\end{eqnarray}
where $D_F$, $D_\beta$ are given by Eqs.\ (\ref{DF-QQ}) and
(\ref{Dbeta-QQ}), respectively.

In the $QG$ channel we find in an analogous way
\begin{eqnarray*}
{^{QG}\! D^A} (z) &=&
3\, C_F T_F N_f
\left\{
-(1+2z)\ln z -3(1-z) - \frac{1}{2}(1-2z)
\right\} \\
& &- 2\, C_A T_F N_f
\left\{
- 3 (1+2z)\ln z -9 (1-z)
+ \left[ 1 + 2 \zeta (2) \right] (1-2z)
\right\}
\nonumber
\end{eqnarray*}
by a convolution of $a$-kernels,
\begin{eqnarray*}
{^{QQ}\! p^a} \OO^\ri {^{QG} p^a} =  -(1+2z) \ln z -3(1-z),
\end{eqnarray*}
together with ${^{QG} p^a} = 1-2 z$ the ER--BL-representation:
\begin{eqnarray}
{^{QG}\! D}^A
&=& 3\, C_F T_F N_f
\left\{
{^{QQ} v}^a \OO^\re {^{QG} v}^a
- \frac{1}{2} {^{QG} v}^a
\right\} \\
& &- 2\, C_A T_F N_f
\left\{
3\, {^{QQ} v}^a \OO^\re {^{QG} v}^a
+ \left[ 1 + 2 \zeta (2) \right] {^{QG} v}^a
\right\}.
\nonumber
\end{eqnarray}
A more complicated expression arises in the $GQ$ channel:
\begin{eqnarray*}
{^{GQ} D}^A(z)
&=& C_F^2
\left\{(2-z)\ln (1-z)-(4+z)\ln z-6(1-z)
- \frac{3}{2}(2-z)
\right\}\\
&&- C_F \frac{\beta_0}{2}
\left\{
2(2-z) \ln(1-z)-(2-z)\ln z+ \frac{3}{2} z
- \frac{1}{6} (2-z)
\right\} , \\
& &- C_F C_A
\left\{
\frac{3+2z-z^2}{z} \ln(1-z) - 2\ln z - 6(1 - z) -
\left[\frac{7}{3} -2 \zeta (2) \right] (2 - z)
\right\} , \nonumber
\end{eqnarray*}
which is characterized by new terms containing $(2-z) \ln (1-z)$ and
$[\ln (1-z)]/z$. Such contributions can be generated by convolutions
of $b$-kernels with $a$- and $c$-kernels, respectively\footnote{Note that
we can eliminate one of them e.g.\ ${^{GQ} p^a} {\displaystyle \OO^\ri}
\left[{^{QQ} p}\right]_+ =\left[ {^{GG} p^A}\right]_+{\displaystyle \OO^\ri}
{^{GQ} p^a} - 3 {^{GG} p^a} {\displaystyle \OO^\ri} {^{GQ} p^a}/2 + 3
{^{GQ} p^a}/2 $.}
\begin{eqnarray*}
&&{^{GQ} p^a} \OO^\ri \left[{^{QQ} p}\right]_+  =
2(2-z)\ln(1-z) -(2-z)\ln z +3/2 z,
\\
&&{^{GG} p^A} \OO^\ri {^{GQ} p^a} =
2(2-z) \ln(1-z)-2  (4+z) \ln z-12(1-z), \\
&& {^{GG} p^A}  \OO^\ri {^{GQ} p^c} =
2\frac{(1-z)^2}{z}\ln(1 - z) + 2 (4 + z)\ln z + 10 (1 - z),
\\
&&{^{GG} p^a} \OO^\ri {^{GQ} p^a}  -2(2+z) \ln z- 6(1-z),
\quad {^{GQ} p^a} = 2-z.
\end{eqnarray*}
The result can be written as
\begin{eqnarray}
{^{GQ} D}^A
&=& C_F^2
\left\{ \frac{1}{2}
{^{GG} v}^A  \OO^\re {^{GQ} v}^a
- \frac{3}{2} {^{GQ} v}^a
\right\}
- C_F \frac{\beta_0}{2}
\left\{
{^{GQ} v}^a \OO^\re \left[ {^{QQ} v} \right]_+
- \frac{1}{6} {^{GQ} v}^a
\right\} , \\
& &- C_F C_A
\left\{
\frac{3}{2}  {^{GG} v}^A \OO^\re {^{GQ} v}^c
+ \left[
2  {^{GG} v}^A  - \frac{1}{2} {^{GG} v}^a
\right] \OO^\re {^{GQ} v}^a
- \left[ \frac{7}{3} - 2 \zeta (2) \right] {^{GQ} v}^a
\right\} . \nonumber
\end{eqnarray}

The remaining entry reads in the DGLAP representation
\begin{eqnarray*}
{^{GG} D}^A(z)
&=& C_A^2
\Bigg\{
4(1-z)\ln(1 - z)-2(5+3z)\ln z +
\frac{4}{3} \left[\frac{2-3z+2z^2}{1 - z}\right]_+ -
\frac{41}{2} (1-z) \\
& &-2 \delta(1-z)
\Bigg\}
- C_A \frac{\beta_0}{2}
\left\{
2(1 + z)\ln z + \frac{10}{3} \left[\frac{2-3z + 2z^2}{1 - z}\right]_+ +
6(1 - z) + 2\delta(1 - z)
\right\} \nonumber\\
&&- C_F T_F N_f
\left\{
-4(1+z)\ln z -10(1-z) +\delta(1-z)
\right\}. \nonumber
\end{eqnarray*}
and can be expressed by convolutions of $a$- and $b$-type kernels,
\begin{eqnarray*}
&&{^{GG} p^A} \OO^\ri {^{GG} p^a} =
4 (1-z) \ln(1-z)-4(2+z) \ln z-16 (1-z)
\\
&& {^{GG} p^a} \OO^\ri {^{GG} p^a}=-4(1+z)\ln z -8 (1-z),\
{^{GG} p^A} = 2 \frac{2-3z+2z^2}{1 - z},\ {^{GG} p^a}=2(1-z),
\end{eqnarray*}
from which the desired kernel follows:
\begin{eqnarray}
{^{GG} D}^A
&=& C_A^2
\left\{
\left[
 {^{GG} v}^A + \frac{1}{2} {^{GG} v}^a
\right] \OO^\re {^{GG} v}^a
+ \frac{2}{3}  {^{GG} v}^A
- \frac{1}{4} {^{GG} v}^a - 2 \delta(x - y)
\right\}
\nonumber\\
&-& C_A \frac{\beta_0}{2}
\left\{
- \frac{1}{2} {^{GG} v}^a \OO^\re {^{GG} v}^a
+ \frac{5}{3} {^{GG} v}^A
+ {^{GG} v}^a + 2 \delta(x - y)
\right\} \\
&-& C_F T_F N_f
\left\{
{^{GG} v}^a \OO^\re {^{GG} v}^a
- {^{GG} v}^a + \delta(x - y)
\right\}. \nonumber
\end{eqnarray}
These results provide us the missing information to obtain the whole
kernel from Eq.\ (\ref{def-str-NLO}) in the flavor singlet sector
for odd parity.

\subsubsection{Parity even sector.}

In analogous way we treat now the parity even sector, starting with the
calculation of the matrix $\mbox{\boldmath$\cal A$}^V$. For brevity we
present here only the results for the difference between vector and
axial-vector case, i.e.\  $\mbox{\boldmath$\cal
A$}^\delta = \mbox{\boldmath$\cal A$}^V - \mbox{\boldmath$\cal A$}^A$
defined in Eq.\ (\ref{def-A-matrix}). This difference is induced by the
$c$-kernels. Thus, it is not surprising that ${^{QG}\!{\cal A}^\delta}$
consists of terms  $1/z \ln^iz \ln^j(1-z)$  and  $z^2 \ln^iz \ln^j(1-z)$
with $0\le i + j\le 2$:
\begin{eqnarray}
\label{def-A-QQ-delta}
{^{QQ}\!{\cal A}^\delta}&=&
-2 C_F T_F N_f \left\{
\frac{1}{9z} (1-z)(31 + 142z - 5z^2)
+ \frac{4}{3z} (1 + 7z + 7 z^2 - z^3) \ln z
\right\},
\\
{^{QG}\!{\cal A}^\delta}&=&
\label{def-A-QG-delta}
2 C_F T_F N_f \Bigg\{(1 - z)(10-7z)
+ 2 (1 - z)^2 \left[ - 2 \zeta(2) + 3 \ln\frac{z}{1 - z} +
\ln^2 \frac{z}{1 - z}\right]
\nonumber\\
&& -(1 - 4z + (1 - 2z)\ln z)\ln z\Bigg\} +
2 C_A T_F N_f \Bigg\{
- \frac{1 - z}{18 z} (88 + 175 z + 367 z^2) \\
&&-
\frac{1}{3 z} (4 + 13 z + 88 z^2) \ln z + (1 - 2 z) \ln^2 z \nonumber\\
&&- (1 - z)^2
\left[
4 \zeta(2) + \frac{4}{3} \ln(1 - z) + 2 \ln^2 (1-z)
\right] \Bigg\},
\nonumber\\
\label{def-A-GQ-delta}
{^{GQ}\!{\cal A}^\delta}&=&
C_F^2 \left\{
- 3(1 - z) - \left[3 z + (2-z)\ln z \right] \ln z
- 2\frac{(1 - z)^2}{z} \ln^2(1 - z)
\right\} \nonumber\\
&&+ C_F C_A \Bigg\{
\frac{1}{z} (17 + 3z - 21 z^2 + z^3)
+ \frac{1}{3z} (31 + 24 z + 57 z^2 - 4 z^3) \ln z \\
&&+ \frac{1}{z} (2 - 2 z + z^2) \ln^2 z
+ \frac {2 (1 - z)^2}{z}
\left[ - 3 - 2 \ln z + \ln (1 - z) \right] \ln (1-z)
\Bigg\} \nonumber\\
&&+ C_F\beta_0 \big\{ 1 - z + (2 - z) \ln z \big\} ,
\nonumber\\
\label{def-A-GG-delta}
{^{GG}\!{\cal A}^\delta}&=&
C_A^2 \Bigg\{
\frac{2(1 - z)}{9z} (112 - 2 z + 112 z^2)
+ \frac{2}{3z} (22 - 18 z + 81 z^2 - 11 z^3) \ln z
\nonumber\\
&&- \frac{2 (1-z)^3}{3 z}\left[6 \zeta(2) + 11 \ln(1-z)-3\ln^2 z
+ 6 \ln z \ln(1-z) \right]\Bigg\}
\\
&&+ 2 C_F T_F N_f \left\{ 4 (2 - z)(1 - z) + 2(3 - z) \ln z\right\}
\nonumber
\end{eqnarray}

Taking these results and the corresponding ones for the axial-vector case
in Eqs.\ (\ref{def-A-QQ-A})-(\ref{def-A-GG-A}) as well as the definitions
(\ref{LIM-S-G})-(\ref{LIM-S-bH}) together with (\ref{def-T-AB-V}) for the
$\mbox{\boldmath$G$}$ kernels, we obtain from  the NLO kernel
\cite{FurPet82} the diagonal terms
\begin{eqnarray}
\label{get-D-as-LIM-3}
\mbox{\boldmath$D$}^V (z)
= \mbox{\boldmath$P$}^{(1)V}(z)
-
\left\{ \mbox{\boldmath$\cal A$}^A +\mbox{\boldmath$\cal A$}^\delta+
\mbox{\boldmath$G$}^{V}
\right\}(z).
\end{eqnarray}
Again we observe a cancellation of all double log terms. As to be expected, the
convolution formulae from the previous subsection are not able to express
all terms in $\mbox{\boldmath$D$}^V$ as convolutions. What is missing are
convolutions of $c$-type kernels with $a,b$ and $c$ ones. We show in the
following, that this is enough to restore $\mbox{\boldmath$D$}^V$ for
non-forward kinematics.

For the pure singlet part in the $QQ$ channel we have
\begin{eqnarray*}
2 C_F T_F N_f \left\{
\frac{2}{3} {^{QQ}\! p^a} \OO^\ri {^{QQ} p^a}-
 2 {^{QQ} p^a} + (1-z)\frac{17+46 z+17z^2}{3z} +4(1+z)\frac{1+8z+z^2}{3z}
\ln z
\right\},
\end{eqnarray*}
where the convolution of $a$ kernels is
$ {^{QQ}\! p^a} \OO{^{QQ} p^a} = -2 {^{QQ} p^a}(z)- {^{QQ} p^a}(-z) \ln z$.
Making use of
\begin{eqnarray*}
{^{QQ}\! p^c} \OO^\ri {^{QQ} p^c}
=
 -(1-z)\frac{11+38z+11z^2}{27z} -
   (1+z)\frac{1+8z+z^2}{9z}\ln z, \quad {^{QQ} p^c}= \frac{(1 - z)^3}{3z}.
\end{eqnarray*}
we can restore the ER-BL representation
\begin{eqnarray}
\label{D-QQ-e}
{^{QQ}\! D}^V
&=& {^{\rm NS}\! D}+ 4\, C_F T_F N_f \left\{
\frac{1}{3} {^{QQ}\!v}^a \OO^\re {^{QQ}\!v}^a
- 6{^{QQ}\!v}^c \OO^\re {^{QQ}\!v}^c
- {^{QQ}\!v}^a + \frac{7}{6}{^{QQ}\!v}^c
\right\},
\end{eqnarray}
For the $QG$ entry we find
\begin{eqnarray*}
{^{QG}\! D}^V(z)&=& 2 C_F T_F N_f\Bigg\{
- \frac{1}{2} {^{QQ}\! p^a} \OO^\ri {^{QG} p^a}-\frac{3}{4}{^{QQ}\! p^a}
 -3 {^{QQ}\! p^c}-(1-z) -(1-2z+2z^2)\ln z
\nonumber\\
&&+ 2(1-z)^2 \ln (1-z)\Bigg\} +
2 C_A T_F N_f\Bigg\{
\frac{130}{3} {^{QQ}\! p^a} \OO^\ri {^{QG} p^a}+
  \left(\frac{55}{9} - 2 \zeta(2) \right){^{QG}\! p^a}
\\
&&- \left(\frac{301}{18}+ 4\zeta (2)\right)  {^{QG} p^c} +
8(1-z)\frac{7+ 130z+70z^2}{9z}+ \frac{8}{3} (20+44 z+5z^2)\ln z
\nonumber\\
&&+
 \frac{16}{3}(1-z)^2 \ln(1-z)\Bigg\}.
\nonumber
\end{eqnarray*}
With the help of
\begin{eqnarray*}
\left[ {^{QQ}\!p} \right]_+  \OO^\ri {^{QG}\!p}^c& =&
1-z+(1-2z+2z^2)\ln z- 2(1-z)^2 \ln (1 - z),
\\
{^{QQ}\!p}^c \OO^\ri {^{QG}\!p}^c& =&
-(1-z)\frac{1+19z+10 z^2}{9z} -
 \frac{1}{3} (3+6z+z^2)\ln z,
\end{eqnarray*}
we can express the remaining terms as convolutions and obtain so
the ER-BL representation:
\begin{eqnarray}
{^{QG}\!D}^V
&=& - C_F T_F N_f
\left\{
2\left[ {^{QQ}\!v} \right]_+  \OO^\re {^{QG}\!v}^c
+ {^{QQ}\!v}^a \OO^\re {^{QG}\!v}^a
+ \frac{3}{2} {^{QG}\!v}^a + 6 {^{QG}\!v}^c
\right\}
\nonumber\\
&+& 2\, C_A T_F N_f
\Bigg\{
- \left[ \frac{8}{3} \left[{^{QQ}\!v} \right]_+
+ 56 {^{QQ}\!v}^c \right]
\OO^\re {^{QG}\!v}^c + \frac{130}{3} {^{QQ}\!v}^a
\OO^\re {^{QG}\!v}^a
\\
&&\hspace{2cm}+ \left[ \frac{55}{9} - 2 \zeta (2) \right] {^{QG}\!v}^a
- \left[\frac{301}{18} + 4 \zeta (2) \right] {^{QG}\!v}^c
\Bigg\}. \nonumber
\end{eqnarray}
A similar situation appears in the $GQ$ and $GG$ channel. Beside the known
convolutions we need only the following ones
\begin{eqnarray*}
 {^{GG}\!p}^A
\OO^\ri {^{GQ}\!p}^c &=&
5(1 - z) + (4 + z) \ln z + \frac{(1-z^2)}{z} \ln(1-z),
\\
{^{GG}\!p}^c \OO^\ri {^{GQ}\!p}^c &=&
-(1-z)\frac{10+19z+z^2}{3z} -
  \frac{1+6z+3z^2}{z} \ln z,
\\
{^{GG}\!p}^A
\OO^\ri {^{GG}\!p}^c &=&
10(1-z) + (6 + 3 z + z^2)\ln z+ \frac{(1 -z)^3}{z} \ln(1 - z),
\\
{^{GG}\!p}^c \OO^\ri {^{GG}\!p}^c &=&
-(1-z)\frac{11+38z+11z^2}{3z} -(1+z)\frac{1+8z+z^2}{z} \ln z,
\end{eqnarray*}
and thus find the following diagonal kernels
\begin{eqnarray}
{^{GQ}\!D}^V
&=& C_F^2
\left\{
-{^{GG}\!v}^A
\OO^\re \left[\frac{1}{2} {^{GQ}\!v}^a
+ 3 {^{GQ}\!v}^c \right]
-5 {^{GG}\!v}^a  \OO^\re {^{GQ}\!v}^a
- 3 {^{GQ}\!v}^a
\right\} \nonumber\\
&&- C_F \beta_0
\left\{ {^{GG}\!v}^A  \OO^\re
\left[\frac{1}{2} {^{GQ}\!v}^a +  {^{GQ}\!v}^c \right]
+ \frac{3}{4} {^{GG}\!v}^a  \OO^\re {^{GQ}\!v}^a
+ \frac{5}{3} {^{GQ}\!v}^a
\right\} \\
&&+ C_F C_A
\Bigg\{
- {^{GG}\!v}^A \OO^\re
\left[ {^{GQ}\!v}^a - \frac{3}{2} {^{GQ}\!v}^c \right]
-\frac{25}{6} {^{GG}\!v}^a \OO^\re {^{GQ}\!v}^a
+ 9 {^{GG}\!v}^c \OO^\re {^{GQ}\!v}^c
\nonumber\\
&&\hspace{2cm} - \left( \frac{43}{9}
+ 2 \zeta(2) \right) {^{GQ}\!v}^a
+ \left( \frac{8}{9} - 4 \zeta(2) \right) {^{GQ}\!v}^c
\Bigg\} , \nonumber\\
{^{GG}\!D}^V
&=& C_A^2
\Bigg\{
 {^{GG}\!v}^A
\OO^\re \left[ {^{GG}\!v}^a + \frac{11}{3} {^{GG}\!v}^c \right]
- 14 {^{GG}\!v}^a \OO^\re {^{GG}\!v}^a
+ 12{^{GG}\!v}^c \OO^\re {^{GG}\!v}^c \nonumber\\
&+& \frac{2}{3} {^{GG}\!v}^A
- \frac{131}{12} {^{GG}\!v}^a
+ \frac{91}{18} {^{GG}\!v}^c - 2 \delta(x - y)
\Bigg\} \\
&-& C_A \frac{\beta_0}{2}
\left\{
- \frac{1}{2} {^{GG}\!v}^a \OO^\re {^{GG}\!v}^a
+ \frac{5}{3}  {^{GG}\!v}^V
+ 3 {^{GG}\!v}^a + \frac{13}{3} {^{GG}\!v}^c + 2 \delta(x - y)
\right\} \nonumber\\
&+& C_F T_F N_f
\left\{
{^{GG}\!v}^a \OO^\re {^{GG}\!v}^a
+ \frac{4}{3} {^{GG}\!v}^c -\delta(x-y)
\right\} . \nonumber
\end{eqnarray}
It is worth mentioning that our result for the evolution kernels in the
parity even singlet sector possesses the correct conformal moments in both
the physical and unphysical sectors. This is to be contrasted with an
explicit momentum fraction space calculation at LO and quark bubble
insertions in NLO kernels for the mixed channels \cite{BelMue98a} where the
improved kernels do not appear.

\section{Explicit representations of the kernels.}
\label{sec-rep}

In this section we give the explicit form of the ER-BL type kernels.
The analytical continuation to the whole $(x,y)$-plane allows to extend
our results to the skewed kinematics and extract from them the so-called
skewed DGLAP kernels, which are responsible for the evolution of SPDs
in the DGLAP region. The exclusive convolutions we need are straightforward
to handle and details are collected in appendix \ref{app-convol}. To
represent the final result in the shortest possible manner, we expanded
the output of the convolutions in terms of powers of logs. Then we rewrite
their arguments in terms of the ratio $x/y$ where the remaining ones posses
only an $x$-dependence. It turns out that these terms for the (non-)diagonal
entries are (anti)symmetric with respect to the interchanges $x\to \bar x,\
y\to \bar y$, so they have the support $0\leq x,y \leq 1$.

Now we should comment on the $\mbox{\boldmath $G$}$-kernels
constructed in section \ref{subsec-Con-G}. Although they are diagonal, this
property is lost for the unphysical moments when we map the second
$\theta$-structure to the first one by means of symmetry. A slight change
in the definition of ${{^{AB}\! H}^I}$ and ${{^{AB}\! \overline{H}}^I}$ in
Eqs.\ (\ref{kernel-S-H}) and (\ref{kernel-S-bH}), by subtracting a symmetric
function, i.e.\
\begin{eqnarray*}
2\left[\theta(y-x) \pm \theta(x-y) \right]
\left[
{^{AB}\!f^I}{\rm Li_2}(x)\pm
\left\{x\to {\bar x} \atop y\to {\bar y}\right\}
\right],
\end{eqnarray*}
from the common $\theta$-structure and adding them to the second one,
ensures that both parts are separately diagonal. Thus, in the following we
use instead of the definitions (\ref{kernel-S-H}) and (\ref{kernel-S-bH})
the definitions:
\begin{eqnarray}
{^{AB}\! h^I} &=&
\pm  2\,  {^{AB}\!\overline{f}^I}\ln{\bar x} \ln{y} -
2\,{^{AB}\!f^I} \left[{\rm Li_2}(x) + {\rm Li_2}(\bar y) \right],
\\
{^{AB}\!\overline{h}^I} &=&
\left(\mp {^{AB}\!\overline{f}^I} +{^{AB}\! f^I}\right)
\left[2 {\rm Li_2}\left(1 - \frac{x}{y}\right)+\ln^2{y} \right]
+2\,{^{AB}\! f^I}\left[{\rm Li_2}({\bar y})- \ln{x} \ln{y}\right]
\pm 2\, {^{AB}\!\overline{f}^I} {\rm Li_2}({\bar x}),
\nonumber
\end{eqnarray}
where again the upper (lower) sign corresponds to the (non)diagonal entries.
Together with the addenda given in eqs.\
(\ref{def-DH-QG-A})-(\ref{def-DH-GG-A}) as well as eqs.\
(\ref{def-DH-QG-V})-(\ref{def-DH-GG-V}) the kernels are separately diagonal
w.r.t.\ the Gegenbauer polynomials.

The LO kernels are defined in Eqs.\
(\ref{decomp-V-I})-(\ref{def-f-functions-cAnotB}) and the NLO kernels we
write for convenience as
\begin{eqnarray}
\label{decomp-V-I-NLO}
\mbox{\boldmath$V$}^{(1)I} (x, y)
=
\left(
\begin{array}{ll}
{^{QQ}\!V^{(1)I}}(x, y)  & {^{QG}V^{(1)I}}(x, y) \\
{^{GQ}\!V^{(1)I}}(x, y)  & \left[{^{GG}\!V^{(1)I}}(x, y)\right]_+
\end{array}
\right) +
\left(
\begin{array}{ll}
0 & \hfil 0 \hfil \\
0  & {^{GG}\!V^{(1)I}_{11}}
\end{array}
\right) \delta(x-y),
\end{eqnarray}
where contrary to our previous definitions
$\left[{^{AA}\!V^{(1)I}}\right]_+$ denotes now the {\em conventional}
regularization (\ref{def-con-+pre}) applied {\em also} in the $GG$ channel.
The contributions concentrated in $x=y$ are  simply fixed
 by the lowest conformal moment.  We have in the $GG$ channel:
\begin{eqnarray}
{^{GG}\!V^{(V)I}_{11}} &=&
- \frac{N_f}{108}\left( 35 C_A + 74 C_F \right),
\\
{^{GG}\!V^{(A)I}_{11}} &=&
C_A^2\left(\frac{95}{27} - \frac{14}{3} \zeta(2) + 2\zeta(3)  \right) +
\frac{N_f}{54} (29 C_A- 28 C_F).
\nonumber
\end{eqnarray}
The $QQ$ channel contains a contribution coming from the non-singlet kernel
and the so-called pure singlet part. Analogous to the forward case
we define the $\pm$ kernels in the flavor non-singlet sector to be
\begin{eqnarray}
& &\hspace{-2cm}
{{^{QQ}\! V}^{(1)\pm}} =
{{^{QQ}\! V}^{(1)V\pm}} = {{^{QQ}\! V}^{(1)A\mp}}=
\nonumber\\
& &\hspace{-1cm}
\Bigg[C_F^2\Bigg\{\theta(y-x)\Bigg[
\left(\frac{4}{3}-2\zeta(2)\right) \qqv +3 \frac{x}{y}-
\left(\frac{3}{2} \qqv-\frac{x}{2{\bar y}}\right)\ln\frac{x}{y}
-(\qqv-\qqbv) \ln\frac{x}{y}
\nonumber\\
& &\hspace{-1cm}
\times  \ln\left(1-\frac{x}{y}\right)
+\left(\qqv+\frac{x}{2{\bar y}}\right)\ln^2\frac{x}{y}
\Bigg]
- \frac{x}{2{\bar y}}\ln{x}\left(1+ \ln{x} -2\ln{\bar x}\right)
\Bigg\}
-C_F \frac{\beta_0}{2}\theta(y-x)
\\
& &\hspace{-1cm}
\times\Bigg\{
\frac{5}{3} \qqv + \frac{x}{y}+ \qqv \ln\frac{x}{y}
\Bigg\}
-C_F\left(C_F - \frac{C_A}{2}\right)\theta(y-x)\Bigg\{
\frac{4}{3} \qqv + 2 \frac{x}{y} + \HQQ(x,y)
\nonumber\\
& &\hspace{-1cm}
\mp \bHQQ({\bar x},y)\Bigg\}
+ \Bigg\{{x\to {\bar x} \atop  y\to {\bar y}}\Bigg\}\Bigg]_+
+ \frac{1 \pm 1}{2}
C_F \left(C_F-\frac{C_A}{2}\right)
\left[\frac{13}{2}-6\zeta(2) + 4\zeta(3)\right] \delta(x-y).
\nonumber
\end{eqnarray}
The kernels ${{^{QQ}\! V}^{(1)\pm}}$ are responsible for the evolution of
the sum (difference) and difference (sum) of quark and anti-quark
distributions in the parity even (odd) sector. Our results for the entries
of the parity odd sector read:
\begin{eqnarray}
& &\hspace{-0.5cm}{{^{QQ}\! V}^{(1)A}} =
\\
& &
{{^{QQ}\! V}^{(1)-}}
-2 C_F T_F N_f \Bigg\{\theta(y-x)
\frac{x}{\bar y}\ln\frac{x}{y} \left(1-\ln\frac{x}{y}
\right) -
\frac{x}{\bar y}\ln{x} \left(1-\ln{x}\right)
+ \left\{x\to {\bar x} \atop  y\to {\bar y}\right\}
\Bigg\},
\nonumber\\
& &\hspace{-0.5cm}{{^{QG}\! V}^{(1)A}} =
\\
& &
N_f T_F C_F \Bigg\{\theta(y-x)\Bigg[
2\left(5-2 \zeta(2)\right) \qgv + \frac{5x}{y{\bar y}}
+\left(
2\qgv -\frac{3 x}{{\bar y}^2} +\frac{2x}{y{\bar y}}
\right)\ln\frac{x}{y}
-2\Big(\qgv+\qgbv
\nonumber\\
& &
+ \frac{1}{y {\bar y}} \Big)\ln\left(1-\frac{x}{y}\right)
- \qgbv \ln^2\frac{x}{y}
+ \left(\qgv + \qgbv\right) \ln^2\left(\frac{y}{x}-1\right)
\Bigg]
-\qgv \ln{x} \big(2+2 \ln{\bar x}
\nonumber\\
& &
-\ln{x}\big)
+\frac{3x }{{\bar y}^2} \ln{x}
\Bigg\}
+N_f T_F C_A \Bigg\{\theta(y-x)\Bigg[
-4\qgv-\frac{6x}{y{\bar y}}
-2\left(\qgv -\frac{2 x}{{\bar y}^2} \right)  \ln\frac{x}{y}
+2\Bigg(\qgv
\nonumber\\
& &
+\qgbv+\frac{1}{y{\bar y}}\Bigg)\ln\left(1-\frac{x}{y}\right)
+\left(\qgv-\frac{3 x}{{\bar y}^2} \right)\ln^2\frac{x}{y}
-\left(\qgv+\qgbv\right)\ln^2\left(1-\frac{x}{y}\right)
\nonumber\\
& &
-\HQGo(x,y) -\bHQGo({\bar x},y)
\Bigg]
+\frac{6 x {\bar y} -4x y}{y{\bar y}^2}\ln{x}
+\qgv\ln{x} \left(2-\ln{x}\right)+\frac{3 x}{{\bar y}^2} \ln^2{x}
\Bigg\}
-\left\{x\to {\bar x} \atop  y\to {\bar y}\right\},
\nonumber\\
& &
\hspace{-0.5cm}{{^{GQ}\! V}^{(1)A}} =
\\
& &
\frac{C_F^2}{2} \Bigg\{\theta(y-x)\Bigg[
-3 \frac{x(1+x)}{y}
+\frac{x^2 y + 2x({\bar x}-x){\bar y}}{y\bar y}\ln\frac{x}{y}
-\frac{x({\bar x} +{\bar y}) + {\bar x} y}{y{\bar y}}
\ln\left(1-\frac{x}{y}\right)
+\gqv
\nonumber\\
& & \times \ln^2\frac{x}{y}
-\left(\gqv + \gqbv \right)\ln^2\left(1-\frac{x}{y}\right)
\Bigg]
-\frac{x^2 y+(1-{\bar x}^2){\bar y}}{y\bar y} \ln{x}-\gqv \ln^2{x}
\Bigg\}
-C_F\frac{\beta_0}{2}
\nonumber\\
& &
\times \Bigg\{\theta(y-x)\Bigg[
-\frac{2}{3}\gqv +  2\frac{x}{y}
+\left(\gqv + \gqbv\right)  \ln\left(1-\frac{x}{y}\right)
\Bigg]
+ \gqv \ln{x}
\Bigg\} + C_F C_A
\nonumber\\
& &
\times\Bigg\{\theta(y-x)\Bigg[
\left(\frac{13}{3}-2\zeta(2)\right)\gqv -\frac{x}{y}
-\left( \frac{x({\bar x}-x)}{y}+ 5\frac{x{\bar x}}{\bar y} \right)
\ln\frac{x}{y}
+\left(
\frac{{\bar x} - 2{\bar x}^2}{\bar y} +  \frac{x - 2x^2}{y}
\right)
\nonumber\\
& &
\times \ln\left(1-\frac{x}{y}\right)
-\frac{1}{2}\left(\gqbv+3\frac{x^2}{\bar y}\right) \ln^2\frac{x}{y}
+\frac{1}{2}\left(\gqv +\gqbv \right)\ln^2\left(\frac{y}{x}-1\right)
-\frac{1}{2} \HGQo(x,y)
\nonumber\\
& &+\frac{1}{2} \bHGQo({\bar x},y)
+\left(
5 \frac{x {\bar x}}{\bar y} + \frac{x}{y} +\gqv
\left(-2+\frac{1}{2} \ln{x}-\ln{\bar x}\right)
+\frac{3}{2} \frac{x^2}{\bar y} \ln{x}
\right) \ln{x}
\Bigg\}
-\left\{{x\to {\bar x}\atop y\to {\bar y}}\right\},
\nonumber\\
& &
\hspace{-0.5cm}{{^{GG}\! V}^{(1)A}}=
\\
& &
C^2_A \Bigg\{\theta(y-x)\Bigg[
\left(\frac{2}{3}-2\zeta(2)\right)\ggv  -\frac{15}{4} \frac{x^2}{y^2}
-6 \frac{x{\bar x}}{y{\bar y}}
+\frac{x(x+{\bar y})}{y^2 {\bar y}}
-\left(
2\frac{x}{y{\bar y}}-6\frac{x{\bar x}}{{\bar y}^2} +
\frac{x^2}{{\bar y}^2}
\right)
\nonumber\\
& &
\times \ln\frac{x}{y}
+\left(\ggv + 2 \frac{x^2}{{\bar y}^2}\right) 	\ln^2\frac{x}{y}
-\left(\ggv-\ggbv\right)\ln\frac{x}{y} \ln\left(1-\frac{x}{y}\right)
-\frac{1}{2} \HGGo(x,y)
\nonumber\\
& &
+\frac{1}{2}\bHGGo({\bar x},y)
\Bigg]
+\left(3\frac{({\bar x}-x)x}{y{\bar y}}
-6 \frac{x {\bar x}}{{\bar y}^2}
+ \frac{x^2}{{\bar y}^2}
- 2 \frac{x^2}{{\bar y}^2} \left(\ln{x} + \ln{\bar x} \right)
+\frac{3x-y}{{\bar y}^2}\ln{\bar x}\right)\ln{x}
\Bigg\}
\nonumber\\
& &
+C_A  \frac{\beta_0}{2}\theta(y-x) \Bigg\{
-\frac{5}{3}\ggv -\frac{5}{2}\frac{x^2}{y^2}
+\frac{x({\bar y}+x y)}{y^2 {\bar y}}
+\frac{x^2}{{\bar y}^2}\ln{y}
+\frac{{\bar x}^2}{y^2}\ln{\bar x}
\Bigg\}
+C_F N_f T_F \Bigg\{\theta(y-x)
\nonumber\\
& & \times
\Bigg[
2\frac{x(1-2x{\bar y})}{y^2 {\bar y}}
+2\frac{x^2}{y{\bar y}} \ln\frac{x}{y}
- \frac{x}{{\bar y}^2} \ln\frac{x}{y}\left(2 +x \ln\frac{x}{y}\right)
\Bigg]
+\frac{x}{{\bar y}^2} \ln{x}\left(2 +x \ln{x}\right)
\Bigg\}
	+\left\{{x\to \bar x  \atop y\to \bar y}\right\}.
\nonumber
\end{eqnarray}
Except of the pure singlet part, where the convolution of two $c$-kernels
generates Spence functions which die out in the forward limit, we
recover in the parity even sector a similar structure as observed in
the previous cases:
\begin{eqnarray}
& &\hspace{-0.7cm}{{^{QQ}\! V}^{(1)V}}=
\\
& &\hspace{-0.4cm}
{{^{QQ}\! V}^{(1)+}} +
2 C_F T_F N_f \Bigg\{\theta(y-x)\Bigg[
\frac{x(3-8x{\bar y}) }{y}
+\frac{x(5-8x)}{\bar y} \ln\frac{x}{y}
+\left(\frac{x}{\bar y} -4x{\bar x}\right)
\nonumber\\
& &\hspace{-0.4cm}
\times\ln^2\frac{x}{y}
+8x{\bar x}\left(
{\rm Li}_2(\bar x)-{\rm Li}_2(\bar y) + \ln{\bar x}\ln{y}
\right)
\Bigg] -\frac{290}{9} x{\bar x}
-\left( \frac{x (5-8x)}{\bar y}+\frac{2x(9-19{\bar x})}{3}\right)
\nonumber\\
& &\hspace{-0.4cm}
\times  \ln{x}
-\left(\frac{x}{\bar y} -4 x{\bar x}\right) \ln^2{x}
+4 x{\bar x} \ln{x} \ln{\bar x}
\Bigg\}
+\left\{x\to {\bar x} \atop  y\to {\bar y}\right\},
\nonumber\\
& &\hspace{-0.7cm}{{^{QG}\! V}^{(1)V}}=
\\
& &\hspace{-0.4cm}
N_f T_F C_F \Bigg\{\theta(y-x)\Bigg[
2(5-2\zeta(2))\qgvv - 4 \qgv + \frac{x}{y{\bar y}}
+2\left(\qgvv+\qgbvv+\frac{2x-2y+x y}{2y{\bar y}^2}\right)
\nonumber\\
& &\hspace{-0.4cm}
\times \ln\frac{x}{y}
-2\left(\qgvv+\qgbvv+\frac{1}{y{\bar y}}\right)
\ln\left(1-\frac{x}{y}\right)
+\qgbv \ln^2\frac{x}{y}
+(\qgvv+\qgbvv)\ln^2\left(\frac{y}{x}-1\right)
\Bigg]
\nonumber\\
& &\hspace{-0.4cm}
-2\left(
\qgvv+\qgbvv+\qgbv +\frac{x(y+4{\bar y}^2)}{2y{\bar y}^2}
\right)\ln{x}
-2 \qgvv \ln{\bar x} \ln{x}+\left(\qgvv-\qgbvv-\qgbv\right)
\nonumber\\
& &\hspace{-0.4cm}
\times \ln^2{x}
\Bigg\}
+N_f T_F C_A  \Bigg\{\theta(y-x)
\Bigg[
\frac{6x(3-4x)}{\bar y} - \frac{2x(11-16 x)}{y} + \frac{4x(1-3x)}{y^2}
\nonumber\\
& &\hspace{-0.4cm}
-2\left(\qgvv - 5 \qgbvv + \frac{5-7 x}{{\bar y}^2} +
\frac{2 x (3-2 x-12{\bar x} y)}{y {\bar y}}\right)
\ln\frac{x}{y}
+2\Bigg(\qgvv+\qgbvv+\frac{1}{y{\bar y}}\Bigg)
\nonumber\\
& &\hspace{-0.4cm}
\times
\ln\left(1-\frac{x}{y}\right)
+\left(\qgvv-\qgbvv + \frac{1-4x}{{\bar y}^2}\right) \ln^2\frac{x}{y}
-\left(\qgvv+\qgbvv\right)\ln^2\left(1-\frac{x}{y}\right)
\nonumber\\
& &\hspace{-0.4cm}
-\HQGe + \bHQGe\Bigg]
+2 \left(
\qgvv -5 \qgbvv + \frac{5-7x}{{\bar y^2}} +
\frac{x (5-6x -22 y +28 x y)}{y{\bar y}}
\right) \ln{x}
\nonumber\\
& &\hspace{-0.4cm}
-\left(\qgvv-\qgbvv + \frac{1-4x}{\bar y^2}\right) \ln^2{x}
\Bigg\}
-\left\{x\to {\bar x} \atop  y\to {\bar y}\right\},
\nonumber\\
& &\hspace{-0.7cm}{{^{GQ}\! V}^{(1)V}}=
\\
& &\hspace{-0.4cm}
C_F^2 \Bigg\{\theta(y-x)\Bigg[
3\gqv-\frac{9}{2} \frac{x}{y}
-\left(
\frac{1}{2}\frac{x (2+x)}{\bar y}  - \frac{x{\bar x}}{y{\bar y}}
\right)\ln\frac{x}{y}
-\left(\frac{3}{2} \gqvv+\frac{3}{2} \gqbvv +\frac{x{\bar x}}{y{\bar y}}\right)
\ln\left(1-\frac{x}{y}\right)
\nonumber\\
& &\hspace{-0.4cm}
- \frac{1}{2}\gqv \ln^2\frac{x}{y}
-\frac{1}{2}\left(\gqvv+\gqbvv\right)\ln^2\left(1-\frac{x}{y}\right)
\Bigg]
- \frac{1}{2} \left(\frac{x (2-5x)}{y} -3 \frac{x^2}{\bar y} \right) \ln{x}
+ \frac{1}{2}\gqv \ln^2{x}
 \Bigg\}
\nonumber\\
& &\hspace{-0.4cm}
-C_F\frac{\beta_0}{2}\theta(y-x)\Bigg\{
\frac{10}{3} \gqvv + 2 \frac{x{\bar x}}{y}   +
\left(\gqvv+\gqbvv\right) \ln\left(1-\frac{x}{y}\right) -
\gqv \ln{x}+\gqbv \ln{\bar x}
\Bigg\}
\nonumber\\
& &\hspace{-0.4cm}
C_F C_A\Bigg\{\theta(y-x)\Bigg[
\left(\frac{4}{9} -2 \zeta(2) \right)\gqvv  - \frac{26}{9}  \frac{x^2}{y}
+\frac{x(3+4 x^2)}{y}  - \frac{x}{9} (105 -246 x +188 x^2)
\nonumber\\
& &\hspace{-0.4cm}
-\left(
\frac{x{\bar x}}{y{\bar y}}-\frac{x(6-11x+8 x^2)}{\bar y}+
4x{\bar x}({\bar x}-x)
\right)\ln\frac{x}{y}
+ \frac{x{\bar x}}{y{\bar y}} \ln\left(1-\frac{x}{y}\right)
+\frac{1}{2}\Bigg(2\gqvv +2\gqv
\nonumber\\
& &\hspace{-0.4cm}
+\gqbv-3\frac{x^2}{\bar y} -2\Bigg)\ln^2\frac{x}{y}
+\frac{1}{2}\left(2-\gqv-\gqbv\right)\ln^2\left(\frac{y}{x}-1\right)
-\frac{1}{2} \HGQe(x,y) -\frac{1}{2}\bHGQe({\bar x},y)
\Bigg]
\nonumber\\
& &\hspace{-0.4cm}
+\left( \frac{x{\bar x}}{y{\bar y}} - \frac{x(6 -11 x + 8 x^2)}{\bar y} +
\frac{x(18 - 63 x + 62 x^2)}{3} \right)\ln{x}
-\gqvv \ln{x}\ln{\bar x}- \frac{x^2(1-4 y)}{2y \bar y} \ln^2{x}
\Bigg\}
\nonumber\\
& &\hspace{-0.4cm}
-\left\{x\to {\bar x} \atop  y\to {\bar y}\right\},
\nonumber\\
& &\hspace{-0.7cm}{{^{GG}\! V}^{(1)V}}=
\\
& &\hspace{-0.4cm}
C_A^2\Bigg\{\theta(y-x)\Bigg[
\left(\frac{2}{3}-2\zeta(2)\right)\ggvv
-\frac{1}{2} \ggvc + \frac{x(4+5x)}{4y^2}
-\frac{x(10 -19 x+ 16 x^2)}{y}
\nonumber\\
& &\hspace{-0.4cm}
+\frac{x (6 - 13 x + 8 x^2) }{{\bar y}}
+\left(
\frac{2x[1 -2{\bar x} (x y + {\bar x}{\bar y})]}{y{\bar y}}
+\frac{x(2-3 x -8 {\bar x}^2)}{{\bar y}^2}
\right) \ln\frac{x}{y}
+\left(\ggvv+2 \frac{x^2}{{\bar y}^2}\right)
\nonumber\\
& &\hspace{-0.4cm}
\times \ln^2\frac{x}{y}
-(\ggvv-\ggbvv)\ln\frac{x}{y} \ln\left(1-\frac{x}{y}\right)
-\frac{1}{2} \HGGe(x,y) -\frac{1}{2} \bHGGe({\bar x},y)
\Bigg]
-2\frac{x^2}{{\bar y}^2} \ln^2{x}
\nonumber\\
& &\hspace{-0.4cm}
+\frac{{\bar x}[x^2+{\bar x}^2 -2{\bar x}(1+2x){\bar y}]
+{\bar y}}{{\bar y}^2}  \ln{x}\ln{\bar x}
+\left(
\frac{3x(x^2+{\bar x}^2)({\bar y}-y) -4 x^2 y}{{y\bar y}}
-\frac{x(2-3 x - 8 {\bar x}^2)}{{\bar y}^2}
\right)
\nonumber\\
& &\hspace{-0.4cm}
\times \ln{x} \Bigg\}
+ C_A \frac{\beta_0}{2} \theta(y-x)\Bigg\{
-\frac{5}{3}\ggvv- \frac{13}{3}\ggvc -\frac{11}{2} \frac{x^2}{y^2}
+\frac{x(x+ {\bar y})}{y^2{\bar y}}
+\frac{x^2}{{\bar y}^2} \ln{y}
+\frac{{\bar x}^2}{y^2} 	\ln{\bar x} \Bigg\}
\nonumber\\
& &\hspace{-0.4cm}
+ C_F  N_f T_F\Bigg\{ \theta(y-x) \Bigg[
-\frac{20}{3} \ggvc -12 \frac{x^2}{y^2} +
\frac{2 x(2+y +3 x y - 4y^2)}{y^2{\bar y}}
-\frac{2 x(x-y+2x y)}{y{\bar y}^2} \ln\frac{x}{y}
\nonumber\\
& &\hspace{-0.4cm}
-\frac{x^2}{{\bar y}^2}\ln^2\frac{x}{y}
\Bigg]
-\frac{2 x (1-3x -2x {\bar y})}{{\bar y}^2} \ln{x}
+\frac{x^2}{{\bar y}^2} \ln^2{x}
	\Bigg\} +\left\{x\to {\bar x} \atop  y\to {\bar y}\right\}.
\nonumber
\end{eqnarray}

\noindent

Now we will derive the NLO skewed DGLAP kernels which govern the evolution
of SPDs in the kinematic range of $\zeta<z<1$ and $-1+\zeta<z< 0$ with
$z=\frac{t+\eta}{1+\eta}$ and $\zeta=\frac{2\eta}{1+\eta}$. We chose this
set of variables such that we have the closest resemblance to the usual
DGLAP kinematics. In fact for DVCS and vector meson production $\zeta =
x_{\rm Bj}$.

After the continuation of the ER-BL kernels to the whole
$x-y$ plane by the replacement of the $\theta$-function structure, we
derive the kernels in the following way which is suggested by the
non-zero support of the $\theta$-functions after the replacement
$x \rightarrow z/\zeta, \bar x \rightarrow 1 - z/\zeta \equiv\bar
\frac{z}{\zeta}, y \rightarrow 1/\zeta$ and $\bar y \rightarrow 1 -
1/\zeta\equiv\bar \frac{1}{\zeta}$ for the above mentioned kinematical regime:
\begin{eqnarray}
\mbox{\boldmath$P$}^{(1)I} (z, \zeta)
=
\left(
\begin{array}{ll}
\frac{1}{\zeta}\left({^{QQ}\!f^{(1)I}} \left(\frac{z}{\zeta},
\frac{1}{\zeta}\right) - {^{QQ}\!f^{(1)I}} \left({\bar \frac{z}{\zeta}},
{\bar \frac{1}{\zeta}}\right)\right)
& \frac{1}{\zeta^2}\left({^{QG}\!f^{(1)I}} \left(\frac{z}{\zeta},
\frac{1}{\zeta}\right) + {^{QG}\!f^{(1)I}} \left({\bar \frac{z}{\zeta}},
{\bar \frac{1}{\zeta}}\right)\right) \\
{^{GQ}\!f^{(1)I}} \left(\frac{z}{\zeta},\frac{1}{\zeta}\right)
+ {^{GQ}\!f^{(1)I}} \left({\bar \frac{z}{\zeta}},{\bar \frac{1}{\zeta}}\right)
& \frac{1}{\zeta}\left({^{GG}\!f^{(1)I}} \left(\frac{z}{\zeta},
\frac{1}{\zeta}\right) - {^{GG}\!f^{(1)I}} \left({\bar \frac{z}{\zeta}},
{\bar \frac{1}{\zeta}}\right)\right)
\end{array}
\right),
\end{eqnarray}
where the functions ${^{AB}\!f^{(1)I}}(x,y)$ are the coefficients of
the $\theta(y-x)$-term appearing in the ER-BL kernels.
Note that we omitted the additional factors of $1/z$ in the $GG$ and
$GQ$ kernels so as to take into account the fact that in the forward
limit, the skewed gluon distribution turns into $zG(z,Q^2)$ rather
than $G(z,Q^2)$. Thus, we can take the "+"-prescription from the
ER-BL kernels, where contributions concentrated in $x=y$ turns into
end-point concentrated ones, i.e.\   $\delta(x-y) \to \delta(1-z)$.

Following our afore mentioned prescription, the result for the parity odd
and even skewed DGLAP kernels are up to end-point concentrated terms, which
can be easily restored, given by:
\bigskip
\bea
& &{^{QQ}{P}}^{\pm}_{NS} = \nonumber\\
& & C_F^2 \Bigg\{\frac{1}{2}\lozembze \left[ \frac{1}{\bz}
\left(1-\frac{z\bz}{\zeta\be}\right)+\frac{z}{\bz}\embze\right ]
+\frac{1}{2}\lozz\left[2\frac{z}{\bz}\embze +\frac{z}{\zeta\be}\right]
\nonumber\\
& & - {^{QQ}{p}}\left[\loembze\lobzobe+\lobz\loz\right]
- 2\loembze\embze\left[\frac{1 - \bz - \be - \frac{z\zeta}{4}}{\zeta\bz}
\right]\nonumber\\
& & -\frac{1}{2}\loz\left[4\frac{z}{\zeta\be}
+3\frac{z}{\bz}\embze\right]+3\bzoe+\left(\frac{4}{3}-\frac{\pi^2}{3}\right)
{^{QQ}{p}}
\Bigg\}\nonumber\\
& & -\frac{C_F\beta_0}{2}\Bigg\{ \frac{5}{3}{^{QQ}{p}}
+ \bzoe +\loembze\frac{1}{\bz}\left[1-\frac{z\bz}{\zeta\be}\right]
+ \loz\left[\frac{z}{\bz}\embze + \frac{z}{\zeta\be}\right]\nonumber\\
& &-C_F\left(C_F-\frac{C_A}{2}\right)\Bigg\{2\bzoe + \frac{4}{3}{^{QQ}{p}}
+ {^{QQ}h}\mp{^{QQ}\bar h} \Bigg\} ,
\eea
\bea
& &{^{QQ}{P}}^A_S =  \nonumber\\
& & + 2 C_FT_FN_F\Bigg\{\emze\loembze\left[\loembze - 1\right]
+\frac{z}{\zeta\be}\loz \left[ \loz - 1 \right ] \Bigg\}\nonumber\\
& & + {^{QQ}{P}}^{-}_{NS} ,
\eea
\bea
& &{^{QG}{P}}^A_S = \nonumber\\
& & + C_AN_FT_F \Bigg\{ -\left(\frac{3}{2}\pqgpo
+ \frac{4}{\be^2}\emze + 3\right)\lozembze
- \frac{1}{2}\pqgpo\left[\lozbzobe + \lozbz\right]\nonumber\\
& & - \lozz \zoe\left(1 + \frac{3}{\be^2}\right)
+ 2 \loembze\left[\pqgpo + 2 +\frac{3}{\be^2}\emze\right]
+ 2\loz\zoe\left(1 + \frac{2}{\be^2}\right)\nonumber\\
& & + 2 \left[\lobzobe + \lobz\right]\left(\frac{1}{2}\pqgpo
-\frac{1}{\be}\right) -2\pqgpo +\frac{6}{\be} -\frac{1}{2} {^{QG}h}^A
- \frac{1}{2} {^{QG}{\bar h}}^A\Bigg\}\nonumber\\
& & + C_FN_FT_F\Bigg\{ \zoe\left[\lozembze-\lozz\right]
+ \frac{1}{2}\pqgpo\left[\ln^2\left(\frac{\be}{\bz}-1\right)
+ \lozbzoz - \lozz\right]\nonumber\\
& & + \lozme\frac{1}{\be}\embze\left[\frac{7}{\zeta} + 3\zeta
- 8\right]+ \loe\left[3\emze -2\zoe\frac{1+\be}{\be}\right]
+ 4\lobz\frac{\bz}{\be^2}[1+\be]\nonumber\\
& & - \loz\left[\zoe\left(2+\frac{5}{\be}\right)
+3\frac{z}{\be^2}\right] +\left(5-\frac{\pi^2}{3}\right){^{QG}{p}}^A
- \frac{5}{\be}\Bigg\} ,
\eea
\bea
& &{^{GQ}P}^A_S = \nonumber\\
& & + C^2_F \Bigg\{ \frac{1}{2}\lozembze\left[\pgqpo
- \frac{z^2}{\zeta}\right] -\frac{1}{2}\pgqpo\left[\lozbzobe + \lozbz\right]
+\frac{z^2}{2\zeta}\lozz\nonumber\\
& & -\frac{1}{2}\lozme\embze\left[z - \zeta + 3
- 5\zoe\right] + \frac{1}{2}\loe\left[\embze\emze(4-\zeta)
+z\left(2-\zoe\right)\right]\nonumber\\
& & -\frac{1}{2}\loz\left[5\zoe\embze\left(1-\frac{4}{5}\zeta\right)
+ z\left(2+\frac{1}{\be}\right)\right] -\lobz\left[\embze(2+z-\zeta)
+ \zeta\right]\nonumber\\
& & -\frac{3}{2}{^{GQ}p}^A -\frac{3}{2}z -\frac{3}{2}\embze \Bigg\}
-\frac{2}{3}C_FN_FT_F\Bigg\{ {^{GQ}p}^A\left[\lobzobe + \lobz
- \frac{2}{3}\right] + 2z + 2\embze\Bigg\}\nonumber\\
& & + C_FC_A\Bigg\{ \frac{1}{2}\lozembze\left[3\pgqpo\be
-\frac{z^2}{\zeta}\left(1+3\be\right)\right]
+ \frac{1}{2}\lozz\left[\frac{z^2}{\zeta\be}(4 - \zeta)
-\pgqpo\right]\nonumber\\
& & + \frac{1}{2}{^{GQ}p}^A\left[\ln^2\left(\frac{\be}{\bz}-1\right)
+ \lozbzoz\right]
+\loembze\embze\left[4-5z+3\frac{\be}{\zeta}\embze\right]\nonumber\\
& & +\frac{1}{6}\left[\lobzobe + \lobz\right]\left(5z+\embze(5+z)\right)
- \loz\left[4\zoe\embze + z\left(1-\zoe - \bzoe\right)\right]\nonumber\\
& & + \left(\frac{28}{9}-\frac{\pi^2}{3}\right){^{GQ}p}^A
+ \frac{8}{3}\left(1 + z - \bzoe\right) + \frac{1}{2}{^{GQ}{\bar h}}^A
- \frac{1}{2}{^{GQ}h}^A\Bigg\} ,
\eea
\bea
& &{^{GG}P}^A_S = \nonumber\\
& & \frac{2}{3}C_AN_FT_F\Bigg\{ \loembze\zeta\emze^2
- \frac{z^2}{\zeta\be^2}\loz - \frac{8}{3}{^{GG}p}^A
+ \frac{z\embze}{\bz}\left[\embze +z\right]\nonumber\\
& & -\frac{3}{2}\pgga + 2\frac{z}{\be}
- \frac{\zeta}{\be^2}(1-z\zeta)\Bigg\}\nonumber\\
& & + C_FN_FT_F\Bigg\{ \lozembze\zeta\emze^2
- \frac{z^2}{\zeta\be^2}\lozz - 2\loembze\left[\embze
- \zeta\right]\emze\nonumber\\
& & - 2\loz\frac{1}{\be}\left[\zoe\embze + \frac{z(2 - z)}{\be}\right]
- 8\frac{z\bz}{\be^2} + 4\frac{\zeta}{\be^2}(1-z^2)
- 2\left(\frac{\zeta}{\be}\right)^2\bz \Bigg\}\nonumber\\
& & + C^2_A\Bigg\{\frac{1}{2}\lozembze\left[1+9z-4\zeta
-\frac{2}{\be}(4-3z^2) -\frac{8}{\zeta}\embze^2
+ \frac{\bz}{\be^2}(7-6z)\right]\nonumber\\
& & + \lozz\left(\frac{z}{\be}\right)^2\left[4\frac{\be}{\zeta}
+2\zeta + \frac{\be^2}{\bz}\right]
+ \frac{1}{2}\pggpo\left[\ln^2\left(\frac{\be}{\bz}-1\right)
-\lozbzobe- 2\loz\lobz\right]\nonumber\\
& & - \loembze\emze\left[2\frac{\zeta}{\be} + \frac{31}{6}z
+ \frac{5}{6}\zeta\right]+ \loz\frac{z}{\be^2}\left[6+2\be
-\frac{31}{6}\zoe\right]\nonumber\\
& & + \left(\frac{67}{18}-\frac{\pi^2}{3}\right){^{GG}p}^A
+\frac{5}{6}\left(\frac{\zeta}{\be}\right)^2\bz
+\frac{11}{3}\frac{z\bz}{\be^2} - \frac{11}{6}\frac{\zeta}{\be^2}(1-z^2)
+ \frac{1}{2}{^{GG}{\bar h}}^A - \frac{1}{2}{^{GG}h}^A \Bigg\} ,
\eea
\bea
& &{^{QQ}{P}}^V_S = \nonumber\\
& & - 2 C_FT_FN_F\Bigg\{ \emze\left(1-4\frac{z}{\zeta^2}\right)\lozembze
+ 8\frac{z}{\zeta^2}\emze\Bigg [\loembze(2\ln(\zeta)-\loz)\nonumber\\
& & - 2\left({\rm Li_2}\left(1-\zoe\right)
- {\rm Li_2}\left(1-\frac{1}{\zeta}\right)\right)\Bigg ]
+ \lozz\frac{z}{\zeta^2}\left[3 - 4\zoe +\frac{1}{\be}\right]
- \loembze\emze\left[3 - 8\zoe\right]\nonumber\\
& & + \loz\frac{1}{\be}\zoe\left[5-8\zoe\right] -3\bzoe
+ 8\frac{1-z^2}{\zeta\be} - 16\frac{z\bz}{\zeta^2\be}  \Bigg\}
+ {^{QQ}{P}}^{+}_{NS} ,
\eea
\bea
& &{^{QG}P}^V_S = \nonumber\\
& & + C_AN_FT_F\Bigg\{ - 2\lozembze\left[\pqgpo + \frac{3}{2}
+\frac{1}{\be^2}\emze\left(1+6\zoe-4\frac{z}{\zeta^2}\right)\right]
\nonumber\\
& & - \frac{1}{2}\pqgpe\left[\lozembze + \lozbzobe + \lozbz
- \lozz\right] + \lozz\frac{2}{\be^2}
\Bigg[\emze\left(1+6\zoe-4\frac{z}{\zeta^2}\right) \nonumber\\
& &-\frac{3}{2}\Bigg]
+ 2\loembze\left[6\pqgpe -7\pqgc +2+\frac{1}{\be^2}\emze\left(3
-2\zeta+20z-8\zoe\right)\right]\nonumber\\
& & + \left[\lobzobe + \lobzoz\right]\left[\pqgpe-\frac{2}{\be}\right]
+ \loz\frac{2}{\be^2}\left[ 1+\zeta - \emze\left(1+4z+8\zoe\right)
+2\be\zoe\right]\nonumber\\
& & -2(\pqgpe+\pqgc) -32\frac{\bz}{\zeta\be}\emze
+ \frac{2}{\be}\left(3+8\frac{\bz}{\zeta}\right) +
 \frac{1}{2} {^{QG}{\bar h}}^V-\frac{1}{2}{^{QG}h}^V\Bigg\}\nonumber\\
& & + C_FN_FT_F\Bigg\{ -\lozembze\frac{z}{\zeta}
+\frac{1}{2}\pqgpe\left[\ln^2\left(\frac{\be}{\bz}-1\right)
+ \lozbzoz\right] - \lozz\frac{1}{\be^2}\emze\nonumber\\
& & +\lozme\frac{1}{\be}\embze\left[4z - \zeta -\frac{1}{\zeta}\right]
- \loe\left[1-\zoe+2\frac{z}{\be}\right]
+ \loz\Bigg[\zoe + 3\frac{z}{\be}\embze \nonumber\\
& &+ \left(\frac{z}{\be}\right)^2\Bigg]
+ 4\lobz\frac{\bz}{\be^2}(2z -\zeta)
-\frac{\pi^2}{3}{^{QG}p}^V - 20\frac{z\bz}{\be^2}
+8\frac{\bz}{\be^2} + 6\frac{\zeta\bz}{\be^2} + \frac{5}{\be}\Bigg\} ,
\eea
\bea
& &{^{GQ}P}^V_S = \nonumber\\
& & + C^2_F\Bigg\{ +\frac{1}{2}\lozembze\left[\pgqpe-2
+\frac{z^2}{\zeta}\right] -\frac{1}{2}\frac{z^2}{\zeta}\lozz
- \frac{1}{2}\pgqpe\left[\lozbzobe + \lozbz\right]\nonumber\\
& & - \frac{1}{2}\loembze\emze\left[3+z\frac{\zeta}{\be}\left(3
-\frac{1}{\zeta}\right)\right] -2\lobz\left[z\embze
+\frac{3}{2}\left(1+\frac{\bz^2}{\be}\right)\right]\nonumber\\
& & + \frac{1}{2}\loe\left[5z\embze+3\bzoe(1+\be)\right]
+ \loz z\left[1+\frac{z}{\be}\left(1+\frac{1}{2\zeta}\right)\right]
- \frac{3}{2}\embze(1+3z) \nonumber\\
& &-\frac{3}{2}\frac{z\bz}{\be}\Bigg\}
- \frac{2}{3}C_FN_FT_F\Bigg\{\pgqpe\left[\lobzobe + \lobz\right]
+\frac{10}{3}\pgqpe -2\pgqpo + 2z +2\embze\Bigg\}\nonumber\\
& & + C_FC_A\Bigg\{ +\frac{1}{2}\lozembze\left[\emze^2
\left(4-3\zeta+8\zoe\right)-2+\frac{z^2}{\zeta}\right]
+\frac{1}{2}\pgqpe\Bigg[\ln^2\left(\frac{\be}{\bz}-1\right) \nonumber\\
& & + \lozbzoz\Bigg]
+ \frac{1}{2}\lozz\left(\zoe\right)^2\left[\frac{3}{\be} + \be
+ 8\emze - \left(\frac{\zeta}{z}\right)^2\pgqpe\right]
+\left[\lobzobe + \lobzoz\right] \nonumber\\
& &\times\left[\frac{17}{6}\pgqpe - \bzoe(2-\zeta)\right]
+ \loembze\Bigg[\emze\left(2z+3\zeta-5\zoe+8z\frac{\be}{\zeta}\emze\right)
+\frac{z\bz}{\zeta\be} \nonumber\\
& &-\left(\zoe\right)^2\Bigg]
+ \loz\left[\frac{11}{6} -\frac{11}{3}\frac{z^2}{\zeta}
+ \frac{11-18z-11z^2}{6\be} -\frac{1}{\be}\zoe\emze\left(4+\frac{8}{3}\zeta
-8\zoe\right)\right] \nonumber\\
& &+ 4\frac{z\bz}{\zeta\be}\left(2+z-2\zoe\right)+ \frac{20}{3}z
+\frac{16}{9} - \bzoe\left(7-\frac{118}{9}\bz\right)
-\frac{\pi^2}{3}{^{GQ}p}^V +\frac{11}{3}\embze
-\frac{1}{2} {^{GQ}{\bar h}}^V \nonumber\\
& &- \frac{1}{2}{^{GQ}h}^V \Bigg\} ,
\eea
\bea
& &{^{GG}P}^V_S = \nonumber\\
& & + \frac{2}{3}C_AN_FT_F\Bigg\{\loembze\zeta\emze^2
- \frac{z^2}{\zeta\be^2}\loz -\frac{5}{3} {^{GG}p}^V - \frac{13}{3}
{^{GG}p}^c - \frac{9}{2}{^{GG}p}^a\nonumber\\
& & -\left(\frac{\zeta}{\be}\right)^2\bz\Bigg\}
+ C_FN_FT_F\Bigg\{\zeta\emze^2\lozembze
- \frac{z^2}{\zeta\be^2}\lozz\nonumber\\
& & + 2\loz\left[\frac{z\bz}{\be^2}-2\frac{z^2}{\zeta\be^2}\right]
+ 2\loembze\left[(\zeta-2z)\emze-\be\embze^2\right]\nonumber\\
& & -\frac{20}{3}\pggc -12\pgga -4\pqgpo +\frac{8}{\be}\embze
-4\bz\left(\frac{\zeta}{\be}\right)^2\Bigg\}\nonumber\\
& & + C^2_A\Bigg\{\lozembze\left[1-\frac{1}{\bz}
-2\zeta+5z-2\left(\zoe\right)^2\left((2+\zeta)\emze + 1
+ \zeta\right)\right]\nonumber\\
& & + \pggpe\left[\loembze\ln\left(\frac{\be}{\bz}-1\right)
- \loz\lobz\right]\nonumber\\
& & + \lozz\left[\pggpe - \frac{1}{\bz}-\frac{1-3z}{\be^2}
+\frac{2}{\be}\left(\zoe\right)^2\left(\left(3-\frac{1}{\be}\right)\emze
+1\right)\right]\nonumber\\
& & + \loembze\left[4\left(\zoe\right)^2\left[2\zoe\be-\bzoe-2\right]
+ \zoe\left(4+\frac{55}{6}z\right) -\frac{7}{3}z -\frac{5}{6}\zeta
+ 2\zeta\bzoe\right]\nonumber\\
& & + \loz\frac{1}{\be^2}\left[2\zoe\emze\left(2-4\emze-\zeta^2\right)
+\left(\zoe\right)^2\be\left(4z+\frac{5\zeta^2}{6\be}
-\frac{31}{6}\zeta\right)\right] \nonumber\\
& &+\left(\frac{67}{18}
-\frac{\pi^2}{3}\right){^{GG}p}^V
+\frac{67}{9}\pggc +\frac{11}{2}\pgga +8\frac{1}{\be^2}\zoe\left(1
-z^2-\frac{z\bz}{\zeta}\right) -4\frac{\bz}{\be^2}(2-\zeta\bz) \nonumber\\
& &+\frac{5}{6}\left(\frac{\zeta}{\be}\right)^2\bz
-\frac{1}{2} {^{GG}{\bar h}}^V -  \frac{1}{2}{^{GG}h}^V\Bigg\} ,
\eea
where the ${^{AB}h}^I$ and ${^{AB}{\bar h}}^I$ are given by
\bea
& &{^{AB}h}^I = \nonumber\\
& & + 2{^{AB}p}^I(z,\zeta)\left[\ln(\zeta)\loembze
+ {\rm Li_2}\left(1-\frac{1}{\zeta}\right)
- {\rm Li_2}\left(1-\zoe\right) -\frac{\pi^2}{6}\right]\nonumber\\
& & + 2{^{AB}k_1}^I(z,\zeta)\left[\loembze\left(\ln\left(\zoe\right)
- \ln(\zeta)\right) - 2\left({\rm Li_2}\left(1-\frac{1}{\zeta}\right)
- {\rm Li_2}\left(1-\zoe\right)\right)\right] , \nonumber\\
& &{^{AB}{\bar h}}^I= \nonumber\\
& & + 2{^{AB}p}^I(\zeta-z,\zeta)\Bigg[\ln(\zeta)\loembze
+\frac{1}{2}\loe\left(\ln\left(\zoe\right) + \loz
- \ln\left(\frac{\be}{\zeta}\right)\right)
+ {\rm Li_2}\left(1-\frac{1}{\zeta}\right)\nonumber\\
& & - {\rm Li_2}\left(1-\zoe\right) -\frac{\pi^2}{6}
- \ln(\be+z)\left(\loembze + \loz\right) - {\rm Li_2}(\zeta-z)
- {\rm Li_2}\left(-\frac{z}{\be}\right)\Bigg]\nonumber\\
& & + 2{^{AB}k_2}^I(z,\zeta)\left[\loembze\left(\ln\left(\zoe\right)
- \ln(\zeta)\right) - 2\left({\rm Li_2}\left(1-\frac{1}{\zeta}\right)
- {\rm Li_2}\left(1-\zoe\right)\right)\right] .
\eea
The LO skewed kernels appearing in the above formulas are given by:
\bea
{^{QQ}p}^I &=& \frac{1+z^2-\zeta(1+z)}{\be\bz}\, ,
{^{QG}p}^A = 2\frac{2z-1-z\zeta}{\be^2}\, ,
{^{QG}p}^V = 2\frac{z^2+\bz^2-z\zeta}{\be^2} , \nonumber\\
{^{GQ}p}^A &=& \frac{z(2-z)-\zeta}{\be}\, ,
{^{GQ}p}^V = \frac{1 + \bz^2 - \zeta}{\be}\, , \nonumber\\
{^{GG}p}^A &=& \frac{\zeta^2(1+z^2)+2z(z\be-\zeta)}{\be^2\bz}
+4\frac{z\bz}{\be^2} - \frac{2\zeta}{\be^2}(1-z^2) ,\nonumber\\
{^{GG}p}^V &=& \frac{1}{\bz}\left(z^2+\embze\right)
+2\left(\bzoe\right)^2 +\frac{2}{\be}\left(\frac{1}{2}-z^2\right)\embze\, ,
{^{QG}p}^c = 2\left(\bzoe\right)^2 , \nonumber\\
{^{GG}p}^a &=& 2\frac{z\bz}{\be^2} - \frac{\zeta}{\be^2}(1-z^2)\, ,
{^{GG}p}^c = \frac{\bz^3}{\be^2},
\eea
where in comparison to our previous conventions we included here a factor
2 in the LO kernels of the $QG$-channel. The ${^{AB}k}^I_i$ were found to
be:
\bea
& &{^{QQ}k}_1 = \frac{z}{\zeta} + \frac{z}{\bz}\, ,
{^{QQ}k}_2 = \frac{z}{\be\zeta}\left(1-\frac{\zeta}{\be+z}\right)\, ,
{^{QG}k}^A_1 = -2\zoe\, ,{^{QG}k}^A_2 = -2\frac{z}{\zeta\be^2} ,
\nonumber\\
& &{^{GQ}k}^A_1 = \frac{z^2}{\zeta}\, ,
{^{GQ}k}^A_2 = -\frac{z^2}{\zeta\be}\, ,
{^{GG}k}^A_1 = 2\frac{z^2}{\zeta} + \frac{z^2}{\bz}\, ,
{^{GG}k}^A_2 = -\frac{z^2}{\be^2\zeta}\left(2-\frac{\zeta}{\be+z}\right) ,
\nonumber\\
& &{^{QG}k}^V_1 = -2\left[\zoe -2\left(\zoe\right)^2
+4\emze\frac{z}{\zeta^2}\right]\, ,
{^{QG}k}^V_2 = 2\frac{z}{\zeta\be^2}\left[1+\emze\left(\frac{4}{\zeta}
-6\right)\right] ,
\nonumber\\
& &{^{GQ}k}^V_1 = - \frac{z^2}{\zeta}
\left[1-\frac{1}{\zeta}\left(6-4\zoe\right)\right]\, ,
{^{GQ}k}^V_2 = \frac{z^2}{\zeta\be}\left[1+\frac{\be}{\zeta}
\left(6-4\zoe\right)\right] ,
\nonumber\\
& &{^{GG}k}^V_1 = \frac{z^2}{\bz} +2\left(\zoe\right)^2
+ 2\emze\frac{z^2}{\zeta}\left(1+\frac{2}{\zeta}\right) , \nonumber\\
& &{^{GG}k}^V_2 = \left(\frac{z}{\zeta\be}\right)^2
\left[10 + 6z -9\zeta + \frac{1}{\be+z}\left(\zeta(1+z)
\left(1-4\frac{z}{\zeta^2}\right) - 4\be\right)\right] .
\eea
The set of explicit formulae given in this section represents the
main result of the present paper.

\section{Conclusions.}
\label{sec-Con}

In this paper we have given a detailed description of the formalism for
construction of the two-loop exclusive evolution kernels in the flavor
singlet sectors and presented explicit results. Our formalism allowed us to
avoid complicated NLO calculations and was based on three main ingredients:
(i) the known form of conformal symmetry breaking counterterms for
renormalization of conformal operators in NLO transformed into the language
of ER-BL kernels; (ii) supersymmetric constraints which allowed us to
construct the contribution of cross ladder diagrams; (iii) reduction
formulae and known two-loop splitting functions which completely constrained
the diagonal part of NLO ER-BL kernels. The kernels predicted here are
numerically checked by direct comparison of their Gegenbauer moments to the
anomalous dimensions of conformal operators whose correctness was shown by
supersymmetric \cite{BelMueSch98} and superconformal \cite{BelMul99s}
constraints. Moreover, the predicted $\beta_0$ terms have been checked by a
direct calculation of diagrams with quark bubble insertions. Note also that
the predictions of the conformal operator product expansion rotated to the
$\overline{\rm MS}$ scheme coincide with the NLO coefficient functions. This
proofs again our results of conformal symmetry breaking in $QQ$ and $QG$
channels.

As a byproduct, our understanding of the general structure of ER-BL type
kernels implies the discovery of a simple structure for the DGLAP kernels,
which results from the topology of Feynman graphs. Indeed, the only new functions
appearing in NLO is ${^{AB}\!G}(z)$ arising from the crossed ladder
diagrams. Everything else can be represented as convolutions of LO kernels
and the ones obtained in the forward limit of the conformal anomalies. This is
interesting, but unfortunately not very restrictive, since the conformal
algebra does not constrain these kernels.

With the ER-BL kernels we have calculated, we open up a new
way to effectively perform the evolution of singlet distribution
amplitudes and skewed parton distributions by direct numerical integration
of two-loop evolution equations. The numerical algorithms are still to be
developed but a priori it is clear that they will provide a superior
alternative to the ones which rely on the orthogonal reconstruction
approach.

A.B. was supported by the Alexander von Humboldt foundation at the
initial stages of the work. A.F. was supported by the E.U. contract
\# FMRX-CT98-0194.

\appendix

\section{Uniqueness of extension.}
\label{app-extension}

In Eq.\ (\ref{Extension}) we gave a simple recipe for the extension of the
ER-BL kernels into the whole region. Based on the holomorphic properties
of the Fourier transform of $\mbox{\boldmath $\gamma$}(t,t',\eta)$
we give here a more complicated method
which proves the uniqueness of the procedure \cite{MueRobGeyDitHor94}.
Due to the scaling relation (\ref{scaling-rel}) we can restrict ourselves to
$\eta=1$.

First we perform a Fourier transform of the anomalous dimension
$\mbox{\boldmath $\gamma$}(t,t',1)$ with respect to $t$, restricted to
the ER-BL region $|t,t'|\leq 1$. Due to the known support properties (see
fig.\ \ref{fig-sup-lr}b) it is sufficient to restrict only the variable
$t'$, i.e.\ $|t'| \leq 1$ (the anomalous dimension matrix vanishes then for
$|t|>1$). Taking into account the support restrictions arising from the
connection (\ref{conv-LCP->LCF}) with the anomalous dimension of light-ray
operators, we observe that
\begin{eqnarray}
\mbox{{\boldmath ${\widetilde \gamma}$}} (\lambda, t')_{|t'|\leq 1}
&=& \int dt\, e^{i\lambda t}  \mbox{{\boldmath ${\gamma}$}} (t, t')_{|t'|\leq 1}
\\
&=&\int_0^1 dy \int_0^1 dz
\left(
\begin{array}{cc}
{^{QQ}\!\gamma}(y,z) &  i\lambda {^{QG}\!\gamma}(y,z)\\
{ \frac{1}{i\lambda} ^{GQ}\!\gamma}(y,z) & {^{GG}\!\gamma}(y,z)
\end{array}
\right)
e^{i\lambda \left\{t' (1 - y - z) + y - z \right\}}
\nonumber
\end{eqnarray}
is an entire function in $\lambda$ and $t'$ for $|t'| \leq 1$. Thus we can
perform the analytical continuation which coincide with the Fourier transform
of  $\mbox{{\boldmath ${\gamma}$}} (t, t',1)$. This proves the uniqueness
of the extension procedure.

\section{Forward limit.}
\label{app-LIM}

We give now a more heuristic derivation of the forward limit procedure
(\ref{def-LIM}) of the generalized ER-BL kernels. The moments of the kernels
are related to the anomalous dimensions of local twist-2 operators by
\begin{eqnarray}
\label{Mom-ERBL}
\int_0^1 dx\, \left( {x^j\atop  x^{j-1}} \right)
\left(
{{^{QQ}\!V}\ {^{QG}\!V}
\atop
{^{GQ}\!V}\ {^{GG}\!V}}
\right)(x,y)
=
-\frac{1}{2} \sum_{k=0}^{j}
\left(
{{^{QQ}\!\gamma}\ {^{QG}\!\gamma}
\atop
{^{GQ}\!\gamma}\  {^{GG}\!\gamma}}
\right)_{jk} \left( {y^k\atop  y^{k-1}} \right).
\end{eqnarray}
On the other hand the diagonal entries of the anomalous dimensions are
given by the moments of the DGLAP kernels
\begin{eqnarray}
\label{Mom-DGLAP}
\int dz\, z^j
\left(
{{^{QQ}\!P}\ {^{QG}\!P}
\atop
{^{GQ}\!P}\ {^{GG}\!P}}
\right)(z)
=
- \frac{1}{2}\left(
{{^{QQ}\!\gamma}\ {^{QG}\!\gamma}
\atop
{^{GQ}\!\gamma}\  {^{GG}\!\gamma}}
\right)_{jj}.
\end{eqnarray}
To extract the diagonal ones from Eq.\ (\ref{Mom-ERBL}), we substitute
$y$ by $1/\eta$ and multiply each entry on both sides with a
sufficient power of $\eta$. After extension of the kernels, we
can rescale the integration variable and find in the limit $\eta \to 0$:
\begin{eqnarray}
\label{Mom-ERBL-Ext}
\lim_{\eta\to 0}
\int \frac{dx}{|\eta|}\, x^j
\left(
{\ \ {^{QQ}\!V}\ \frac{1}{\eta}{^{QG}\!V}
\atop
\frac{\eta}{x} {^{GQ}\!V}\ \frac{1}{x}{^{GG}\!V}}
\right)^{\rm ext}\left(\frac{z}{\eta},\frac{1}{\eta}\right)
=
-\frac{1}{2}	\left(
{{^{QQ}\!\gamma}\ {^{QG}\!\gamma}
\atop
{^{GQ}\!\gamma}\  {^{GG}\!\gamma}}
\right)_{jj}.
\end{eqnarray}
Interchanging integration and limit and comparison with
Eq.\ (\ref{Mom-DGLAP}) provides us with the desired formula:
\begin{eqnarray}
P(z) = {\rm LIM} V(x,y) \equiv \lim_{\eta\to 0} \frac{1}{|\eta|}
\left(
{\ \ {^{QQ}\!V}\ \frac{1}{\eta}{^{QG}\!V}
\atop
\frac{\eta}{z} {^{GQ}\!V}\  \frac{1}{z}{^{GG}\!V}}
\right)^{\rm ext}\left(\frac{z}{\eta},\frac{1}{\eta}\right).
\end{eqnarray}

Let us look more closely at the forward limit of the +-prescription. In the
QQ-channel the formal prescriptions for exclusive and inclusive channels
are simple related to each other:
\begin{eqnarray}
\label{def-QQ+pre}
{\rm LIM} \left[ {^{QQ}\! V} (x,y)\right]_+
= {\rm LIM} {^{QQ}\! V} (x,y)
- \delta(1-z) \int d z {\rm LIM} {^{QQ}\! V} (x,y)
= \left[ {^{QQ}\! P} (z)\right]_+ .
\end{eqnarray}
However, in the GG-channel an extra $z$ factor appears which induce a
finite term concentrated at $z = 1$:
\begin{eqnarray}
{\rm LIM} \left[ {^{GG}\! V} (x,y)
- \delta(x - y) \int_0^1 dz {^{GG}\! V} (z,y)
\right]
&=& z {^{GG}\! P}(z) - \delta(1-z) \int d z z {^{GG}\! P}(z) \\
&=& z \left[ {^{GG}\! P} (z)\right]_+
+ \delta(1-z) \int d z (1-z) {^{GG} P} (z).
\nonumber
\end{eqnarray}
In order to have the same simple correspondence between the +-prescriptions for
exclusive and inclusive kernels as given in Eq.~(\ref{def-QQ+pre}) for the
QQ channel, we redefine, as implicitly done in our previous papers
\cite{BelFreMue99}, the standard +-definition by a finite part:
\begin{eqnarray}
\label{def-GG+pre}
{\ }\left[ {^{GG}\! V} (x,y) \right]_+
= {^{GG}\! V} (x,y)
- \delta(x-y) \left[ \int dz {^{GG}\! V} (z,y)
+ \int dz (1 - z) {^{GG}\! P} (z) \right].
\end{eqnarray}

If the kernel $v(x,y)$ contains only the usual $\theta$-structure, i.e.
$\theta(y-x)$, the limit yields
\begin{eqnarray}
\label{def-LIM-mat}
{\rm LIM} \mbox{\boldmath$v$}(x,y) &=&
	\theta(1-z)\theta(z) {\rm LIM} \mbox{\boldmath$F$}(x,y),
\\
{\rm LIM} \mbox{\boldmath$F$}(x,y) &\equiv&
\lim_{\eta\to 0} \frac{1}{|\eta|}
\left(
{\ \ ({^{QQ}\!F}-{^{QQ}\!\overline{F}})\
 \frac{1}{\eta}({^{QG}\!F}+{^{QG}\!\overline{F}})
\atop
\frac{\eta}{z} ({^{GQ}\!F}+{^{GQ}\!\overline{F}})\
\frac{1}{z}({^{GG}\!F}-{^{GG}\!\overline{F}})}
\right)\left(\frac{z}{\eta},\frac{1}{\eta}\right).
\nonumber
\end{eqnarray}
In the case of $\theta(y-\bar{x})$-structure, we have to replace in the
above equation $\theta(1-z)\theta(z)$ by $\theta(1+z)\theta(-z)$.

\section{Exclusive convolutions.}
\label{app-convol}

In this appendix we present the exclusive convolution formula.
Due to the two different regions $y > x$ and $y < x$ appearing in the
kernels, the convolution looks more complicated than for the inclusive case.
Fortunately, it is sufficient to consider one of the regions and for the
convolution of two regular kernels,
\begin{eqnarray}
{^{{A}{B}\!}v}_i(x,y) =
\theta(y-x) {^{AB}\!f}(x,y) -	(-1)^{\nu(A)+\nu(B)}
\left\{{x\to \bar x \atop y \to \bar y} \right\}
\quad\mbox{with}\quad i=\{1,2\},
\end{eqnarray}
we use their symmetry and define the convolution of the $f_i(x,y)$
functions by
\begin{eqnarray}
 {^{{A}{B}\!}v}_1\OO^\re{^{{B}{C}\!}v}_2 (x,y)
=
\theta(y-x)  {^{{A}{B}\!}f}_1\OO^\re  {^{{B}{C}\!}f}_2(x,y)  -
(-1)^{\nu(A)+\nu(C)}
\left\{{x\to \bar x \atop y \to \bar y} \right\},
\end{eqnarray}
where
\begin{eqnarray}
{^{{A}{B}\!}f}_1\OO^\re  {^{{B}{C}\!}f}_2(x,y) &\equiv&
\int_x^y dz\, {^{{A}{B}\!}f}_1(x,z) {^{{B}{C}\!}f}_2(z,y)   -
(-1)^{\nu(B)+\nu(C)}
\int_y^1 dz\, {^{{A}{B}\!}f}_1(x,z) {^{{B}{C}\!}f}_2(\bar z,\bar y)
\nonumber\\
& & -(-1)^{\nu(A)+\nu(B)}
\int_0^x dz\, {^{{A}{B}\!}f}_1(\bar x,\bar z) {^{{B}{C}\!}f}_2(z,y).
\end{eqnarray}
In the case that the kernels need a regularization, e.g.\ the conventional
one
\begin{eqnarray}
{^{AB}\!f}_{+}(x,y) =
{^{AB}\!f}(x,y) - \delta(x-y) \int_0^y dz {^{AB}\!f}(z,y),
\end{eqnarray}
we have to arrange the integrals in such a way that each of them is defined.
Since the regularization  only appears in the diagonal channels, we have
to treat the following two cases for convolutions in the mixed channels:
\begin{eqnarray}
{^{AB}\!f}_{1}\otimes {^{BB}\!f}_{2,+} &=&
\int^{y}_{x} dz \left\{f_1(x,z)-f_1(x,y)\right\}f_2(z,y)
+\int^1_y dz \left[f_1(x,z) - f_1(x,y)\right] f_2(\bar z, \bar y)
\nonumber\\
&-&\int^x_0 dz \left\{ f_1(\bar x, \bar z) + f_1(x,y)
\right\} f_2(z,y),
\\
f^{AA}_{1,+}\otimes f^{AB}_{2} &=&
\int^{y}_{x} dz \left[
f_1(x,z)\left\{ f_2(z,y)-f_2(x,y)\right\}
+\left\{f_1(x,z)-f_1(\bar z, \bar x)\right\}f_2(x,y)\right]
\nonumber\\
&-& \int^1_y dz
\left[f_1(x,z) f_2(\bar z,\bar y) + f_1(\bar z,\bar x) f_2(x,y)\right]
\\
&+&\int^x_0 dz \left[
f_1(\bar x,\bar z) \left\{f_2(z,y)-f_2(x,y)\right\}
+ \left\{f_1(\bar x,\bar z) - f_1(z,x)\right\} f_2(x,y) \right],
\nonumber
\end{eqnarray}
where $A\not= B$ and $f_{i}\equiv \left\{f^{AA}_{i}, f^{AB}_{i}\right\} $
and each term is separately integrable. The corresponding equations for the
convolution of the diagonal kernels follows by changing the sign in the
third or second integral, respectively. For the convolution of two kernels
with "+"-prescription we write
\begin{eqnarray}
\left[{^{{AA}\!}v}_1\right]_+ \OO^\re \left[{^{{AA}\!}v}_2\right]_+ (x,y)
=
\left[\theta(y-x) {^{{AA}\!}f}_1 \OO^\re {^{{AA}\!}f}_2 (x,y) +
 \left\{{x\to \bar x \atop y\to \bar y } \right\}  \right]_+ ,
\end{eqnarray}
where the convolution on the r.h.s.\ is given by
\begin{eqnarray}
f^{AA}_{1}\otimes f^{AA}_{2}
&=&\int^{y}_{x} dz \left\{
\left[f_1(x,z)-f_1(x,y) \right]\left[f_2(z,y)-f_2(x,y)\right]
+\left[f_1(x,z)-f_1(\bar z, \bar x)\right]f_2(x,y) \right\}
\nonumber\\
&+& \int^1_y dz \left\{
\left[f_1(x,z)-f_1(x,y)\right]f_2(\bar z,\bar y) -
\left[f_1(\bar z, \bar x)-f_1(x,y)\right]f_2(x,y) \right\}
\\
&+&\int^x_0 dz \left\{
\left[f_1(\bar x, \bar z)-f_1(x,y) \right]\left[f_2(z,y)-f_2(x,y)\right]
+\left[f_1(\bar x, \bar z) - f_1(z,x)\right] f_2(x,y)\right\}.
\nonumber
\end{eqnarray}


\begin{thebibliography}{99}
\bibitem{MueRobGeyDitHor94}
D. M\"uller, D. Robaschik, B. Geyer, F.-M. Dittes, J. Ho{\v r}ej{\v s}i,
Fortschr. Phys. 42 (1994) 101.
\bibitem{Ji96}
X. Ji, Phys. Rev. D 55 (1997) 7114.
\bibitem{Rad97}
A.V. Radyushkin, Phys. Rev. D 56 (1997) 5524.
\bibitem{ColFraStr96}
J.C. Collins, L. Frankfurt, M. Strikman, Phys. Rev. D 56 (1997) 2982.
\bibitem{Rad96}
A.V. Radyushkin, Phys. Lett. B 385 (1996) 333.
\bibitem{ColFre98}
J.C. Collins, A. Freund, Phys. Rev. D 59 (1999) 074009.
\bibitem{JiOsb98}
X. Ji, J. Osborne, Phys. Rev. D 58 (1998) 094018.
\bibitem{GeyDitHorMueRob88}
F.-M. Dittes, B. Geyer, D. M\"uller, D. Robaschik, J. Ho{\v r}ej\v{s}i,
Phys. Lett. B 209 (1988) 325.
\bibitem{EfrRad78}
A.V. Efremov, A.V. Radyushkin, Theor. Math. Phys. 42 (1980) 97;
Phys. Lett. B 94 (1980) 245.
\bibitem{BroLep79}
S.J. Brodsky, G.P. Lepage, Phys. Lett. B 87 (1979) 359; Phys. Rev. D
22 (1980) 2157.
\bibitem{Cha80a}
M.K. Chase, Nucl. Phys. B 174 (1980) 109.
\bibitem{Ohr81}
Th. Ohrndorf, Nucl. Phys. B 186 (1981) 153.
\bibitem{BaiGro81}
V.N. Baier, A.G. Grosin, Nucl. Phys. B 192 (1981) 476.
\bibitem{BukFroKurLip85}
A.P. Bukhvostov, G.V. Frolov, E.A. Kuraev, L.N. Lipatov, Nucl. Phys.
B 258 (1985) 601.
\bibitem{Sar84}
M.H. Sarmadi, Phys. Lett. B 143 (1984) 471.
\bibitem{DitRad84}
F.-M. Dittes, A.V. Radyushkin, Phys. Lett. B 134 (1984) 359.
\bibitem{MikRad85}
S.V. Mikhailov, A.V. Radyushkin, Nucl. Phys. B 254 (1985) 89.
\bibitem{FraFreGuzStr97}
L. Frankfurt, A. Freund, V. Guzey, M. Strikman, Phys. Lett. B 418 (1998)
345; (E) ibid. B 429 (1998) 414.
\bibitem{MusRad99}
I.V. Musatov, A.V. Radyushkin, {\it Evolution and models for skewed
parton distributions}, hep-ph/9905376.
\bibitem{Beletal97}
A.V. Belitsky, B. Geyer, D. M\"uller, A. Sch\"afer, Phys. Lett. B 421
(1998) 312.
\bibitem{ManPilWei97}
L. Mankiewicz, G. Piller, T. Weigl, Eur. J. Phys. C 5 (1998) 119.
\bibitem{Cha80}
M.K. Chase, Nucl. Phys. B 167 (1980) 125.
\bibitem{ShiVys81}
M.A. Shifman, M.I. Vysotsky, Nucl. Phys. B 186 (1981) 475.
\bibitem{Shu99}
A.G. Shuvaev, Phys. Rev. D 60 (1999) 116005.
\bibitem{BalBra89}
I.I. Balitsky, V.M. Braun, Nucl. Phys. B 311 (1989) 541.
\bibitem{KivMan99}
N. Kivel, L. Mankiewicz, Phys. Lett. B 443 (1999) 308; Nucl. Phys. B
??? (1999) ???.
\bibitem{Mue97a}
D. M\"uller, Phys. Rev. D 58 (1998) 054005.
\bibitem{BelMue97a}
A.V. Belitsky, D. M\"uller, Phys. Lett. B 417 (1997) 129.
\bibitem{BelMue98a}
A.V. Belitsky, D. M\"uller, Nucl. Phys. B 527 (1998) 207.
\bibitem{Mue94}
D. M\"uller, Phys. Rev. D 49 (1994) 2525.
\bibitem{BelMue98c}
A.V. Belitsky, D. M\"uller, Nucl. Phys. B 537 (1998) 397.
\bibitem{BelMueNieSch98a}
A.V. Belitsky, D. M\"uller, L. Niedermeier, A. Sch\"afer, Phys. Lett.
B 437 (1998) 160.
\bibitem{BelMueNieSch98b}
A.V. Belitsky, D. M\"uller, L. Niedermeier, A. Sch\"afer, Nucl. Phys.
B 546 (1999) 279.
\bibitem{MikRad86}
S.V. Mikhailov, A.V. Radyushkin, Nucl. Phys. B 273 (1986) 297.
\bibitem{Mik97}
S.V. Mikhailov, Phys. Lett. B 416 (1997) 421.
\bibitem{GeyRobBorHor85}
B. Geyer, D. Robaschik, M. Bordag, J. Ho{\v r}ej{\v s}i, Z. Phys. C 26
(1985) 591.
\bibitem{BraGeyHorRob87}
T. Braunschweig, B. Geyer, J. Ho{\v r}ej{\v s}i, D. Robaschik, Z. Phys.
C 33 (1987) 275.
\bibitem{AdlColDun77}
S.L. Adler, J.C. Collins, A. Duncan, Phys. Rev. D 15 (1977) 1712.
\bibitem{ColDunJog77}
J.C. Collins, A. Duncan, S.D. Joglekar, Phys. Rev. D 16 (1977) 438.
\bibitem{Nie77}
N.K. Nielsen, Nucl. Phys. B 120 (1977) 212.
\bibitem{Symanzik71}
K. Symanzik, Comm. Math. Phys. 23 (1971) 49;\\
B. Schroer, Nuovo Cim. Lett. 2 (1971) 627;\\
G. Parisi, Phys. Lett. B 39 (1972) 643.
\bibitem{BE53_1}
H. Bateman, A. Erd\'elyi, {\it Higher transcendental functions}, V. 1,
Mc Graw-Hill, New York (1953).
\bibitem{Mue91a}
D. M\"uller, Z. Phys. C 49 (1991) 293.
\bibitem{BelMul99s}
A.V. Belitsky, D. M\"uller, {\it Superconformal constraints for QCD
conformal anomalies}, preprint YITP-SB-99-63.
\bibitem{FloKouLac81}
E.G. Floratos, C. Kounnas, R. Lacaze, Nucl. Phys. B 192 (1981) 417.
\bibitem{MerNee96}
R. Mertig, W.L. van Neerven, Z. Phys. C 70 (1996) 637.
\bibitem{Vog96}
W. Vogelsang, Phys. Rev. D 54 (1996) 2023; Nucl. Phys. B475 (1996) 47.
\bibitem{FurPet82}
W. Furmanski, R. Petronzio, Phys. Lett. B97 (1980) 437.
\bibitem{Dok77}
Yu.L. Dokshitzer, Sov. Phys. JETP 46 (1977) 641.
\bibitem{BelMueSch98}
A.V. Belitsky, D. M\"uller, A. Sch\"afer, Phys. Lett. B 450 (1999) 126.
\bibitem{CurFurPet80}
G. Curci, W. Furmanski, R. Petronzio, Nucl. Phys. B 175 (1980) 27.
\bibitem{BelFreMue99}
A.V. Belitsky, A. Freund, D. M\"uller, Phys. Lett. B 461 (1999) 270;\\
A.V. Belitsky, D. M\"uller, Phys. Lett. B 464 (1999) 249.
\end{thebibliography}
\end{document}